\UseRawInputEncoding
\documentclass[12pt,draftcls,onecolumn]{IEEEtran}
\usepackage{amsfonts}
\usepackage{amssymb}
\usepackage{amsmath}
\usepackage{mathrsfs}
\usepackage{subfigure}
\usepackage{cite}
\usepackage{graphicx}
\usepackage{epstopdf}
\usepackage{color}
\usepackage{colortbl,booktabs}
\usepackage{cite}
\usepackage{multirow}
\usepackage{algorithm}
\usepackage{algorithmic}

\makeatletter
\newenvironment{breakablealgorithm}
  {
   \begin{center}
     \refstepcounter{algorithm}
     \hrule height.8pt depth0pt \kern2pt
     \renewcommand{\caption}[2][\relax]{
       {\raggedright\textbf{\ALG@name~\thealgorithm} ##2\par}%
       \ifx\relax##1\relax 
         \addcontentsline{loa}{algorithm}{\protect\numberline{\thealgorithm}##2}%
       \else 
         \addcontentsline{loa}{algorithm}{\protect\numberline{\thealgorithm}##1}%
       \fi
       \kern2pt\hrule\kern2pt
     }
  }{
     \kern2pt\hrule\relax
   \end{center}
  }
\makeatother

\makeatletter

\newcommand\old[1]{}
\newtheorem{definition}{Definition}
\newtheorem{lemma}{Lemma}
\newtheorem{corollary}{Corollary}

\newtheorem{theorem}{Theorem}
\newtheorem{remark}{Remark}

\newtheorem{example}{Example}
\newtheorem{assumption}{Assumption}

\newcommand{\RNum}[1]{\uppercase\expandafter{\romannumeral #1\relax}}
\makeatother

\begin{document}
\date{\today}
\title{Workflow-based Fast Data-driven Predictive Control with Disturbance Observer in Cloud-edge Collaborative Architecture}
\author{Runze Gao, Qiwen Li, Li Dai, Yufeng Zhan and Yuanqing Xia$^{*}$
\thanks{R. Gao, Q. Li, L. Dai, Y. Zhan, Y. Xia are with the School of Automation, Beijing Institute of Technology, Beijing 100081, China. ({\footnotesize {\em Corresponding
author: Yuanqing Xia}). Email address:
gaorunze$\_$bit@163.com (R. Gao), penguinlee@bit.edu.cn (Q. Li), daili1887@bit.edu.cn (L. Dai), yu-feng.zhan@bit.edu.cn (Y. Zhan), xia$\_$yuanqing@bit.edu.cn (Y. Xia)
}}}
\maketitle
\begin{abstract}
Data-driven predictive control (DPC) has been studied and used in various scenarios, since it could generate the predicted control sequence only relying on the historical input and output data. Recently, based on cloud computing, data-driven predictive cloud control system (DPCCS) has been proposed with the advantage of sufficient computational resources. However, the existing computation mode of DPCCS is centralized. This computation mode could not utilize fully the computing power of cloud computing, of which the structure is distributed. Thus, the computation delay could not been reduced and still affects the control quality. In this paper, a novel cloud-edge collaborative containerised workflow-based DPC system with disturbance observer (DOB) is proposed, to improve the computation efficiency and guarantee the control accuracy. First, a construction method for the DPC workflow is designed, to match the distributed processing environment of cloud computing. But the non-computation overheads of the workflow tasks, such as receiving and sending data, serialization and deserialization, are relatively high and affect the overall processing speed of workflow. Therefore, a cloud-edge collaborative control scheme with DOB is designed. The low-weight data could be truncated to reduce the non-computation overheads. Meanwhile, we design an edge DOB to estimate and compensate the uncertainty caused by the truncation operation in cloud workflow processing, and obtain the cloud-edge composite control variable. The uniformly ultimately bounded stability of the DOB is also proved. Third, to execute the workflow-based DPC controller and evaluate the proposed cloud-edge collaborative control scheme with DOB in the real cloud environment, we design and implement a practical workflow-based cloud control experimental system based on container technology. Finally, a series of evaluations show that, the computation times are decreased by 45.19$\%$ and 74.35$\%$ for two real-time control examples, respectively, and by at most 85.10$\%$ for a high-dimension control example.
\end{abstract}
\section{Introduction}

With the development of Internet of Things and cyber-physical system, the scale and complexity of the control system increase rapidly \cite{xia2012networked}. This brings challenges for the control systems, of which the computing resources are limited and fixed. Cloud computing is a powerful solution, which outsources the workloads to virtual servers to improve the computation efficiency \cite{marinescu2017cloud, senyo2018cloud, alexandru2020cloud}. Cloud computing has advantages of sufficient computational resources, utilizing global information, handling large-scale data \cite{tanaka2017directed}, which has been used in multi-agent, smart grids, autonomous vehicles, etc \cite{liu2017predictive, xia2020cloud, wang2021cloud}. Based on cloud computing, cloud control system is concerned as a new paradigm to design and implement the control framework \cite{xia2015cloud, mahmoud2017interaction, xia2022brief}. Since the introduction of cloud computing, more controlled plants are involved, and more data are produced and stored. It is hard even impossible to build the exact model of each plant, and the model-based control methods might be not applied. In the meanwhile, data-driven predictive control (DPC) method is developed to provide the control service without the need of the prior knowledge of system model. Therefore, data-driven predictive cloud control system (DPCCS) is proposed, to deal with the control problems only relying on the historical input and output data in the cloud control environment \cite{gao2017new, gao2021design}.

\begin{figure}[!ht]
  \centering
  \includegraphics[width=5in, trim=0 40 0 0,clip]{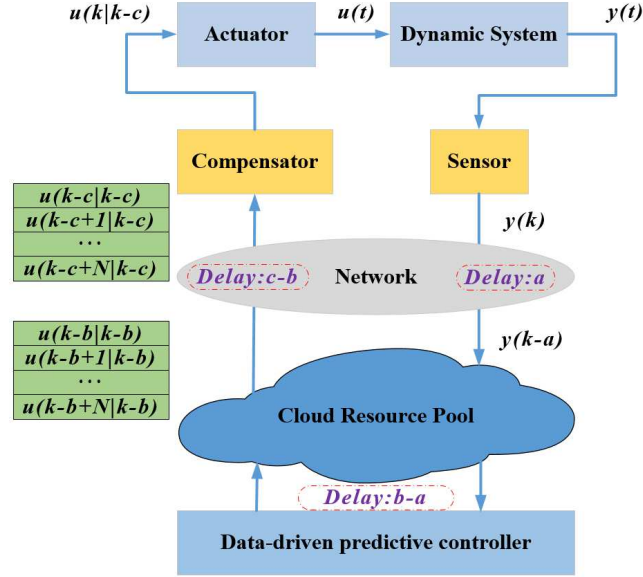}\\
  \caption{Typical Structure of the DPCCS}\label{Data-driven Predictive Cloud Control System}
\end{figure}

The typical structure of DPCCS is shown in Fig. \ref{Data-driven Predictive Cloud Control System}. The latest input and output data are collected and uploaded to the cloud server. The DPC controller is deployed in the cloud server and produces the data-driven cloud predictive control sequence based on the received data. Then, the cloud predicted control sequence is downloaded to the local controlled plant and a complete control circle is finished. Time delay occurs in all the stages of DPCCS, including the upload, computation and download courses. To reduce the influence of time delay, a compensator is designed to select a proper control variable, from the received cloud predicted control sequence based on the total time delay. The total time delay is measured as the difference between the recorded times of sending the sensed data and receiving the cloud signals.

However, the traditional scheme with time delay compensator is a passive solution to compensate the influence of time delay. In the traditional DPCCS, the computation mission of the native DPC method is deployed in a single cloud server directly, on which the computation mode is centralized. This kind of computation mode could not make full use of the parallel computing ability of cloud computing, while the distributed processing structure is one of the core advantages of cloud computing. Besides, only the power of cloud computing has been considered, the computing ability of edge device and the potential of the cloud-edge collaborative scheme have not been studied for DPCCS. The above reasons lead that time delay has not been reduced significantly, and would still have serious effects on the control quality. The detailed shortcomings of the traditional DPCSS are listed in the below.
\begin{enumerate}
  \item The time delay compensator works only after time delay occurring, but has no effect on actively reducing the computation time.
  \item When the total time delay is greater than a control period, a later ordered variable would be selected as the control input by the time delay compensator. But the variable ranking behind would bring lager control disturbance because of the accumulation of prediction errors.
  \item In real-time control, the computation period for calculating control algorithm is constrained. Thus, the amount of the data and the scales of the Hankel matrices are restricted, which would affect the real-time control quality, especially for the high-dimension problem.
\end{enumerate}

\begin{figure}[h]
  \centering
  \includegraphics[height=2in]{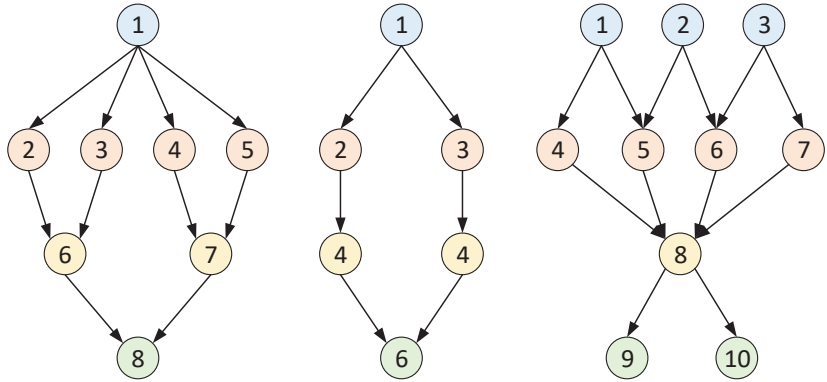}\\
  \caption{Examples of the Cloud Workflow Structure}\label{Examples of the Cloud Workflow Structure}
  \vspace{-0.9em}
\end{figure}

In this work, a novel cloud-edge collaborative workflow-based DPC system is proposed to actively reduce the computation delay, by utilizing the distributed ability of cloud computing. Cloud workflow processing has proved effective for improving the computation efficiency of scientific missions, such as deep learning, genetic calculation, etc, as it decomposes a complex, data-intensive application into smaller tasks and executes the tasks in the cloud environment \cite{ye2021shws, chen2016parallel, xiao2020malfcs, gao2021fast}. As shown in Fig. \ref{Examples of the Cloud Workflow Structure}, cloud workflows are represented by the directed acyclic graphs (DAGs). The nodes represent the computational tasks, and the directed edges between the nodes determine the interdependencies between the tasks. Therefore, we propose the workflow-based DPC method, in which the DPC method with the centralized computation mode is restructured into the workflow form, based on the distributed numerical computation algorithm. When the workflow-based cloud controller starts, the tasks in the DPC workflow could be executed in parallel to reduce the computation delay in cloud control system. 

\begin{figure}[h]
  \centering
  \includegraphics[width=4in]{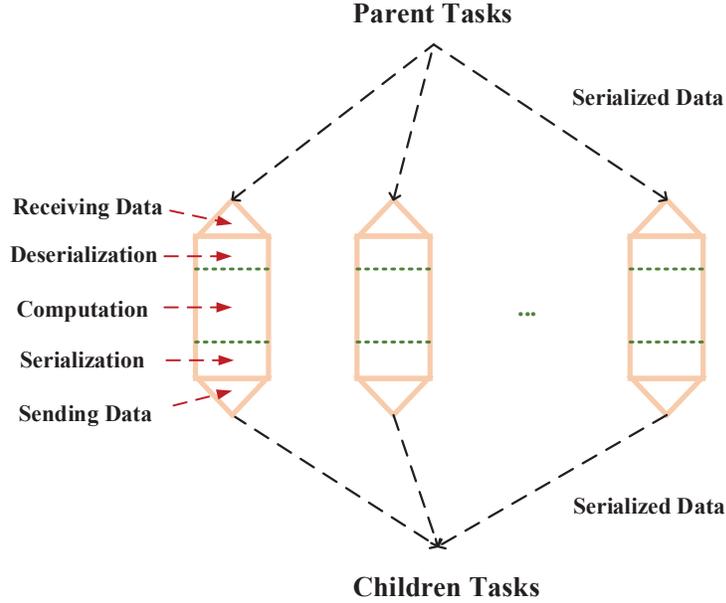}\\
  \caption{Structure of a Middle-tier Task in Workflow}\label{Structure of a Workflow Task}
  \vspace{-0.9em}
\end{figure}

However, in the DPC workflow tasks, the non-computation overheads are relatively high. As shown in Fig. \ref{Structure of a Workflow Task}, a middle-tier task is made up by the computation part and four non-computation parts, including the receiving data, deserialization, serialization and sending data parts. The computation part is the designed main function of the DPC workflow task, which is divided from the complete DPC mission. The parts of receiving and sending data are used to connect with the parent and children tasks. In the computation part, the data formats are '\emph{float}', '\emph{double}', etc. But, these data formats could not be delivered in the network space. Thus, the operations of serialization and deserialization are required, to convert the data formats in the computation part with the serialized data formats, such as '\emph{bytes}', '\emph{string}', '\emph{json}', etc. Since DPC is a control method relying on analyzing data, there are a lot of data transmitting between the DPC workflow tasks. Therefore, the overheads of receiving and sending data, serialization and deserialization are heavy, and have no benefits for reducing the completion time of the workflow-based DPC mission.

Next, to reduce the non-computational overheads and guarantee the control accuracy in the meanwhile, a cloud-edge collaborative scheme with disturbance observer (DOB) is proposed. In the DPC method, there are a lot of matrix operations on the large-scale Hankel matrices. The involved Hankel matrices are high-dimension and low-rank \cite{gao2021fast}. That is to say, most of the matrix information could be truncated. Then, the data amount of network transmission, serialization and deserialization could be decreased by the truncation operation, and so the non-computational overheads are reduced. However, the truncation operation would bring some uncertainties in the workflow processing. The uncertainty event would cause adverse effects on the control performance, which is not what we expected. DOB is a disturbance estimation module, in which an auxiliary dynamical system estimates the effects of uncertain acting on a system \cite{chen2015disturbance}. Once the uncertain is estimated, it is possible to compensate for it by using the right magnitude of control effort. Therefore, we propose the cloud-edge collaborative workflow-based DPC system with DOB to solve this problem. In this system, the workflow-based DPC controller is deployed in the cloud environment, and a DOB is designed close to the edge controlled plant, The DOB is used to estimate the uncertainty produced in the workflow processing, and generate the cloud-edge composite control variable based on the received cloud control signals.

Finally, a workflow-based cloud control experimental system in practice is also required to conduct the DPC workflow, and evaluate the results of computation acceleration and control. In the cloud environment, the tasks in a workflow would be scheduled to different processing nodes, in which the required computing resources are encapsulated from a shareable resource pool. Cloud workflow processing based on virtual machine (VM) is the major approach at present \cite{zhu2018scheduling, mao2019learning}. VM technology provides a virtual operation system to access the computing resources, and create the isolated spaces for the tasks. However, the operation costs of VM are quite high, including the creation, startup, configuring times and configuring complexity. For example, the creation and startup stages of VM cost several minutes and 30-40 s, respectively. To reduce the costs, container technology is created and applied \cite{kaur2017container, goldschmidt2018container, mellado2020container}. This technology enables that the smaller isolated processes could be created as well as started in an instant, and released automatically when the tasks are finished. Compared with VM technology, container also has potential benefits on the configuring complexity, computing resource usage, overhead cost, migration speed across hosts and template reuse cost, as summarized in Table \ref{Benefits of Container Technology} \cite{ranjan2020energy}.


\begin{table}[htbp]
	\centering
	\caption{Benefits of Container Technology}
	\begin{tabular}{|l|c|c|}
		\toprule  
		\textbf{Parameter}&\textbf{Container technology}&\textbf{VM technology} \\
		\midrule  
		Kernel state&Booting needed&Operational \\
        \hline
        Creation time&$<$ 1 s&10 min \\
        \hline
        Start/stop time&$<$ 50 ms&30-40 and 5-10 s \\
        \hline
        Startup overhead&Low&Relatively high \\
        \hline
        Configuring complexity&Midden&High \\
        \hline
        CPU/memory usage&Low&High \\
        \hline
        Overhead costs&Low&High \\
        \hline
        Migration&Quick&Slow \\
        \hline
        Template reuse cost&Low&High \\
		\bottomrule  
	\end{tabular}
\label{Benefits of Container Technology}
\end{table}

Motivated by the above reasons, the main contributions of this paper on the novel cloud-based DPC method are summarized in the following.
\begin{enumerate}
  \item A workflow-based DPC method is proposed to tackle the issue of accelerating the data-intensive control computation missions. Through the truncated distributed singular value decomposition (SVD) algorithm, we develop a method of restructuring the DPC algorithm into DAG, which is the cloud workflow form.
  \item A cloud-edge collaborative control scheme of the cloud accelerated controller and edge DOB-based uncertainty compensator is designed. To further improve the computation speed, the truncation precision could be adjusted. The upper bound of cloud control variable error is proved based on the truncation precision. Then, DOB is designed to estimate and compensate the uncertainty. The uniformly ultimately bounded (UUB) stability of the error dynamics of this DOB is guaranteed. The practical algorithm of this cloud-edge collaborative control scheme is also provided.
  \item A practical containerised workflow-based cloud control experimental system is designed and implemented to carry out the workflow-based DPC missions. Specifically, the initialization, execution and communication schemes of the system are designed and established.
  \item Finally, three control examples, consisting of two real-time systems and one high-dimension numerical system, are conducted to verify the performance of the proposed workflow-based DPC method and the containerised  worklfow-based cloud control experimental system. In the real-time control examples, which are the ball-beam system and vehicle tracking system, the computation times are reduced by 45.19$\%$ and 74.35$\%$, respectively. In the high-dimension numerical example, the computation time is reduced by 85.10$\%$ at most.

\end{enumerate}

The rest of this paper is organized as follow. Section \ref{Preliminary of Data-driven Predictive Control} presents the preliminary of the DPC method. Section \ref{Overview of workflow-based data-driven predictive cloud control} provides the overview of the proposed cloud-edge collaborative containerised workflow-based DPC system with DOB. The computational features analyses and the construction method of the DPC workflow are presented in Section \ref{Establishment method of data-driven predictive control workflow}. The DOB and the cloud-edge collaborative control scheme with DOB are proposed in \ref{DOB-based cloud-edge collaborative control scheme}. The design and establishment of the containerised workflow-based cloud control experimental system are provided in Section \ref{Design and implementation of containerised cloud control testbed}. The evaluations and discussions are provided in Section \ref{Experiment and discussion}. The final section presents the conclusion and the future work.

\section{Preliminary of Data-driven Predictive Control}\label{Preliminary of Data-driven Predictive Control}

In this preliminary, the DPC method is recalled \cite{xia2013data, huang2008dynamic}. Suppose the measurements of the inputs $\{ u(n), n = 0, 1,\ldots, 2N+j-2\}$ and outputs $\{y(n), n = 1, 2,\ldots, 2N+j-1\}$ are available. The data block Hankel matrices for $u(n)$, represented as $U_p$ and $U_f$, with $N$-block rows and $j$-block columns are defined as
\begin{eqnarray}\label{dpc1}
  \!\!\!\!\!\!U_p \!\!\!\!&=& \!\!\!\!\!\!\left[
            \begin{array}{cccc}
              \!\!u(0) & \!\!u(1) & \!\!\ldots & \!\!u(j-1) \\
              \!\!u(1) & \!\!u(2) & \!\!\ldots & \!\!u(j) \\
              \!\!\vdots & \!\!\vdots & \!\!\ddots & \!\!\vdots \\
              \!\!u(N-1) & \!\!u(N) & \!\!\ldots & \!\!u(N+j-2) \\
            \end{array}
          \right], \\
  \!\!U_f \!\!\!\!&=& \!\!\!\!\!\!\left[
            \begin{array}{cccc}
              \!\!\!u(N) & \!\!\!u(N+1) & \!\!\!\ldots & \!\!\!u(N+j-1) \\
              \!\!\!u(N+1) & \!\!\!u(N+2) & \!\!\!\ldots & \!\!\!u(N+j) \\
              \!\!\!\vdots & \!\!\!\vdots & \!\!\!\ddots & \!\!\!\vdots \\
              \!\!\!u(2N-1) & \!\!\!u(2N) & \!\!\!\ldots & \!\!\!u(2N+j-2) \\
            \end{array}
          \right].
\end{eqnarray}

Each block element in the above Hankel matrices is a column vector of the input. Similarly, the data Hankel matrices for $y(n)$, represented as $Y_p$ and $Y_f$, can be written. Following the above subspace notations, the data column vectors over a prediction horizon for the inputs and outputs, respectively, are similarly defined as
\begin{equation}\label{dpc2}
  u_f(k) \!=\! \left[\!
          \begin{array}{c}
            u(k) \\
            u(k\!+\!1) \\
            \vdots \\
            u(k\!+\!N\!-\!1) \\
          \end{array}\!
        \right]\!, \, y_f(k) \!=\! \left[\!
          \begin{array}{c}
            y(k+1) \\
            y(k\!+\!2) \\
            \vdots \\
            y(k\!+\!N) \\
          \end{array}\!
        \right]\!, \,
\end{equation}
\begin{equation}\label{dpc3}
  u_{p}(k) \!=\! \left[\!
          \begin{array}{c}
            u(k-\!N) \\
            u(k-\!N\!+\!1) \\
            \vdots \\
            u(k\!-\!1) \\
          \end{array}\!
        \right]\!, \, y_{p}(k) \!=\! \left[\!
          \begin{array}{c}
            y(k\!-\!N\!+\!1) \\
            y(k\!-\!N\!+\!2) \\
            \vdots \\
            y(k) \\
          \end{array}\!
        \right]\! \,.
\end{equation}
Then, we define the intermediate variable $w_p$ as
\begin{equation}\label{dpc4}
  w_p(k) = \left[
             \begin{array}{c}
               y_{p}(k) \\
               u_{p}(k) \\
             \end{array}
           \right]\!.
\end{equation}

Using regression analysis approach \cite{xia2013data}, we obtain
\begin{equation}\label{dpc5}
  y_{f}(k)=L_{w}w_{p}(k)+L_{u}u_{f}(k)+L_{e}e_{f}(k)
\end{equation}
where $e_{f}(k)$ is the column vector consisting of the noise $\{ e(n), n = k, k+1,\ldots,k+N-1\}$, and $L_w$, $L_u$, $L_e$ are the corresponding coefficient matrices with proper dimensions. Since $e_f(k)$ consists of white noise, the optimal prediction $\widehat{y}_f(k)$ is given as
\begin{equation}\label{dpc6}
  \widehat{y}_{f}(k)=L_{w}w_{p}(k)+L_{u}u_{f}(k).
\end{equation}
Then, $L_{w}$ and $L_{u}$ are required to calculate $\widehat{y}_f(k)$. In the DPC method, the Hankel matrices are used to compute $L_{w}$ and $L_{u}$. Similar with $w_p$, let
\begin{equation}\label{dpc7}
  W_p=\left[
           \begin{array}{c}
             Y_p \\
             U_p \\
           \end{array}
         \right].
\end{equation}

The Hankel matrix of the predicted outputs, $\widehat{Y}_{f}$ is computed as
\begin{equation}\label{dpc8}
  \widehat{Y}_{f}=L_{w}W_{p}+L_{u}U_{f}.
\end{equation}
By solving the following least squares problem
\begin{equation}\label{dpc9}
  \min_{L_{w},L_u}\|Y_{f}-\left[
                           \begin{array}{cc}
                             \!\!\!L_{w} & \!\!\!L_{u} \\
                           \end{array}\!\!
                         \right]\!\left[
                                  \!\begin{array}{c}
                                    W_{p} \\
                                    U_{f} \\
                                  \end{array}\!\!
                                \right]\!
\|^{2}_{F}
\end{equation}
where $\|\cdot\|_{F}$ means $F$-norm, the coefficient matrices $L_w$ and $L_u$ could be computed by
\begin{eqnarray}\label{dpc10}
  \left[
                           \!\!\begin{array}{cc}
                            L_{w} & \!\!\!\!L_{u} \\
                          \end{array}\!\!
                        \right] \!\!\!\!&=& \!\!\!\!Y_{f}\left[
                                 \!\!\begin{array}{c}
                                   W_{p} \\
                                   U_{f} \\
                                 \end{array}\!\!
                               \right]^{\dag}.
\end{eqnarray}
where $\dag$ is the Morre-Penrose pesudo-inverse. Since $\left[
                                 \!\!\begin{array}{c}
                                   W_{p} \\
                                   U_{f} \\
                                 \end{array}\!\!
                               \right]$ may be not a square matrix, the Morre-Penrose pesudo-inverse is used. Finally, consider an unconstrained MPC problem with the objective function
\begin{equation}\label{dpc11}
  J=(r_{f}-\widehat{y}_{f}(k))^{T}(r_{f}-\widehat{y}_{f}(k))+u_{f}(k)^{T}(\lambda I)u_{f}(k)
\end{equation}
where $r_f$ is the reference value of the output, $\lambda$ is an adjustable coefficient and $I$ is a unit matrix. Substituting (\ref{dpc6}) into (\ref{dpc11}) and taking derivative with respect to $u_f$, the data-driven predictive control law without constraints can be computed as
\begin{equation}\label{dpc12}
  u_{f}(k)=(\lambda I+L_{u}^{T}L_{u})^{-1}L_{u}^{T}(r_{f}-L_{w}w_{p}(k)).
\end{equation}

\section{Overview}\label{Overview of workflow-based data-driven predictive cloud control}

\begin{figure*}[t!]
  \centering
  \includegraphics[width=6in]{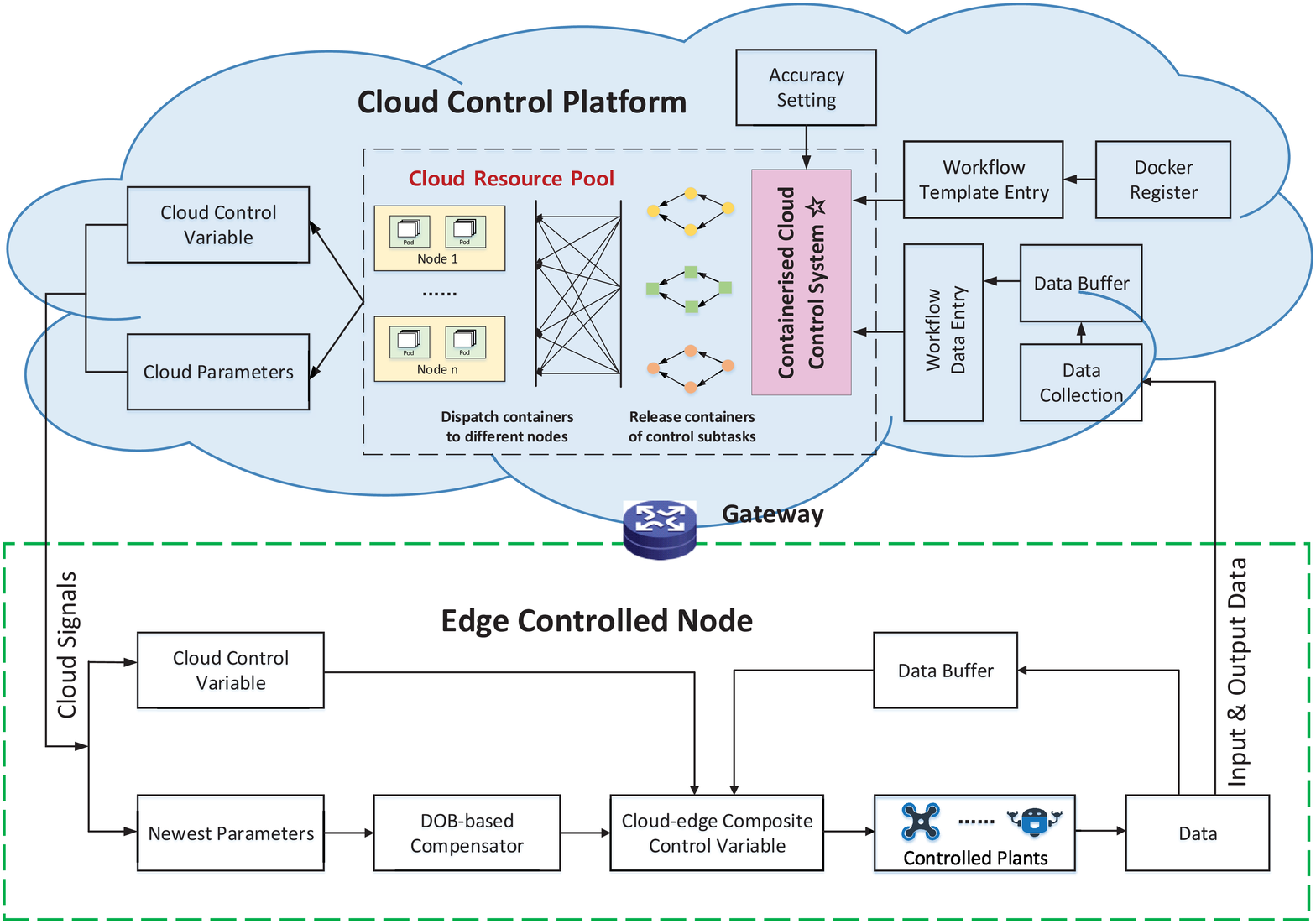}\\
  \caption{Structure of the Workflow-based Fast DPC System with DOB in Cloud-edge Collaborative Architecture}\label{Architecture of Data-driven Predictive Cloud Control System}
\end{figure*}

The structure of the workflow-based fast DPC system with DOB in cloud-edge collaborative architecture is provided as Fig. \ref{Architecture of Data-driven Predictive Cloud Control System}. This structure is made up of cloud control platform and the edge controlled node. Some important modules are described in Table \ref{Descriptions of part defined modules}.The cloud control platform is in the upper part of the structure. After receiving the input and output data from the edge controlled node, data are stored in the cloud data buffer. In the execution of the DPC workflow, the data buffer is kept as an unchanged scale by deleting the oldest data and adding the latest data. Then, the stored data are transmitted to the workflow data entry. In the meanwhile, the template of the DPC workflow is pulled to the workflow template entry from the Docker register. Then, the data and workflow structure as well as the task images are received by the containerised cloud control system, which is designed in Section \ref{Design and implementation of containerised cloud control testbed}. The truncation accuracy of the DPC workflow can be set to adjust the non-computation cost and overall computation efficiency. This system creates containers and dispatches them to the different computing nodes. Finally, the cloud signals consisting of the control variable and necessary parameters are sent to the edge controlled node.

\begin{table}[htbp]
	\centering
	\caption{Descriptions of part defined modules in Fig. \ref{Architecture of Data-driven Predictive Cloud Control System}}
    \renewcommand{\multirowsetup}{\centering}
	\begin{tabular}{|c|c|}
		\toprule  
		\textbf{Module}&\textbf{Description} \\
		\hline  
        \hline
        \multirow{2}{2.5cm}{Cloud Resource Pool}&\multirow{2}{7.5cm}{An abstract concept for describing all computing resource including physical server, VM and container}\\
        & \\
        \hline
        \multirow{2}{2.5cm}{Docker Register}&\multirow{2}{7.5cm}{A sharing warehouse for storing reusable images of pre-defined tasks in DPC method}\\
        & \\
        \hline
        \multirow{2}{2.5cm}{Workflow Template}&\multirow{2}{7.5cm}{The template consisting of the task images and the topology structure between the tasks}\\
        & \\
        \hline
        \multirow{2}{2.5cm}{Workflow Template Entry}&\multirow{2}{7.5cm}{A middle module to provide the template of the DPC workflow pulled from Docker Register}\\
        & \\
        \hline
        \multirow{2}{2.5cm}{Workflow \\ Data Entry}&\multirow{2}{7.5cm}{A middle module to provide the input and output data of the identified plant for the DPC workflow}\\
        & \\
		\bottomrule  
	\end{tabular}
\label{Descriptions of part defined modules}
\end{table}

The edge controlled node is in the lower part of this structure. In the edge controlled node, the received cloud parameters are used to build the DOB, which is described in Section \ref{DOB-based cloud-edge collaborative control scheme}. The uncertainty caused by the truncation operation is considered as the input disturbance. Meanwhile, there also exists a data buffer in the edge controlled node, to provide the historical data for calculating the DOB-based compensative control variable. Then, the cloud-edge composite control variable is obtained, based on the cloud and edge DOB-based compensative control variables. Next, the obtained cloud-edge composite control variable is applied to the controlled plant. Finally, the latest input and output data are used to update the edge data buffer. Up to here, the fast and accurate workflow-based data-driven predicted control variable is obtained based on the designed cloud-edge collaborative control scheme.


\section{Construction Method of the DPC Workflow}\label{Establishment method of data-driven predictive control workflow}

To match the distributed environment of cloud computing, the construction method of the DPC workflow is proposed. First, the computational features of all calculation stages are analyzed in detail. The results show that singular value decomposition (SVD) is the most time-consuming calculation stage. Then, we carry out the parallelization of the DPC method based on the truncated SVD algorithm. Later, the DPC workflow in DAG form is constructed. Finally, the computational complexity analysis is provided.

\begin{table*}[htbp]
	\centering
    \scriptsize
	\caption{Averaged calculation time statistics of the control examples with different dimensions}
    \renewcommand{\multirowsetup}{\centering}
	\begin{tabular}{|c|c|c|c|c|c|c|c|c|c|}
		\hline  
        \cline{1-10}
        \multirow{4}{3cm}{\textbf{Dimensions of simulation examples}}&\multicolumn{2}{c|}{\multirow{2}{2.6cm}{\textbf{SVD}}}&  \multicolumn{2}{c|}{\multirow{2}{2.6cm}{\textbf{Pesudo-inverse}}}  & \multicolumn{2}{c|}{\multirow{2}{2.6cm}{\textbf{Coefficient matrices}}} & \multicolumn{2}{c|}{\multirow{2}{2.6cm}{\textbf{Control sequence}}} &\multirow{4}{1.6cm}{\textbf{Total (s)}}  \cr
        &  \multicolumn{2}{c|}{\multirow{2}{2.2cm}{}}  & \multicolumn{2}{c|}{\multirow{2}{2.2cm}{}} &\multicolumn{2}{c|}{\multirow{2}{2.2cm}{}} &\multicolumn{2}{c|}{\multirow{2}{2.2cm}{}} &\cr
        \cline{2-9}
        &\multirow{2}{1.15cm}{Time (s)}&\multirow{2}{1.15cm}{Per.}&\multirow{2}{1.15cm}{Time (s)}&\multirow{2}{1.15cm}{Per.}&\multirow{2}{1.15cm}{Time (s)}&\multirow{2}{1.15cm}{Per.}&\multirow{2}{1.15cm}{Time (s)}&\multirow{2}{1.15cm}{Per.}&\\
        &&&&&&&&& \\
        \hline
        \hline
        $A: 2\times2 \ \ B: 2\times2$& 0.0293 & \textbf{35.1$\%$} & 0.0012 & 1.5$\%$ & 0.0005 & 0.7$\%$ & 0.0004 & 0.5$\%$ & \textbf{0.0934} \\
        \cline{1-10}
        $A: 4\times4 \ \ B: 4\times4$& 0.1655 & \textbf{72.4$\%$} & 0.0038 & 1.7$\%$ & 0.0053 & 2.3$\%$ & 0.0006 & 0.3$\%$ & \textbf{0.2283} \\
        \cline{1-10}
        $A: 8\times8 \ \ B: 8\times8$& 0.5455 & \textbf{85.4$\%$} & 0.0082 & 1.3$\%$ & 0.0183 & 2.9$\%$ & 0.0016 & 0.2$\%$ & \textbf{0.6380} \\
        \cline{1-10}
        $A: 16\times16 \ \ B: 16\times16$& 3.4034 & \textbf{96.5$\%$} & 0.0302 & 0.8$\%$ & 0.0157 & 0.4$\%$ & 0.0065 & 0.2$\%$ & \textbf{3.5234} \\
        \cline{1-10}
        $A: 32\times32 \ \ B: 32\times32$& 8.5256 & \textbf{97.2$\%$} & 0.0586 & 0.6$\%$ & 0.0658 & 0.7$\%$ & 0.0388 & 0.4$\%$ & \textbf{8.7655} \\
        \cline{1-10}
        $A: 64\times64 \ \ B: 64\times64$& 14.4843 & \textbf{94.9$\%$} & 0.1243 & 0.8$\%$ & 0.2961 & 1.9$\%$ & 0.2332 & 1.5$\%$ & \textbf{15.2524} \\
        \hline
        \cline{1-10}
	\end{tabular}
\label{Mean calculation time statistics for topical parameters in Monte Carlo experiments}
\end{table*}

\subsection{Computation amount analysis of the DPC method}

\begin{algorithm}[!h]
    \renewcommand{\algorithmicrequire}{\textbf{Input:}}
	\renewcommand{\algorithmicensure}{\textbf{Output:}}
    \caption{\textbf{Numerical Calculation Algorithm of the DPC Method}}
    \label{Numerical Computation of Data-driven Predictive Control}
    \textbf{Input:}
    The present time $k$, the constrcuted Hankel matrices $W_p$ and $U_f$, the least input and output signals $u(k)$ and $y(k)$ and the control cycle times $T$.\\
    \textbf{Output:} Data-driven predictive control sequence $u_f(k)$.

    \begin{algorithmic}[1]
        \STATE \textbf{\textrm{for}} k $\leftarrow$ 1 to T \textbf{\textrm{do}}
        \STATE \quad Update the Hankel matrices $W_p$ and $U_f$ by the least \\ \quad input and output signals $u(k)$ and $y(k)$.
        \STATE \quad Conduct SVD operation on $V_p = \left[
                                 \!\!\begin{array}{c}
                                   W_{p} \\
                                   U_{f} \\
                                 \end{array}\!\!
                               \right]$ where $W_p$ is \\ \quad provided as the formula (8) and obtain the decomposed \\ \quad results:
                               \begin{equation}
                                    V_p = MSN^{T}\nonumber
                                \end{equation}
        \STATE \quad Calculate the Morre-Penrose pesudo-inverse of $V_p$ by \\ \quad the decomposed results:\begin{equation}
                                                                                                        V_p^{\dag} = NS^{-1}M^{T}.\nonumber
                                                                                                    \end{equation}
        \STATE \quad Calculate the coefficient matrices $L_w$ and $L_u$:
        \begin{eqnarray}
                                \small{\left[
                           \!\!\begin{array}{cc}
                            L_{w} & \!\!\!\!L_{u} \\
                          \end{array}\!\!
                        \right] \!\!\!\!}&=& \small{\!\!\!\!Y_{f}V_p^{\dag}}.\nonumber
\end{eqnarray}
        \STATE \quad Calculate the optimal predictive control sequence\! $u_f(k)$:
    \begin{equation}
        u_{f}(k)=(\lambda I+L_{u}^{T}L_{u})^{-1}L_{u}^{T}(r_{f}-L_{w}w_{p}(k)).\nonumber
    \end{equation}
    \STATE \textbf{\textrm{end for}}
    \end{algorithmic}
\end{algorithm}
\vspace{-0.8em}

From the preliminary, the data-driven predictive control law is calculated by (\ref{dpc10}) and (\ref{dpc12}). First, the coefficient matrices $L_w$ and $L_u$ are calculated as (\ref{dpc11}). Then, the $L_w$ and $L_u$ are used to compute the optimal data-driven predicted control strategy as (\ref{dpc12}). In the practical numerical calculation, the Morre-Penrose pesudo-inverse is obtained based on the SVD algorithm. Letting $V_p = \left[
                                 \!\!\begin{array}{c}
                                   W_{p} \\
                                   U_{f} \\
                                 \end{array}\!\!
                               \right]$ and conducting SVD operation on $V_p$, we obtain
\begin{equation}\label{e18}
  V_p = MSN^{T}
\end{equation}
where $S$ is the singular value and $M, N$ are decomposed coefficient matrices. As the result, the Morre-Penrose pesudo-inverse of $V_p$ could be calculated by the decomposed results as
\begin{equation}\label{e19}
  V_p^{\dag} = NS^{-1}M^{T}.
\end{equation}

The numerical calculation procedure of the DPC method is summarized as Algorithm \ref{Numerical Computation of Data-driven Predictive Control}. To discuss the computational features, the calculation of the DPC method is divided into four stages, which are:
\begin{itemize}
  \item SVD: conduct the SVD operation.
  \item Pesudo-inverse: \!obtain the Morre-Penrose pesudo-inverse.
  \item Coefficient matrices: compute the coefficient matrices.
  \item Control sequence: compute the final control sequence.
\end{itemize}

Further, to analyse the averaged calculation time of each stage, a series of control simulation examples with different dimensions are carried out in a cloud server. The type of the used server is `\emph{ecs.hfc6.xlarge}' with 4 CPU and 8 GB memory in Alibaba cloud. The state-space models of simulation examples are generated randomly with six kinds of dimensions. The matrices of $A$ and $B$ are defined as $2\times2$, $4\times4$, $8\times8$, $16\times16$, $32\times32$, $64\times64$ dimensions, respectively. The matrices of C are set as the unit matrices with the corresponding dimensions and the matrices of D are zero matrices. The scale parameters are set as $N=1000, j=10$ and all the DPC missions are conducted for 100 cycles. The results of the above simulations are recorded in Table \ref{Mean calculation time statistics for topical parameters in Monte Carlo experiments}.

From Table \ref{Mean calculation time statistics for topical parameters in Monte Carlo experiments}, with the dimensions of control examples growing, the total calculation time of the DPC method increases. The calculation time increases from 0.0934 s of 2-dimension to 15.2524 s of 64-dimension. It is hard even impossible to finish the real-time control for the DPC method on the missions with relatively high dimension. The stage of SVD is the major time-costing calculation stage in all groups. When the dimension reaches 16, the stage of SVD occupies over 90$\%$ of the total calculation time. That is to say, the stage of SVD is the main time-costing stage of the DPC method. Therefore, the speedup of SVD is important for the efficiency improvement of the entire DPC calculation.

In this work, the acceleration is achieved by cloud workflow processing method. The construction method of the DPC workflow is proposed in Sections \ref{Parallelization based on the truncated distributed SVD algorithm} and \ref{DAG establishment of the DPC method}. Besides, notice that the sum of the percentages of the four stages are less than 100$\%$. The time of missing parts are spent by the constructions and updates of the Hankel matrices and other non-computation operations. This problem is solved by the design of communication topology switching, which is provided in Section \ref{Communication scheme of containerised workflow-based cloud control system}.

\subsection{Parallelization based on the truncated distributed SVD algorithm}\label{Parallelization based on the truncated distributed SVD algorithm}

To accelerate the computation of the DPC method, we consider it in a distributed parallelization framework. The distributed SVD algorithm can achieve the distributed decomposition in a lightweight way \cite{bjorck2015numerical}, which inspires us to establish the DAG of SVD, for computing the Morre-Penrose pesudo-inverse.

Assume an $m\times n$ matrix $A$ is split column-wise into submatrices $A_1$ and $A_2$ with the sizes $m\times n_1$ and $m\times n_2$, respectively, and $n_1+n_2 =n$. Conduct SVD operations on $A_1$, $A_2$ and obtain the results $A_1=U_1\Sigma_1V_1^{T}$, $A_2=U_2\Sigma_2V_2^{T}$. Then the SVD of $A = [A_1\ A_2]$ can be written as\vspace{-0.1em}
\begin{equation}\label{e20}
  \left[\!
    \begin{array}{cc}
      A_1\! & \!A_2 \\
    \end{array}
  \!\right]
   \!=\! \left[\!
    \begin{array}{cc}
      U_1\Sigma_1\! & \!U_2\Sigma_2 \\
    \end{array}
  \!\right]\left[\!
                                             \begin{array}{cc}
                                               V_1^{T}\! & \!\textbf{0}\! \\
                                               \textbf{0}\! & \!V_2^{T}\! \\
                                             \end{array}
                                           \!\right]
   \\
   \!\!=\! E \left[\!
                                             \begin{array}{cc}
                                               V_1^{T}\! & \!\textbf{0}\! \\
                                               \textbf{0}\! & \!V_2^{T}\! \\
                                             \end{array}
                                           \!\right]\vspace{-0.1em}
\end{equation}
where $E = \left[\!
    \begin{array}{cc}
      U_1\Sigma_1\!\! & \!U_2\Sigma_2 \\
    \end{array}
  \!\right]$. Conduct SVD on $E$ and obtain\vspace{-0.1em}
\begin{equation}\label{e21}
  \left[\!
    \begin{array}{cc}
      A_1\! & \!A_2 \\
    \end{array}
  \!\right] \!=\! U\Sigma \widetilde{V}^{T} \; \left[\!
                                             \begin{array}{cc}
                                               V_1^{T}\! & \!\textbf{0}\! \\
                                               \textbf{0}\! & \!V_2^{T}\! \\
                                             \end{array}
                                           \!\right] \\
   \!=\! U\Sigma V^{T}\vspace{-0.1em}
\end{equation}
where $V^{T} = \widetilde{V}^{T}\left[\!
                                             \begin{array}{cc}
                                               V_1^{T}\! & \!\textbf{0}\! \\
                                               \textbf{0}\! & \!V_2^{T}\! \\
                                             \end{array}
                                           \!\right]$ is the product of two orthogonal matrices and hence is also an orthogonal matrix.

\begin{figure}[!ht]
  \centering
  \includegraphics[width=3.5in]{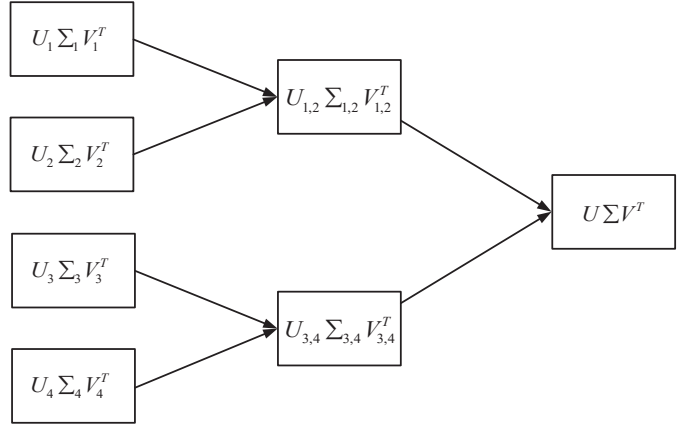}\\
  \caption{The Column-based Distributed SVD Method by MAT Operation}\label{SVD}
\end{figure}

The low rank approximations would be conducted after the individual SVDs are finished, of which the criteria is that only the square matrices corresponding to the singular values are retained. Assume $rank(A)=r$, $rank(A_1)=k$ and $rank(A_2)=l$. Thus, $A_1\!\approx\! U_{1_{k}}\Sigma_{1_{k}}V_{1_{k}}^{T}$ and $A_2\!\approx \!U_{2_{l}}\Sigma_{2_{l}}V_{2_{l}}^{T}$ indicate rank-$k$ and rank-$l$ approximations, respectively. The result is truncated to a rank-$r$ approximation as (\ref{e22}). This operation is therefore named as merge-and-truncate (MAT) operation rather than a simple merging. When there are serval partitions, the MAT operation can be conducted pairwise using a DAG-based strategy as created in Fig. \ref{SVD}.
\begin{equation}\label{e22}
  \left[\!
    \begin{array}{cc}
      A_1\! & \!A_2 \\
    \end{array}
  \!\right] \!\approx \! \left[\!
    \begin{array}{cc}
      U_{1_{k}}\Sigma_{1_{k}}\! & \!U_{2_{l}}\Sigma_{2_{l}} \\
    \end{array}
  \!\right]\!\left[\!
                                             \begin{array}{cc}
                                               V_{1_{k}}^{T}\! & \!\textbf{0}\! \\
                                               \textbf{0}\! & \!V_{2_{l}}^{T}\! \\
                                             \end{array}
                                           \!\right]\! \\
   \!= \!U_r\Sigma_rV_r^{T}.\vspace{-0.35em}
\end{equation}

The above operations are summarized as four self-defined functions, which are provided in Appendix. The matrix $A_{m\times n}$ is first partitioned column-wise and the SVD of each partition is computed in the function \textsc{ParallelSVDbyCols}. The $U, \Sigma, V$ matrices of all partitions are merged using the function \textsc{DoMergeOfBlocks}, which invokes the function \textsc{BlockMerge} as routine. The function \textsc{DoTruncate} carries out the low-rank approximation for $U, \Sigma, V$ by truncating the low-value data in the matrices. The result of truncated SVD is returned by the function \textsc{ParallelSVDbyCols}.

\subsection{DAG Construction of the DPC method}\label{DAG establishment of the DPC method}

\begin{figure*}[!ht]
  \centering
  \includegraphics[width=4in]{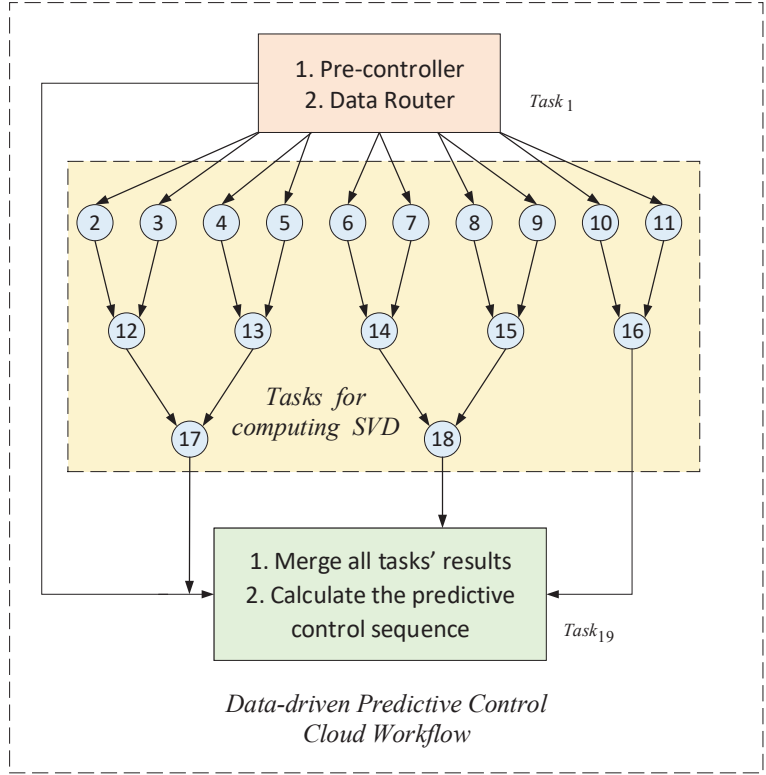}\\
  \caption{The Workflow Structure of the Data-driven Predictive Control Method (A Case of $MPT = 10$)}\label{Workflow Structure of DPC Method}
\end{figure*}

\begin{table}[htbp]
	\centering
	\caption{Relationships between Tasks in Workflow and Functions}
    \renewcommand{\multirowsetup}{\centering}
	\begin{tabular}{|c|c|c|c|}
		\toprule  
		\textbf{Image}&\textbf{Index}&\textbf{Level}&\textbf{Function} \\
        \hline
        \hline
        \multirow{2}{1.5cm}{$Image_A$}&\multirow{2}{1.5cm}{$Task_1$}&\multirow{2}{1cm}{1}&\multirow{2}{6.5cm}{Pre-controller in data preparation stage \\ and data router in the DPC stage} \\
        &&& \\
        \hline
        \multirow{2}{1.5cm}{$Image_B$}&\multirow{2}{1.5cm}{$Task_2-Task_{11}$}&\multirow{2}{1cm}{2} &\multirow{2}{6.5cm}{Compute the truncated SVD of \\each block matrices of $O_i$}\\&&&\\
        \hline
        \multirow{2}{1.5cm}{$Image_C$}&\multirow{2}{1.5cm}{$Task_{12}-Task_{18}$}&\multirow{2}{1cm}{3, 4}&\multirow{2}{6.5cm}{1. Merge the parent tasks' results \\ 2. Compute the new truncated SVD} \\&&&\\
        \hline
        \multirow{2}{1.5cm}{$Image_D$}&\multirow{2}{1.5cm}{$Task_{19}$}&\multirow{2}{1cm}{5}&\multirow{2}{6.5cm}{1. Merge all tasks's results \\ 2. Calculate the predictive control sequence} \\&&&\\
		\bottomrule  
	\end{tabular}
\label{Relationships of Tasks in Workflow and Functions}
\end{table}

Based on the four functions in Algorithm  \ref{Distributed truncated SVD algorithm}, the DPC workflow is established. Fig. \ref{Workflow Structure of DPC Method} provides a structure case of the DPC workflow with the degree of parallelism of SVD stage being $n/col = 10$, where $col$ is the column length of each matrix block. Table \ref{Relationships of Tasks in Workflow and Functions} describes the functions of the tasks in the DPC workflow and the relationships of the tasks. There are four kinds of task images used in this workflow. For instance, the image $Image_{B}$ can be reused ten times in the second level of which the task indexes are $Task_{2}-Task_{11}$ but with different input data. In Fig. \ref{Workflow Structure of DPC Method}, the function and relationship of each kind of the task image can be described as follows.

\begin{enumerate}
  \item The first task $Task_{1}$ occupies two roles in the workflow-based DPC method. In the begining, $Task_{1}$ serves as a pre-controller to prepare the initial data. The data are sent the children tasks to build the first batch of the Hankel matrices. Then, when the collected initial data are enough, $Task_{1}$ would become a data router in the DPC stage. The data received from the controlled plant would be transmitted to the children tasks to update the Hankel matrices. 
  \item The ten slices of Hankel matrices are built and updated in $Task_{2}-Task_{11}$, respectively. In $Task_{2}-Task_{11}$, the sliced blocks are decomposed using the function \textsc{DoMergeOfBlocks}.
  \item $Task_{12}-Task_{18}$ carry out the MAT operations on the results of $Task_{2}-Task_{11}$, i.e., merge the parent tasks' results and compute the new truncated SVDs.
  \item Finally, the export task $Task_{19}$ collects all the results of $Task_1$, $Task_{16}$, $Task_{17}$ and $Task_{18}$ as well as merges them, and later calculates the cloud-based predicted control sequence.
  \vspace{-0.1cm}
\end{enumerate}

In the workflow processing, the MPT is used to represent the maximal parallel tasks in a workflow, which means that at most MPT tasks are running in parallel during the workflow execution. In this case, the MPT is 10 and number of tasks is 19. This subsection only provides one case with a specific granularity. Using the same method, other structures of the DPC workflow could be constructed with different parameters for different kinds and scales of problems and different results would be obtained.

\subsection{Computational complexity analysis}

To study the improvement of this cloud workflow method, the computational complexity of Algorithm \ref{Distributed truncated SVD algorithm} is analyzed. Referring to \cite{bjorck2015numerical}, the floating point operations (flops) number of a complete SVD on an $m\times n$ matrix $A$ is approximated as $6mn^{2}+16n^{3}$. Without loss of generality, we assume all the blocks are divided with the same column width. Partition the matrix $A$ column-wise into $N$ blocks and the size of each block is $m\times s$ where $s = n/N$.

In Algorithm \ref{Distributed truncated SVD algorithm}, the first step is to execute $Image_C$. Thus the flops number of the SVD of each block is $6ms^{2}+16s^{3}$. By the low rank approximation, assume that each SVD is truncated to a $k$-rank matrix. Then at each level of the DAG, the tasks with $Image_D$ are executed. The flops number of $Image_D$ consists of two functions, which are the cost of merging ($2ms^{2}$) and the cost of new truncated SVD ($4mk^{2}+176k^{3}$). The cost of the latter term is made up of the cost of SVD ($6\cdot(2k)\cdot (2k)^{2}+16\cdot(2k)^{3}=176k^{3}$) and the cost of updating the matrix $V$ ($4mk^{2}$).
This task image is required to repeat $N-1$ times. The total flops number is
\begin{eqnarray}\label{e23}\nonumber
  T(m, n, N)=N(6ms^{2}+16s^{3})&&  \\
    && \!\!\!\!\!\!\!\!\!\!\!\!\!\!\!\!\!\!\!\!\!\!\!\!\!\!\!\!\!\!\!\!\!\!\!\!\!\!\!\!\!\!\!\!\!\!\!\!\! +(N-1)(6mk^{2}+176k^{3}).
\end{eqnarray}

Since $k\leq s$, the total number of flops
\begin{equation}\label{e24}
  T(m, n, N) < \frac{12mn^{2}}{N}+\frac{192n^{3}}{N^{2}}.
\end{equation}

Thus, when the degree of parallelism $N$ grows, the proposed method can accelerate the DPC algorithm effectively. Since most of the tasks can be executed in parallel, the total time cost would be reduced further.

\begin{remark}
In the practical establishment of the DPC workflow, we place the last two tasks with $Image_C$ into the export task since the time costs of the two tasks are relatively less. Therefore, there are only $7$ tasks with $Image_C$ in Fig. \ref{Workflow Structure of DPC Method}.
\end{remark}

\section{DOB-based Cloud-edge Collaborative Control Scheme}\label{DOB-based cloud-edge collaborative control scheme}

By the proposed workflow-based method, the computation efficiency of the DPC method could be improved. Furthermore, to reduce the non-computation overheads (e.g., receiving and sending data, serialization and deserialization), the truncation operation is considered to decrease the data amount. Since the involved Hankel matrices are high-dimension and low-rank, the larger singular values occupy the majority of the weight for the control variable computation. The smaller singular values have less influence on the results. Thus, the smaller singular values could be truncated to reduce the inter-communication and processing burdens. That is to say, the computation efficiency could be further improved by adjusting the truncation precision in the distributed SVD algorithm. However, the adjustment also brings the loss of control accuracy of the cloud-based DPC method. In this work, the loss of control accuracy could be considered as an uncertain event in cloud control system. Therefore, this uncertainty could be thought as the input disturbance and estimated by the DOB deployed in the edge controlled node.

For the convenience of understanding, a series of variables are recalled or defined uniformly in this section. We assume that in the non-truncation situation, the coefficient matrices, i.e., the cloud parameters and control variable are obtained. These parameters and variable are accurate and written as $L_w$, $L_u$ and $u(k)$, respectively. The $u(k)$ is the first variable in the predicted control sequence $u_f(k)$. In the truncation situation, the corresponding cloud parameters and control variable are noted as $\widehat{L}_w$, $\widehat{L}_u$ and $\widehat{u}(k)$, respectively. The $u_{cloud}(k)$ is the first term of the cloud predicted control sequence $\widehat{u}_{f}(k)$. Thus, the errors between the above variables of the non-truncation and truncation situations are derived in the below.
\begin{eqnarray}\label{e25}
  L_w &=& \widehat{L}_w + \Delta L_w, \\ \label{e26}
  L_u &=& \widehat{L}_u + \Delta L_u, \\
  u_f(k) &=& \widehat{u}_{f}(k) + \Delta u_f(k), \\ \label{e27}
  u(k) &=& u_{cloud}(k) + \Delta u(k) \label{e28}
\end{eqnarray}
where the $\Delta L_w$, $\Delta L_u$, $\Delta u_f(k)$ and $\Delta u(k)$ are the corresponding errors, respectively.

The error of the cloud control variable $\Delta u(k)$ is considered as the uncertainty in cloud control system. The upper bound of the cloud predicted control sequence error $\Delta u_f(k)$ is dependent on the cloud parameters errors $\Delta L_w$, $\Delta L_u$. Then, the upper bound of the cloud control variable error $\Delta u(k)$ is smaller than that of $\Delta u_f(k)$ because of the scaling relationship of $F$-norm. The detailed derivation is provided in Section \ref{Upper bounds}. The co-design of the DOB and DOB-based cloud-edge composite control law is designed in Section \ref{Disturbance observer and DOB-based composite control law design}. Finally, the upper bound of $\Delta u(k)$ is significant for the guarantee of the UUB stability of the DOB, which is analyzed in Section \ref{Stability analysis of disturbance observer}.

\subsection{Upper bounds of cloud parameter errors $\Delta L_w$, $\Delta L_u$ and cloud control variable error $\Delta u(k)$}\label{Upper bounds}

In this subsection, the upper bounds of the cloud parameter errors $\Delta L_w$, $\Delta L_u$ and cloud control variable error $\Delta u(k)$ are derived. Now, we adopt the truncated distributed SVD algorithm to compute the Morre-Penrose pesudo-inverse of $V_p$ and obtain
\begin{equation}\label{e29}
  V_p = \sum_{i=1}^{r}M_{i}S_{i}N_{i}^{T}
\end{equation}
where $r$ is the rank of $V_p$, $M_{i}$, $S_{i}$, $N_{i}$, $i = 1, 2, \ldots, r$ are the decomposed results of $V_p$, and the singular values are sorted from largest to smallest..

Using the truncated distributed SVD algorithm and setting the truncation precision as $\epsilon_1$, we obtain
\begin{equation}\label{e30}
  \hat{V}_p \approx \sum_{i=1}^{k}M_{i}S_{i}N_{i}^{T}
\end{equation}
where $k$ is the number of singular values satisfying the truncation precision $S_{i}\geq S_{max}*\epsilon_1$ and $S_{max}$ is the maximal singular value of $V_p$.

Thus the approximate Morre-Penrose pesudo-inverse of $\widehat{V}_p$ is computed by
\begin{equation}\label{e31}
  \widehat{V}_p^{\dag} = \sum_{i=1}^{k}N_{i}S_{i}^{-1}M_{i}^{T}.
\end{equation}.

We use this approximate Morre-Penrose pesudo-incerse $\widehat{V}_p^{\dag}$ to compute the intermediate parameters $\widehat{L}_w$, $\widehat{L}_u$ and the data-driven predictive control law $\widehat{u}_f(k)$. However, though the workflow-based DPC method with adjustable truncation precision could improve the computation efficiency significantly, it is obvious that the acceleration method introduces some errors. The relationships between the errors $\Delta L_w$, $\Delta L_u$ and the cloud parameters $\widehat{L}_w$, $\widehat{L}_u$ have been provided in (\ref{e25}) and (\ref{e26}), respectively. Then, based on the Morre-Penrose pesudo-inverse of $\widehat{V}_p$, the $\left[
                                                             \begin{array}{cc}
                                                               \widehat{L}_w & \widehat{L}_u \\
                                                             \end{array}
                                                           \right]$ could be calculated as
\begin{equation}\label{e32}
  \left[
    \begin{array}{cc}
      \widehat{L}_w & \widehat{L}_u \\
    \end{array}
  \right] = \sum_{i=1}^{k}Y_{f}N_{i}S_{i}^{-1}M_{i}^{T}.
\end{equation}

Thus, the input-output relationship of (\ref{dpc6}) would be replaced by
\begin{equation}\label{e33}
  \widehat{y}_f(k) = \widehat{L}_ww_p(k) + \widehat{L}_u\widehat{u}_{f}(k).
\end{equation}

Finally, with the same objective function of (\ref{dpc11}), an approximate cloud predictive control law is provided as
\begin{equation}\label{e34}
  \widehat{u}_{f}(k)=(\lambda I+\widehat{L}_{u}^{T}\widehat{L}_{u})^{-1}\widehat{L}_{u}^{T}(r_{f}-\widehat{L}_{w}w_{p}(k)).
\end{equation}


%


The relationship between the error $\Delta u_f(k)$ and cloud predicted control sequence $\widehat{u}_{f}(k)$ has been provided in (\ref{e27}). Next, the upper bound of the cloud predicted control sequence error $\Delta u_f(k)$ is derived. According to (\ref{e34}), the cloud predicted control sequence $\widehat{u}_{f}(k)$ is dependent on the cloud parameters $\widehat{L}_{w}$ and $\widehat{L}_{u}$. Thus, $\Delta u_f(k)$ is dependent on the errors of the cloud parameters $\Delta L_w$ and $\Delta L_u$. In the below, the upper bounds of $\Delta L_w$ and $\Delta L_u$ are calculated.

By the SVD-based representation, the accurate cloud parameters $L_w$ and $L_u$ in the non-truncation situation is provided by
\begin{equation}\label{e35}
  \left[
    \begin{array}{cc}
      L_w & L_u \\
    \end{array}
  \right] = \sum_{i=1}^{r}Y_{f}N_{i}S_{i}^{-1}M_{i}^{T}.
\end{equation}

Substituting (\ref{e35}) to (\ref{e32}) and calculating $\Delta L_w$ and $\Delta L_u$
\begin{equation}\label{e36}
  \left[
    \begin{array}{cc}
      \Delta L_w & \Delta L_u \\
    \end{array}
  \right] = \sum_{i=k+1}^{r}Y_{f}N_{i}S_{i}^{-1}M_{i}^{T}.
\end{equation}

Thus,
\begin{eqnarray}\label{e37}
  \Delta L_w &=& \sum_{i=k+1}^{r}Y_{f}N_{i}S_{i}^{-1}M_{w,i}^{T}, \\
  \Delta L_u &=& \sum_{i=k+1}^{r}Y_{f}N_{i}S_{i}^{-1}M_{u,i}^{T} \label{e38}
\end{eqnarray}
where $M_{w, i}$ is the first $(m+l)j$ rows of $M_i$ and $M_{u, i}$ is the last $mj$ rows of $M_i$. Since $M_{i}$ and $N_{i}$ are the results of SVD, they are unitary matrices of which the \emph{2}-norms are 1. In this work, $\|\cdot\|$ means the \emph{2}-norm. Defining $S_{min}$ is the minimal singular value of $V_p$, we obtain
\begin{eqnarray}\label{e39}
  \|\Delta L_w\| &\leq& \alpha(r-k)S_{min}^{-1}\|Y_f\| = \epsilon_2, \\
  \|\Delta L_u\| &\leq& \beta(r-k)S_{min}^{-1}\|Y_f\| = \epsilon_3 \label{e40}
\end{eqnarray}
where $\alpha = \|M_{w,i}^{T}\|$, $\beta = \|M_{u,i}^{T}\|$ and $\alpha^{2}+\beta^{2}=1$.

Then, we compute the upper bound of cloud control law error $\Delta u_f(k)$. Based on (\ref{dpc12}) and (\ref{e34}), it holds that
\begin{eqnarray}\label{e48}
  (\lambda I + L_u^{T}L_u)u_f(k) = L_u^{T}r_f - L_u^{T}L_ww_p(k), \\
  (\lambda I + \hat{L}_u^{T}\hat{L}_u)\hat{u}_f(k) = \hat{L}_u^{T}r_f - \hat{L}_u^{T}\hat{L}_ww_p(k). \label{e49}
\end{eqnarray}
Substituting (\ref{e25}) and (\ref{e26}) to (\ref{e48}), we obtain
\begin{eqnarray}\label{e50}
  && [\lambda I + (\hat{L}_u+\Delta L_u)^{T}(\hat{L}_u+\Delta L_u)]u_f(k) = \\\nonumber
  && (\hat{L}_u+\Delta L_u)^{T}r_f - (\hat{L}_u+\Delta L_u)^{T}(\hat{L}_w+\Delta L_w)w_p(k).
\end{eqnarray}

Further, subtracting (\ref{e50}) to (\ref{e49}), we obtain
\begin{eqnarray}\label{e51}
  && [\lambda I + (\hat{L}_u+\Delta L_u)^{T}(\hat{L}_u+\Delta L_u)]\Delta u_f(k) = \\\nonumber
  && -\;(\hat{L}_u^{T}\Delta L_u+\Delta L_u^{T}\hat{L}_u+\Delta L_u^{T}\Delta L_u)\hat{u}_f(k)+\Delta L_u^{T}r_f \\\nonumber
  && -\;(\hat{L}_u\Delta L_w^{T}+\Delta L_u^{T}\hat{L}_w+\Delta L_u^{T}\Delta L_w)w_p(k).
\end{eqnarray}

Then,
\begin{eqnarray}\label{e52}
  && \|[\lambda I + (\hat{L}_u+\Delta L_u)^{T}(\hat{L}_u+\Delta L_u)]\Delta u_f(k)\| = \\\nonumber
  && \|-\;(\hat{L}_u^{T}\Delta L_u+\Delta L_u^{T}\hat{L}_u+\Delta L_u^{T}\Delta L_u)\hat{u}_f(k)+\Delta L_u^{T}r_f \\\nonumber
  && -\;(\hat{L}_u\Delta L_w^{T}+\Delta L_u^{T}\hat{L}_w+\Delta L_u^{T}\Delta L_w)w_p(k)\| \leq \\\nonumber
  && (2\|\hat{L}_u\|\!\|\Delta L_u\|+\|\Delta L_u\|^{2})\|\hat{u}_f(k)\| +\|\Delta L_u\|\!\|r_f\| \\\nonumber
  && +(\|\hat{L}_u\|\!\|\Delta L_w\| \!+\!\|\Delta L_u\|\!\|\hat{L}_w\|\!+\!\|\Delta L_u\|\!\|\Delta L_w\|)\|w_p(k)\|.
\end{eqnarray}

Since $(\hat{L}_u+\Delta L_u)^{T}(\hat{L}_u+\Delta L_u)$ is positive-definite, it follows that $\lambda\|\Delta u_f(k)\| \leq\|[\lambda I + (\hat{L}_u+\Delta L_u)^{T}(\hat{L}_u+\Delta L_u)]\Delta u_f(k)\|$, which further results in
\begin{eqnarray}\label{e54}
\|\Delta u_f(k)\| \leq \frac{1}{\lambda}(2\|\hat{L}_u\|\|\Delta L_u\|+\|\Delta L_u\|^{2})\|\hat{u}_f(k)\| \\\nonumber
   + \frac{1}{\lambda}\|\Delta L_u\|\|r_f\|+\frac{1}{\lambda}(\|\hat{L}_u\|\|\Delta L_w\| \\\nonumber
   + \|\Delta L_u\|\|\hat{L}_w\|+\|\Delta L_u\|\|\Delta L_w\|)\|w_p(k)\|.
\end{eqnarray}

Then, we obtain the upper bound of cloud predicted control sequence error $\Delta u_f(k)$
\begin{eqnarray}\label{e55}
\|\Delta u_f(k)\| \leq \frac{1}{\lambda}(2\epsilon_3\|\hat{L}_u\|+\epsilon_3^{2})\|\hat{u}_f(k)\|+ \frac{1}{\lambda}\epsilon_3\|r_f\|+\frac{1}{\lambda}(\epsilon_2\|\hat{L}_u\| +\epsilon_3\|\hat{L}_w\|+\epsilon_2\epsilon_3)\|w_p(k)\|.
\end{eqnarray}

Finally, since $\|\Delta u(k)\| \leq \|\Delta u_f(k)\|$, the upper bound of cloud control variable error is also derived. Note that $\hat{L}_u, \hat{u}_f(k), r_f, w_p(k)$ are known in the computation of cloud controller. Thus, the upper bound of the cloud control variable error could be obtained.

\begin{remark}
From (\ref{e39}) and (\ref{e40}), the law of parameters setting could be obtained. If the truncation precision $\epsilon_1$ is set low, the computation speed would be improved. But the number of the reserved singular values $k$ becomes smaller and $r-k$ becomes larger. Thus, the right sides of the two inequations, i.e., the upper bounds of the cloud parameter errors become larger, which means the accuracy of the workflow-based DPC is decreased. The derived formulas are in agreement with the experimental results provided in Section \ref{Experiment and discussion}.

\end{remark}

\begin{remark}\label{remark3}
In the practical conduction, the number of the retained singular values also could be used to adjust the truncation accuracy. In the truncation operation of the workflow-based DPC method, there are a large amount of the singular values, which are in the same order of magnitudes. If we adjust the truncation accuracy by increasing or decreasing $\epsilon_1$, the change of the upper bounds of the uncertain would be large. Then, the trade-off between the computation efficiency and control performance would be hard to handle. Therefore, in the third control example: high-dimension numerical system of Section \ref{Experiment and discussion}, the number of the retained singular values was adjusted.

\end{remark}

\subsection{DOB and DOB-based cloud-edge composite control law}\label{Disturbance observer and DOB-based composite control law design}

For the convenience of analysis, we select the first term in the cloud predicted control sequence $\widehat{u}_{f}(k)$ as the cloud control variable $u_{cloud}(k)$, which would be transmitted to the edge controlled node. In the $k$-th time, the first variable of the predicted control sequence, $u_{cloud}(k)$ would be adopted to the plant. Thus, only the corresponding parameters to $\widehat{u}_{f}(k)$ need be sent to the plant. That is to say, based on (\ref{e33}), the nominal system is
\begin{eqnarray}\label{e57}
  \hat{y}(k+1) = \sum_{i=1}^{N}\hat{a}_{i}(k)y(k-N+i) + \sum_{i=0}^{N-1}\hat{b}_{i}(k)u(k-N+i) +  \hat{b}(k)u_{cloud}(k)
\end{eqnarray}
where $\hat{a}_{i}(k), \hat{b}_{i}(k), i = 1,2,..,N$ is the first block-row of $\hat{L}_w$ at sampling time $k$, $\hat{b}(k)$ is the first block-element of $\hat{L}_u$ at sampling time $k$ and $\{u(i), i = k-N, k-N+1, ..., k-1\}$, $\{y(i), i = k-N+1, k-N+2, ..., k\}$ are the input and output data in the edge data buffer, respectively.

The real system considering the uncertainty is given by
\begin{eqnarray}\label{e57}
  y(k+1) \!\!&\!\!=\!\!&\!\! \sum_{i=1}^{N}\hat{a}_{i}(k)y(k-N+i)\!+\! \sum_{i=0}^{N-1}\hat{b}_{i}(k)u(k-N+i) \!+\!  \hat{b}(k)u_{cloud}(k) \!+ \! \hat{b}(k)\Delta u(k)\\\nonumber
   \!\!&\!\!=\!\!&\!\!  \sum_{i=1}^{N}\hat{a}_{i}(k)y(k-N+i)\!+\! \sum_{i=0}^{N-1}\hat{b}_{i}(k)u(k-N+i) \!+\!  \hat{b}(k)u_{cloud}(k) \!+ \! \hat{b}(k)d(k)
\end{eqnarray}
where $d(k)=\Delta u(k)$ is considered as the input disturbance, and $\|d(k)\| = \|\Delta u(k)\| \leq \epsilon_4$.

By \cite{ginoya2015delta}, we design the DOB in the form of
\begin{equation}\label{e59}
  \hat{d}(k) = P(k) + Ly(k)
\end{equation}
where $\hat{d}(k)$ is the estimation of disturbance vector $d(k)$, $L$ is the observer gain matrix to be designed, and $P(k)$ is the auxiliary variable vector updated by
\begin{eqnarray}\label{e60}
  P(k+1) \!\!&\!\!=\!\!&\!\! -L(\sum_{i=1}^{N}\hat{a}_{i}(k)y(k-N+i) + \sum_{i=0}^{N-1}\hat{b}_{i}(k)u(k-N+i) \\\nonumber
   \!\!&\!\!\!\!&\!\!  + \; \hat{b}(k)u_{cloud}(k) + \ \hat{b}(k)\hat{d}(k)).
\end{eqnarray}

Substituting (\ref{e60}) to (\ref{e59}) and calculating $\hat{d}(k+1)$
\begin{eqnarray}\label{e61}
  \hat{d}(k+1) &=& P(k+1) + Ly(k+1), \\
  \hat{d}(k+1) &=& L\hat{b}(k)(d(k)-\hat{d}(k)). \label{e62}
\end{eqnarray}

Defining the estimation error as $\tilde{d}(k)=d(k)-\hat{d}(k)$ and subtracting both sides of (\ref{e62}) from $d(k+1)$ gives
\begin{equation}\label{e63}
  d(k+1)-\hat{d}(k+1) = d(k+1) - L\hat{b}(k)\tilde{d}(k).
\end{equation}

The error dynamics of the proposed observer is provided by
\begin{equation}\label{e64}
  \tilde{d}(k+1) = -L\hat{b}(k)\tilde{d}(k) + d(k+1).
\end{equation}

In this system, the DOB-based cloud-edge composite control law is designed as
\begin{equation}\label{e65}
  u(k) = u_{cloud}(k) + u_{com}(k)
\end{equation}
where $u_{cloud}(k)$ is the cloud control input and $u_{com}(k)$ is the edge DOB-based compensative control variable.

Since the lumped disturbance is considered as the input disturbance, the compensator is designed as
\begin{equation}\label{e66}
  u_{com}(k) = -\hat{d}(k).
\end{equation}

Thus, when $k=N+1,N+2, ...$, we obtain the co-design of cloud-edge composite control law and DOB-based compensator as follows
\begin{equation}\label{e67}
  \left\{
    \begin{array}{ll}
        \hat{d}(k) = P(k) + Ly(k) \\
        u_{com}(k) = -\hat{d}(k) \\
        u(k) = u_{cloud}(k) + u_{com}(k)  \\
        P(k+1) = -L(\sum_{i=1}^{N}\hat{a}_{i}(k)y(k-N+i) \\
        \;\;\;\;\;\; + \sum_{i=0}^{N-1}\hat{b}_{i}(k)u(k-N+i) + \hat{b}(k)u(k)+\hat{b}(k)\hat{d}(k)).
    \end{array}
  \right.
\end{equation}

According to (\ref{e67}), when $k=N+1,N+2, ...$, an alternative expression for calculating $\hat{d}(k)$ is given by
\begin{eqnarray}\label{e68}
  \hat{d}(k) \!&\!\!=\!\!&\! L(y(k) - \sum_{i=1}^{N}\hat{a}_{i}(k-1)y(k-N+i-1) \\\nonumber
   \!&\!\!\!\!&\! - \sum_{i=0}^{N-1}\hat{b}_{i}(k-1)u(k-N+i-1) \\\nonumber
   \!&\!\!\!\!&\! - \ \hat{b}(k-1)u_{cloud}(k-1)).
\end{eqnarray}

Besides, when $k=1,2,..., N$, the data amount in the edge data buffer is not enough to execute the DOB-based compensator. Thus, the DOB-based compensator would not work and the cloud control variable would be used as the practical control input, i.e., $u(k)=u_{cloud}(k)$.

\subsection{Stability analysis of the DOB}\label{Stability analysis of disturbance observer}

Now, we will prove the stability of the designed DOB. Rewrite the error dynamics (\ref{e64}) of the proposed DOB as
\begin{equation}\label{e69}
  \tilde{d}(k+1) = H\tilde{d}(k) + d(k+1)
\end{equation}
where $H = -L\hat{b}(k)$.

\begin{theorem}
The error system (\ref{e69}) are UUB.
\end{theorem}
\begin{IEEEproof}[\textbf{Proof}]
Since $\hat{b}(k)$ is the first block-element of the bounded $\hat{L}_u$, the estimated error $\Delta b(k)$ is also bounded by $\epsilon_{3}$, i.e., the upper bound of $\|\Delta L_u\|$. Therefore, though $\hat{b}(k)$ is a time-varying variable, it is still bounded and the eigenvalues of matrix $H$ could be arbitrary placed in a unit circle by selecting a proper $L$. One can always find a positive-definite matrix $P$ such as
\begin{equation}\label{e70}
  H^{T}PH-P = -Q
\end{equation}
where $Q$ is a given positive-definite matrix. It can be verified that $\delta= \frac{\lambda_{min}(Q)}{\lambda_{max}(P)}<1$, where $\lambda_{min}(Q)$ is the smallest eigenvalue of $Q$ and $\lambda_{max}(P)$ the largest eigenvalue of $P$.

Define a Lyapunov function as
\begin{equation}\label{e71}
  V(\tilde{d}(k))= \tilde{d}(k)^{T}P\tilde{d}(k)
\end{equation} and calculate the increment of $V(k)$ for $k\in [N+1, \infty)$ as
\begin{eqnarray}\label{e73}
  \Delta V(k) \!&\!\!=\!\!&\! V(\tilde{d}(k+1))-V(\tilde{d}(k)) \\\nonumber
   \!&\!\!=\!\!&\! \tilde{d}(k+1)^{T}P\tilde{d}(k+1) - \tilde{d}(k)^{T}P\tilde{d}(k) \\\nonumber
   \!&\!\!=\!\!&\! (H\tilde{d}(k) + d(k+1))^{T}P(H\tilde{d}(k)+ d(k+1))-\;\tilde{d}(k)^{T}P\tilde{d}(k) \\\nonumber
   \!&\!\!=\!\!&\! \tilde{d}(k)^{T}(H^{T}PH-P)\tilde{d}(k)+2\tilde{d}(k)^{T}H^{T}Pd(k+1) +\;d(k+1)^{T}Pd(k+1) \\\nonumber
   \!&\!\!=\!\!&\! -\tilde{d}(k)^{T}Q\tilde{d}(k)+2\tilde{d}(k)^{T}H^{T}Pd(k+1) +\;d(k+1)^{T}Pd(k+1)\\\nonumber
   \!&\!\!\leq\!\!&\! -\lambda_{min}(Q)\|\tilde{d}(k)\|^{2}+2\epsilon_4\|H^{T}P\|\|\tilde{d}(k)\|+\epsilon_4^{2}\|P\|. \nonumber
\end{eqnarray}

Let $\mu = 2\epsilon_4\|H^{T}P\|\|\tilde{d}(k)\|+\epsilon_4^{2}\|P\|$. For $k\in [N+1, \infty)$, $\Delta V(k)$ is obtained as
\begin{eqnarray}\label{dob_prove_plus3}
  \Delta V(k) \!&\!\!\leq\!\!&\! -\lambda_{min}(Q)\|\tilde{d}(k)\|^{2}+\mu \\\nonumber
   \!&\!\!\leq\!\!&\! -\frac{\lambda_{min}(Q)}{\lambda_{max}(P)}\tilde{d}(k)^{T}P\tilde{d}(k)+\mu \\\nonumber
   \!&\!\!=\!\!&\! -\delta V(k)+\mu. \nonumber
\end{eqnarray}

Then, we have
\begin{equation}\label{dob_prove_plus5}
  V(\tilde{d}(k))-V(\tilde{d}(k-1)) \leq -\delta V(\tilde{d}(k-1))+\mu
\end{equation}
which gives
\begin{equation}\label{dob_prove_plus5}
  V(\tilde{d}(k))\leq (1-\delta) V(\tilde{d}(k-1))+\mu.
\end{equation}

For $k\in [N+1, \infty)$, this together with (\ref{dob_prove_plus3}) implies that
\begin{equation}\label{dob_prove_plus6}
  V(\tilde{d}(k))\leq (1-\delta)^{k-N-1} V(\tilde{d}(N+1))+\frac{\mu}{\delta}
\end{equation}
and
\begin{equation}\label{dob_prove_plus7}
  \|\tilde{d}(k)\|^{2}\leq (1-\delta)^{k-N-1}\frac{\lambda_{max}(P)}{\lambda_{min}(P)}\|\tilde{d}(N+1)\|^{2} + \frac{\mu}{\delta \lambda_{min}(P)}
\end{equation}
where $\lambda_{min}(P)$ is the smallest eigenvalue of $P$.

Considering that $\delta\leq1$, we have $\lim_{k\rightarrow\infty}(1-\delta)^{k-N-1} = 0$, which brings that the states of system (\ref{e69}) are UUB.
\end{IEEEproof}

Here, we obtain that the error dynamics of the proposed DOB is UUB. In addition, because of the separability of controller and observer designs and that the stability of the workflow-based DPC method is another question, the stability of cloud controller would be studied in the future. The stability research of the normal DPC method could be found in \cite{berberich2020data}.

\begin{breakablealgorithm}
    \renewcommand{\thealgorithm}{2}
    \renewcommand{\algorithmicrequire}{\textbf{Input:}}
	\renewcommand{\algorithmicensure}{\textbf{Output:}}
    \caption{\textbf{Overall Algorithm of the Cloud-edge Collaborative Control Scheme}}
    \label{Data-driven Predictive cloud control system Algorithm}
    \textbf{Input:}
    The scale parameters $N$ and $j$ of the Hankel matrices, truncation precision $\epsilon_1$ and reference value $r_f$.\\
    \textbf{Output:} Cloud-edge composite control signal $u(k)$ provided by the cloud controller and edge DOB-based compensator.

    \begin{algorithmic}[1]
        \STATE \textbf{\textit{Preparing the Initial Information}}
        \begin{enumerate}
          \item \textit{Set parameters:}
          \begin{itemize}
            \item First, set the scale parameters $N$ and $j$ to determine the size of the Hankel matrices. The truncation precision $\epsilon_1$ and reference $r_f$ also needs to be provided.
          \end{itemize}
          \item \textit{Collect data:}
          \begin{itemize}
            \item Generate a series of initial control input data
           $\{u(0),u(1),\ldots,u(2N+j-1)\}$ by a certain algorithm such as PID or LQR. Then measure the output data of the plant $\{y(1),y(2),\ldots,y(2N+j)\}$ by sensors.
          \end{itemize}
          \item \textit{Create Hankel matrices:}
          \begin{itemize}
            \item Create the initial Hankel matrices $U_f,U_p,Y,Y_{p}$ of the collected input and output data.
          \end{itemize}
        \end{enumerate}
        \STATE \textbf{\textit{Cloud Control Platform}}
        \begin{enumerate}
          \item \textit{Cloud parameters:}
          \begin{itemize}
            \item At the sampling time $k$, build the Hankel matrix $V_p=\left[
                                                                                          \begin{array}{c}
                                                                                            W_p \\
                                                                                            U_f \\
                                                                                          \end{array}
                                                                                        \right]$ where $W_p$ is provided as the formula (\ref{dpc7}).
            \item Conduct SVD operation on $V_p$ as Algorithm \ref{Distributed truncated SVD algorithm} and obtain $M_{i}$, $S_{i}$, $N_{i}$, $i = 1, 2, \ldots, r$ where $r$ is the rank of $V_p$.
            \item Obtain the reserved SVD results: $M_{i}$, $S_{i}$, $N_{i}$, $i = 1, 2, \ldots, k$ where k is the number of the singular values satisfying  $S_{i}\geq S_{max}*\epsilon_1$ and $S_{max}$ is the maximal singular value of $V_p$.
            \item Calculate the cloud parameters $\widehat{L}_w$, $\widehat{L}_u$ as $\left[
    \begin{array}{cc}
      \widehat{L}_w & \widehat{L}_u \\
    \end{array}
  \right] = \sum_{i=1}^{k}YN_{i}S_{i}^{-1}M_{i}^{T}$.
          \end{itemize}
          \item \textit{Cloud control variable:}
          \begin{itemize}
            \item Obtain the control sequence $\widehat{u}_f(k) = [\widehat{u}(k|k) \;\; \widehat{u}(k+1|k) \;\; \ldots \;\; \widehat{u}(k+N|k)]$ as the formula (\ref{e34}) by using $\widehat{L}_w$, $\widehat{L}_u$.
            \item Take the first term of $\widehat{u}_f(k)$ as the cloud control variable $u_{cloud}(k)$.
          \end{itemize}
          \item \textit{Package and send:}
          \begin{itemize}
            \item Pack the cloud control variable $u_{cloud}(k)$ and cloud parameters $\widehat{L}_w$, $\widehat{L}_u$. Then send the information packet to the edge client together.
          \end{itemize}
        \end{enumerate}
        \STATE \textbf{\textit{Edge Controlled Node}}
        \begin{enumerate}
          \item \textit{Receive data:}
          \begin{itemize}
            \item The edge client receives the cloud control variable $u_{cloud}(k)$ and cloud parameters $\widehat{L}_w$, $\widehat{L}_u$.
          \end{itemize}
          \item \textit{Composite control and connect actuator:}
          \begin{itemize}
            \item When $k=1,2,..., N$, let the DOB-based compensator being in idle state and set $u(k)=u_{cloud}(k)$.
            \item When $k=N+1,N+2,...$:
            \begin{enumerate}
                    \item Build the DOB-based compensator based on the cloud parameters $\widehat{L}_w$, $\widehat{L}_u$ and obtain the edge compensative control variable $u_{com}(k)$ as the formula (\ref{e67}).
                    \item Obtain the composite control variable $u(k) = u_{cloud}(k) + u_{com}(k)$.
                  \end{enumerate}
          \item Then, apply $u(k)$ to the dynamic system and detect the corresponding output $y(k)$.
          \end{itemize}
          \item \textit{Package and send:}
          \begin{itemize}
            \item Pack the composite input $u(k)$ and the corresponding $y(k)$ as a combined packet. Then store this packet in the edge buffer and send it to cloud server for the optimization at next step.
          \end{itemize}
        \end{enumerate}
        \STATE \textbf{\textit{Data Update with Feedback}}
        \begin{enumerate}
          \item \textit{Update Hankel matrices:}
          \begin{itemize}
            \item The cloud server receives the latest packet of $u(k)$ and $y(k)$, and updates the Hankel matrices by removing the earliest data as well as adding the newest data to keep the fixed size $N$.
          \end{itemize}
          \item \textit{Feedback:}
          \begin{itemize}
            \item Repeat the course of \textbf{\textit{Cloud Control Platform}} and continue the circulation until receiving a stop command.
          \end{itemize}
        \end{enumerate}
    \end{algorithmic}
    \label{algorithm}
\end{breakablealgorithm}

\subsection{Overall algorithm of cloud-edge collaborative control system}
The overall algorithm of the cloud-edge collaborative control system is summarized in Algorithm \ref{Data-driven Predictive cloud control system Algorithm}. This algorithm procedure is designed as four parts: (1) preparing the initial information; (2) cloud control platform: data processing and cloud control variable computation; (3) edge controlled node: DOB-based compensation, cloud-edge composite control variable computation and executing the control input; (4) data update with feedback. The four steps are performed in this system as a circulation. Notice that all the procedures of cloud control platform are finished in the workflow-based controller. Besides, in this system, we implicitly identify the system model in real time, yet the model is not established throughout the course of control. At each sampling time, this model is updated by real-time data. Then, the predictive control method is deployed in this implicit model. In fact, this model is cancelled as an intermediate process and the control process only relies on data.

\section{Design and Implementation of Containerised Workflow-based Cloud Control Experimental System}\label{Design and implementation of containerised cloud control testbed}

To conduct the workflow-based DPC controller in real cloud environment and evaluate the performance of the cloud-edge collaborative control system, the containerised workflow-based cloud control experimental system is designed and implemented. First, the structure of this experimental system is provided. Second, the initialization procedure and execution form of this system are designed. Finally, the communication scheme of this system is established, including the data transmission in the container network, IP and port mappings between the cloud environment and edge node, and the communication topology switching from the data preparation stage to the DPC stage.

\begin{figure}[!ht]
  \centering
  \vspace{1em}
  \includegraphics[width=4in]{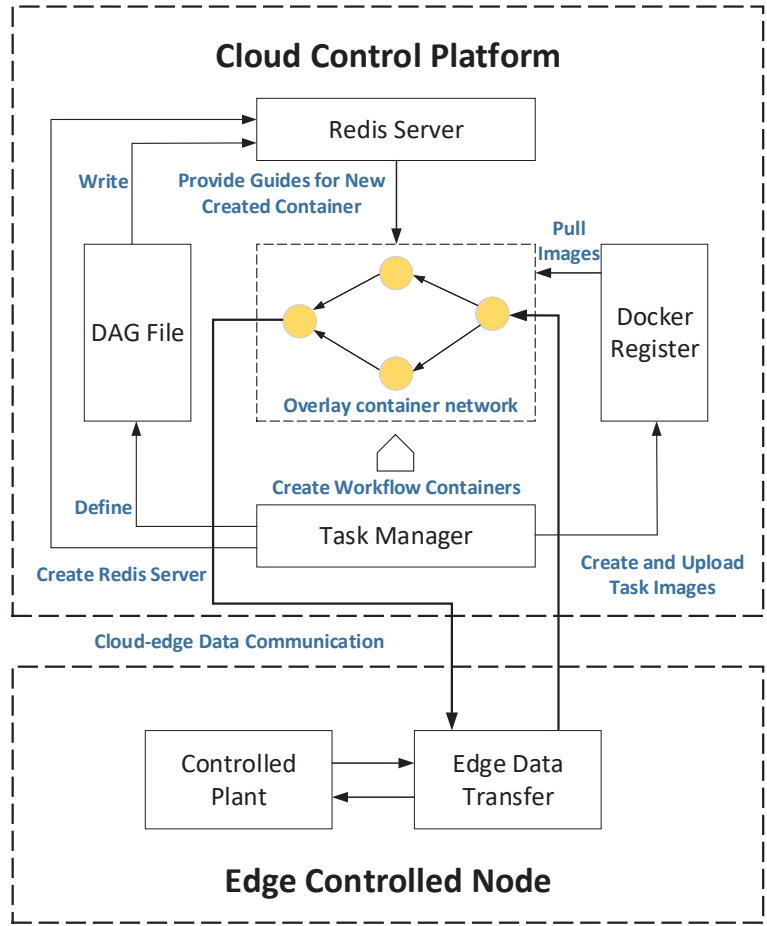}\\
  \caption{Structure of the Containerised Workflow-based Cloud Control Experimental System}\label{The Structure of Containerised Cloud Workflow Processing System}
\end{figure}

\subsection{Structure of the experimental system}\label{Structure of testbed}

The structure of this containerised cloud control system is provided as Fig. \ref{The Structure of Containerised Cloud Workflow Processing System}. This system is made up of the cloud control platform and edge controlled node. In the edge controlled node, an edge data transfer is set to exchange data with the cloud controller. The data transfer also exchanges data with the controlled plant, which could be real dynamical system or mathematical model of simulation.

In the cloud control platform, there are task manager, Docker register, DAG file, remote dictionary storage (redis) server and containerised cloud workflow of which the tasks run in the overlay container network \cite{nelson2016mastering}. First, the redis server is created by the task manager. The redis server could share data online among different hosts via network. Thus, the redis server is used as a data transponder in the cloud environment. Then, the task manager defines the DAG files as `\emph{Map}' structure and writes the file into the redis server. When the workflow containers are created, the redis server would provide guides for the new containers based on the written DAG file, of which the details are provided in Section \ref{Initialization and execution of system}. The guide information contains:
\begin{itemize}
  \item The task index and the level to this task belongs.
  \item The pre- and post- dependence relationships of this task.
  \item The image which would be pulled from Docker register.
\end{itemize}

Besides, the task manager creates the images of workflow task and uploads the images to Docker register, which services as an image hub. Finally, the task images are pulled from Docker register and the workflow containers are created in the across-hosts overlay network. After the initialization guided by the redis server, the workflow-based DPC controller starts running.

\begin{figure*}[!ht]
  \centering
  \vspace{1em}
  \includegraphics[width=6in]{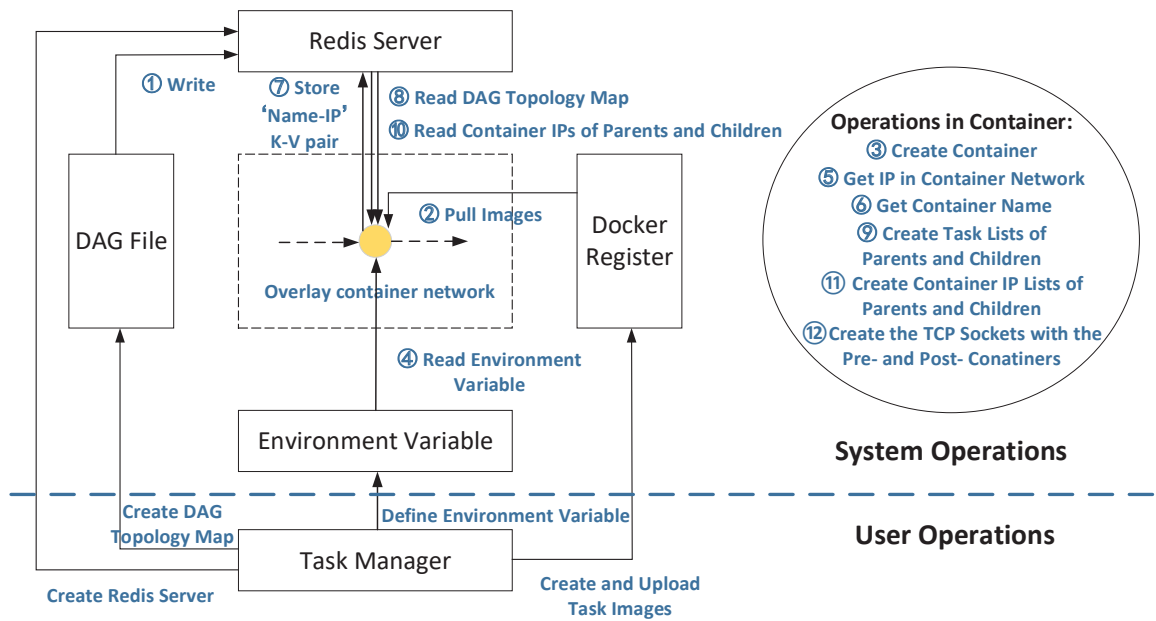}\\
  \caption{Initialization Procedure of the Containerised Workflow-based Cloud Control Experimental System}\label{The Initialization Procedure of Containerised Cloud Control Testbed}
\end{figure*}

\subsection{Initialization and execution of system}\label{Initialization and execution of system}

\subsubsection{Procedure of the system initialization}

In this containerised workflow-based cloud control experimental system, the initialization of containers is an important procedure. When the containers in workflow are created, the containers have no knowledge about the position, level, IP and pre-/post- relationships in the whole workflow of each task. As a result, the workflow-based DPC controller could not be executed. Therefore, the initialization of containers is necessary for starting the workflow-based cloud controller. Fig. \ref{The Initialization Procedure of Containerised Cloud Control Testbed} shows the initialization procedure of the experimental system. The procedure is divided into the user operations and system operations. The user operations includes creating the redis server, defining DAG file, etc which are finished off-line. As for the online part, the system operations are divided into 12 steps, as listed in the below.

\begin{enumerate}
  \item Write the defined DAG file including DAG topology map into the redis server.
  \item Pull the images of tasks to the overlay network environment from Docker register.
  \item Create the workflow containers as the pulled images.
  \item Read the adjustable environment variables as the controller parameters such as the scale parameters $N$ \!and\! $j$.
  \item After the containers start, obtain the IP of each container in the overlay container network.
  \item Obtain the name of each task from the environment variables.
  \item Store the `\emph{Name-IP}' pair as the `\emph{key-value}' mode into the redis server.
  \item The containers read the defined DAG topology map from the redis server.
  \item Create the task lists of the parents and children of each container based on the loaded DAG topology map and task name.
  \item Read the container IPs of parents and children of each task container.
  \item Create container IP lists of the parents and children tasks.
  \item Create the TCP sockets with the pre- and post- containers.
\end{enumerate}

\subsubsection{Pyramid structure of system initialization}

When the containers of the DPC workflow are created, the times of initialization including obtaining necessary information such as DAG map, task index, position, IP, port, etc are different for different containers. Some containers might finish the course earlier, while the others are initialized lately. The inconsistency of initialization would bring new uncertainty for the execution of the DPC workflow. Therefore, in this system, an initialization scheme of which the communication structure seems like pyramid is designed.

\begin{figure}[!ht]
  \centering
  \includegraphics[width=3.5in]{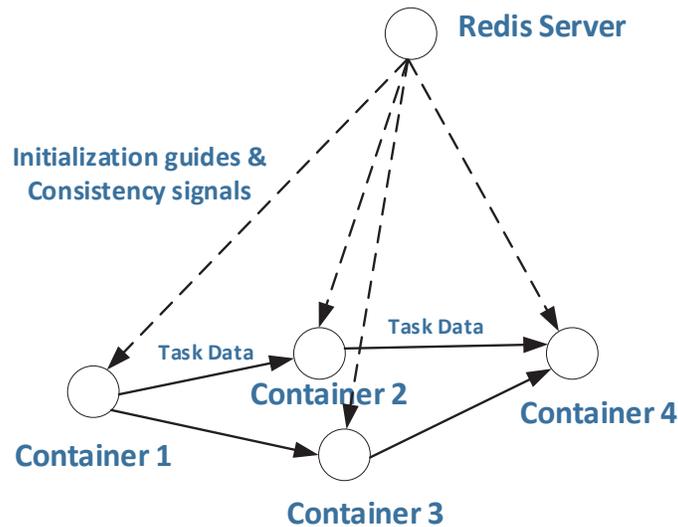}\\
  \caption{Pyramid Communication Structure}\label{The Pyramid Communication Structure}
\end{figure}

The pyramid communication structure is indeed an abstract of the relationship of the redis server and the task containers of a workflow. As shown in Figure \ref{The Pyramid Communication Structure}, the redis server plays the role of guide and controller for the task containers of workflow. By several rounds of `\emph{key-value}' operations, the containers obtain the necessary information to execute the workflow. This stage is the initialization of the designed system. When a container is ready, it would send a `\emph{ready}' signal to the executor (VM or container) in which the redis server runs. When all the `\emph{ready}' signals are collected and acknowledged, this executor would send the `\emph{start}' signal to each container at the same time. The time differences of the containers receiving the `\emph{start}' signal is very small and could be ignored. Thus, the task containers could be considered as starting at the same time.

The practical containerised workflow and the communications between containers are established. Then, the redis server would not take part in the workflow execution unless being invoked. Thus, the lines between the redis server and containers are doted, and this structure is named as pyramid. Up to here, the problem of initialization is solved. The worklflow-based DPC controller is ready to execute the control missions.

\begin{figure}[!h]
  \centering
  \includegraphics[width=4in]{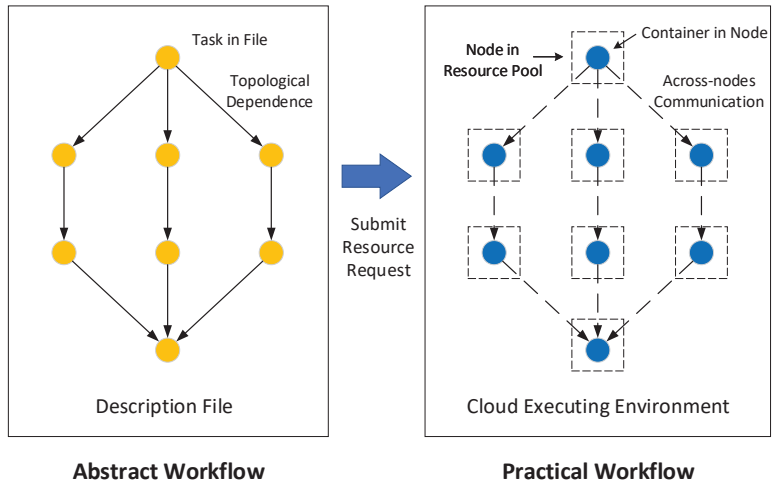}\\
  \caption{Mapping Relationship Between Abstract and Practical Workflows}\label{The Map Relationship Between Abstract and Practical Workflows}
\end{figure}

\subsubsection{Execution of system}

After the initialization procedure is finished, the practical DPC workflow is created and executed. In the cloud environment, the practical containerised DPC workflow is created referring to the abstract workflow template by the the experimental system, as presented in Fig. \ref{The Map Relationship Between Abstract and Practical Workflows}. Notice that Fig. \ref{The Map Relationship Between Abstract and Practical Workflows} shows the mapping relationship of the abstract and practical workflows, but does not represent any specific workflow structure of the DPC method. Each task in the abstract workflow would be translated to an executing container scheduled into a processing node such as VM. The inter-dependencies between the tasks would be translated into the across-nodes communications to transmit the intermediate data. Then, the practical DPC workflow is executed in the overlay container network to conduct the control missions. When the practical DPC workflow is repeated one time, a new cloud control variable and a pair of cloud parameters would be obtained.

\subsection{Communication scheme of containerised workflow-based cloud control experimental system}\label{Communication scheme of containerised workflow-based cloud control system}

In this system, the cloud controller and controlled plant are deployed in different network environments where the problems such as containers communication and the transfers of IP and port, etc arise. Thus, a real-time, stable and high-quality communication scheme is required. The structure of the communication scheme is provided in Fig. \ref{The Structure of Containerised Cloud Workflow Processing System}. There are the overlay container network for the workflow-based cloud controller, IP mapping from cloud network to edge network and port mapping from edge network to cloud network, topology switching scheme in data-driven control in this scheme.

\begin{figure}[!ht]
  \centering
  \includegraphics[width=4.5in]{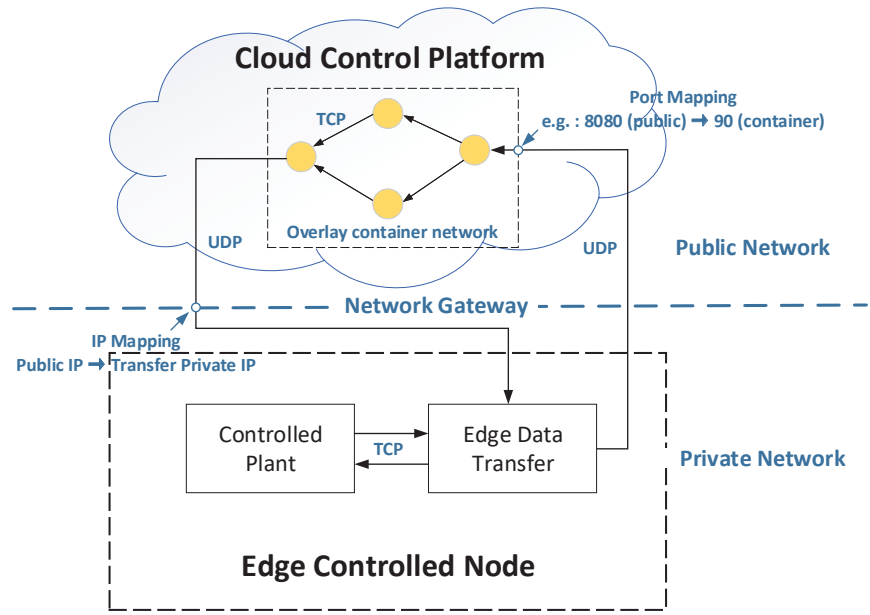}\\
  \caption{Structure of the Communication Scheme}\label{The Structure of Containerised Cloud Workflow Processing System}
\end{figure}

In the cloud-edge communication, the protocol is chosen as user datagram protocol (UDP) because of its high speed and the low enough packet loss rate when crossing different networks. In the cloud environment, since the parallel multiple communications among the task containers, transmission control protocol (TCP) is more suitable as its ability of connection-oriented point-to-point transmission to avoid the communication conflicts. In the edge controlled node, both of TCP and UDP are proper in the communication and TCP is applied in this work.

%
%

\subsubsection{Data transmission in the workflow-based controller via overlay container network}

In the workflow-based problem, the containers are created and scheduled into different processing nodes. In the cloud environment, the processing nodes could be physical machines or VM. In this work, all the processing nodes are set as VMs and connected by network. However, the containers distributed in different processing nodes could not communicate with other containers by the network for VMs. Therefore, a high-level network is required for the temporary containers with different hosts, which is the container overlay network \cite{suo2018analysis}. The container overlay network in the clusters is set as calico protocol, a scalable network to build the connections among containers. A high-level overlay network segment which is prepared for the containers by the calico protocol. When a new container is created, this protocol dispatches a unique overlay IP in the built the overlay network segment to communicate across the existing different hosts.

\subsubsection{IP mapping from cloud network to edge network}

Since the amount of the public IP addresses is limited and each device running in the edge controlled node also needs an IP address, network and port translation (NAPT) technology is adopted in local area network (LAN). In the edge node, all the devices share the same public IP and this public IP is mapped into different edge IPs for the devices in LAN. However, the cloud controller has no knowledge about the hosts of the edge devices at the beginning. Thus, an IP transfer method is designed. The edge controlled plant sends an information group consisting of the input and output signals and internal tuple to the cloud controller. Then, the cloud controller obtains the network host of this edge plant. Furthermore, when the message is received by the cloud controller or edge plant, the receiver would check the attached port of the sender. If the ports of the receiver and sender are matched, the acknowledge process is completed.

\subsubsection{Port mapping from edge network to cloud network}

When the containers run in the cloud control platform, the displayed IP for the edge controlled node is the IP of the hosted VM. Besides, multiple containers share the same IP of virtual machine. Therefore, when the output data is sent from the edge controlled node, the entry container could not be found. To solve this problem, the port mapping is used here. For example, the outside port `\emph{8080}' of host is mapped to the inner port `\emph{90}' of the entry container. Thus, the output data could be transmitted to the workflow-based controller directly.

\subsubsection{Topology switching scheme in data-driven control}\label{Topology switching scheme in data-driven control}

\begin{figure}[!ht]
  \centering
  \includegraphics[width=4.5in]{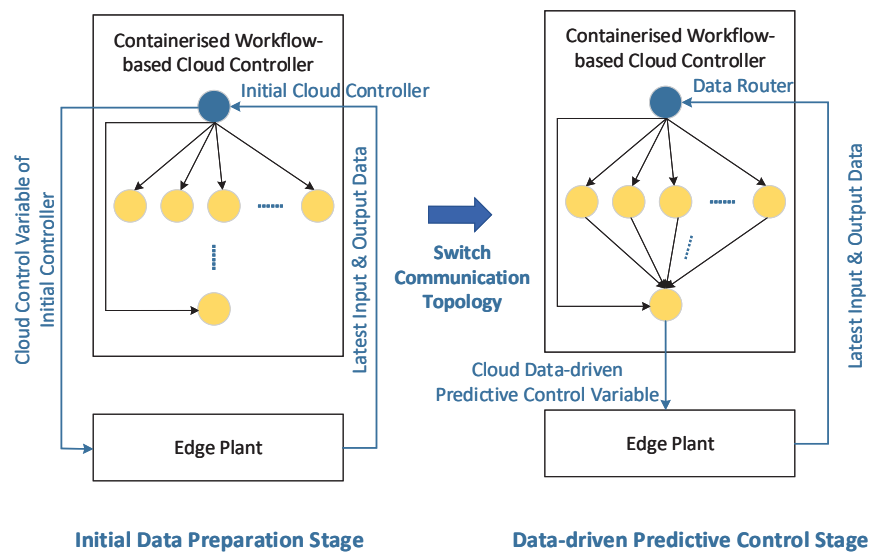}\\
  \caption{The Structure of Containerised Cloud Workflow Processing System}\label{4}
\end{figure}

As described in Algorithm \ref{Data-driven Predictive cloud control system Algorithm}, the initial data preparation occupies an important role in the DPC method. In the stage of preparing information, other basic control algorithm could be applied. But when the initial data is enough, the communication topology needs change because of the different requirements for the communication structures of the basic algorithms and the DPC algorithm. In the stage of initial data preparation, the edge plant connects to the initial cloud controller directly. This initial cloud controller runs in a single container. And this initial cloud controller sends the latest data (the packet of input and output data) from the edge controlled node in each control period. In the same time, the time-costing establishments of the Hankel matrices in the workflow containers are executed. Otherwise, there would exist a considerable time delay in the switching period, since the establishments of the Hankel matrices are required to be finished in one period.

When the communication topology turns to the DPC stage, the initial cloud controller becomes a data router in the cloud environment but not an independent controller. In each control period, the data router still plays the role of the entry container and receives the latest data from the edge controlled node. Then, the data router dispatches the required data to the children tasks in the workflow-based cloud controller. The update method of the Hankel matrices costs relatively low computation burden. Next, the containers of the workflow-based cloud controller execute the tasks owned by themselves. The export container collects the data from its parents and calculates the cloud control variable as well as cloud parameters. Finally, the cloud control variable and cloud parameters are sent to the edge controlled node.

\section{Evaluation and Discussion}\label{Experiment and discussion}

In this section, three control examples, including the ball-beam system, vehicle tracking system and high-dimension numerical system, were conducted to evaluate the effectiveness of the proposed workflow-based cloud control methods and system. First, the experiment setup is described in detail. Second, four DPC methods to be tested are listed. Then, a brief of three control examples is provided. Finally, the detailed results and discussions of the three control examples are presented, respectively.

\subsection{Evaluation setup}

The calculation of the DPC controller, including the existing native DPC and workflow-based DPC are practically carried out in Alibaba Cloud. Furthermore, the workflow-based DPC methods are conducted based on the containerised workflow-based cloud control experimental system established in Section \ref{Design and implementation of containerised cloud control testbed}. The acceleration of control algorithm is achieved in reality by the workflow-based method provided in Section \ref{Establishment method of data-driven predictive control workflow}, rather than relying on the numerical simulator. In the edge controlled node, the cloud-edge composite control variable is generated, as presented in Section \ref{DOB-based cloud-edge collaborative control scheme}. Then, the composite control variable is delivered to the controlled plant, which is simulated by scripting codes. Thus, the evaluation is made up of the practical cloud experiment of the controller and the edge simulation of the controlled plant.

\begin{figure}[!ht]
  \centering
  \includegraphics[width=3.5in]{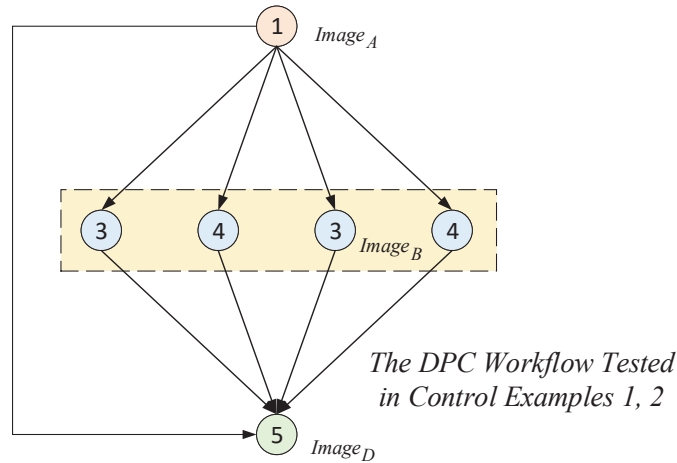}\\
  \vspace{-0.1cm}
  \caption{Workflow Structure Tested in Control Examples 1, 2 ($MPT = 4$)}
  \label{Workflow Structure Tested in the Control Examples 1, 2}
\end{figure}

There are two structure designs of the DPC workflow with different MPT, which are used in the evaluation. The first structure is designed with the MPT being 4 and the task number being 6, as shown in Fig. \ref{Workflow Structure Tested in the Control Examples 1, 2}. This DPC structure is used in the ball-beam system and vehicle tracking system, of which have strong real-time requirement and relatively less data. The second structure applies the design of Fig. \ref{Workflow Structure of DPC Method} with the MPT being 10 and the task number being 19, which is carried out in the high-dimension control example. Besides, there are three kinds of cloud resource pool in the evaluation. The resources and node cluster information are listed in Table \ref{Resources and node cluster information}. The first group is used to carry out the baseline, which the DPC controller is deployed in a single cloud server and calculated in the centralized mode. The second and third groups apply two \emph{ecs.hfc6.4xlarge} nodes and each node has 16 CPU and 32 GB memory. The overlay network in the cluster is set as calico protocol, a scalable network to build the connections between containers. The version of redis is 3.5.3. The containerised cloud control system and the tasks of the DPC workflows are implemented by Python 3.7. The basic scientific libraries used in the implementation of the DPC workflows are Numpy (1.19.5) and Scipy (1.5.4), respectively. The operation system is Ubuntu 20.04.

\begin{table}[!ht]
	\centering
	\caption{Resources and node cluster information}
	\begin{tabular}{|c|c|c|c|}
		\toprule  
		\textbf{Experiment Group}&\textbf{Nodes}&\textbf{Total Resource Pool}&\textbf{Node Type}  \\
		\hline  
        \hline
        Group 1&1 & 4 CPU, 8GB&\emph{ecs.hfc6.xlarge} \\
        \hline
        Group 2&2 & 32 CPU, 64GB&\emph{ecs.hfc6.4xlarge}\\
        \hline
        Group 3&4 & 64 CPU, 128GB &\emph{ecs.hfc6.4xlarge}\\
		\bottomrule  
	\end{tabular}
\label{Resources and node cluster information}
\vspace{-0.5em}
\end{table}

\subsection{Four tested cloud-based DPC methods}

There are four cloud-based DPC methods tested in the evaluation. The former two methods are the native DPC, of which the computation mode of the cloud controller is the traditional centralized structure \cite{gao2021design}, rather than the workflow-based structure. The last two methods are the proposed workflow-based DPC in this paper. The details are provided in the below.

\begin{enumerate}
  \item \textbf{Native DPC:} a native DPC controller is deployed in the cloud server, and no extra component is deployed in the edge controlled node. The first control variable in the received predicted control sequence is applied to the controlled plant.
  \item \textbf{Native DPC with time delay compensator:} a native DPC controller is deployed in the cloud server, and a time delay compensator is deployed in the edge controlled node. The cloud controller sends a predicted control sequence, and the time delay compensator selects a proper element to reduce the influence of time delay.
  \item \textbf{Workflow-based DPC:} a workflow-based DPC controller is deployed in the cloud resource pool, and no extra component is deployed in the edge controlled node.
  \item \textbf{Workflow-based DPC with DOB:} a workflow-based DPC controller is deployed in the cloud resource pool, and a data buffer and a DOB-based compensator are deployed in the edge controlled node to reduce the uncertain caused in the cloud workflow processing.
\end{enumerate}

\subsection{Brief results of the three control examples}

There are three control examples in this section, which are the ball-beam system, vehicle tracking system and high-dimension numerical system. The former two control examples were conducted to detected the computation efficiency of the real-time control performance by the proposed worklow-based DPC methods and system. The third control example is a 78-dimension numerical system, of which the structure of the state-space model is built according to a power network system. This control example is applied to detected the computation efficiency for the high-dimension system.

\begin{enumerate}
  \item \textbf{Control example 1:} the regulating problem of ball-beam system. The DPC workflow with MPT $=4$ was applied in the second group of cloud resource pool. By using the workflow-based method, the computation time of cloud controller was reduced from 25.1917 ms to 13.8064 ms, of which the percentage of reduction was 45.19$\%$.
  \item \textbf{Control example 2:} the tracking problem of vehicle system. The DPC workflow with MPT $=4$ was applied in the second group of cloud resource pool. When the speed was set as 30 km/h, the computation time was reduced from 35.2602 ms to 9.0432 ms. The total time delay, consisting the cloud-edge round-trip communication delay, was reduced from 47 ms to within 20 ms, which is the preset control period. As the result, the native DPC methods could not control the vehicle, while the workflow-based DPC with DOB achieved the stable control performance.
  \item \textbf{Control example 3:} the control problem of high-dimension numerical system. The DPC workflow with MPT $=10$ was applied in the third group of cloud resource pool. The evaluation was repeated with different numbers of the retained singular values, which is another form of determining the truncation precision $\epsilon_{1}$. The computation time was reduced from 367.4199 ms to 54.7559 ms, of which the percentage of reduction was 85.10$\%$. The relationship between the computation efficiency and the number of the retained singular values, and that between the control performance and the number of the retained singular values were also discussed.
\end{enumerate}

\subsection{Control example 1: ball-beam system}

The first control example is the ball-beam system. This example is designed to detect the computation efficiency and control performance for the real-time regulating problem, which means controlling the system output to a constant reference value.

\subsubsection{System model}

\begin{figure}[!htb]
  \centering
  \includegraphics[width=3.5in]{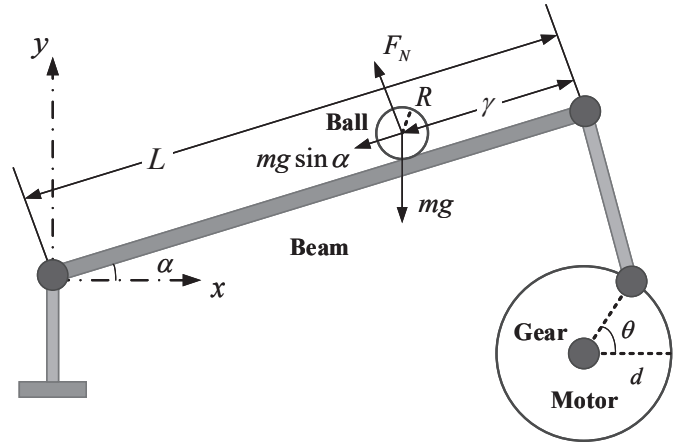}
  \caption{Model of the Ball-beam System}
  \label{Model of the Ball-beam System}
\end{figure}

The structure of the ball-beam system is shown in Fig. \ref{Model of the Ball-beam System}. The main variables and parameters of this ball-beam system are listed in Table \ref{symbol description}. By choosing the ball position $\gamma$ and beam angle $\theta$ as the generalized position coordinates for this system \cite{rahmat2017application}, the Lagrangian equation of motion is given by:
\begin{equation}\label{e77}
  \small{0 = (\frac{J_b}{R^{2}}+m)\ddot{\gamma}+mg\sin\alpha-m\gamma\dot{\alpha}^{2}}
\end{equation}
where $\ddot{\gamma}$ is the acceleration of the ball, $\alpha$ is the beam angle and $\dot{\alpha}$ is the angular velocity of the beam angle. When this system approaches the stable point $\alpha=0$, the local linearization of (\ref{e77}) can be obtained. Since $\alpha$ is small at this point,  $\dot{\alpha}\approx0$ and $\sin\alpha=\alpha$. Therefore, the linear approximation of this system is given as
\begin{equation}\label{e78}
  \small{\ddot{\gamma} = -\frac{mg}{\frac{J_b}{R^{2}}+m}\alpha.}
\end{equation}

\begin{table}[!htb]
    \centering
    \caption{Symbol description and parameter}
    \label{tab1}
    \begin{tabular}{c | ccc | c}
    \hline\hline
    \textbf{Symbol}        & &\textbf{Description} &     & \textbf{Parameter}\\
    \hline
    $L$        & &Length of the bean  &     & 0.4 $m$\\
    \hline
    $d$        & &Radius of the gear  &     & 0.04 $m$\\
    \hline
    $R$ 	 & &Radius of the ball &    & 0.015 $m$\\
    \hline
    $J_b$      &  &Moment of inertia of the ball &    & $9.9\times10^{-6}$ $kg\cdot m^{2}$\\
    \hline
    $m$          &   &Mass of the ball &    & 0.11 $kg$ \\
    \hline
    $g$	    & &Acceleration of gravity  &   & 9.81 $m/s^{2}$\\
    \hline
    $F_N$	    & &Support force on the ball  &   &$-$\\
    \hline
    \hline
    \end{tabular}
    \label{symbol description}
\end{table}

The beam angle $\alpha$ could be expressed related to the gear angle $\theta$ by approximating linear equation $\alpha L= \theta d$. Thus, the relation between $\ddot{\gamma}$ and $\theta$ is
\begin{equation}\label{e79}
  \small{\ddot{\gamma} = -\frac{mdg}{L(\frac{J_b}{R^{2}}+m)}\theta.}
\end{equation}

In this system, the motor gear angle $\theta$ is set as the control input $u$. Letting the state be $x=[x_1\ x_2\ x_3\ x_4]^{T}=[\gamma\;\;\dot{\gamma}\;\;\theta\;\;\dot{\theta}]^{T}$ and output be $y = \gamma$, i.e., the position of the ball, the state-space model is obtained as
\begin{eqnarray}
  \small{\left[\!
    \begin{array}{c}
      \dot{x}_1 \\
      \dot{x}_2 \\
      \dot{x}_3 \\
      \dot{x}_4 \\
    \end{array}\!
  \right] \!\!\!\!\!\!}&=&\small{\!\!\!\!\!\! \left[\!
            \begin{array}{cccc}
              0 & 1 & 0 & 0 \\
              0 & 0 & -\frac{mdg}{L(\frac{J_b}{R^{2}}+m)} & 0 \\
              0 & 0 & 0 & 1 \\
              0 & 0 & 0 & 0 \\
            \end{array}\!
          \right]\!\!\!\left[\!
    \begin{array}{c}
      x_1 \\
      x_2 \\
      x_3 \\
      x_4 \\
    \end{array}\!
  \right]\!\!+\!\!\left[\!
    \begin{array}{c}
      0 \\
      0 \\
      0 \\
      1 \\
    \end{array}\!
  \right]\!\!u} \\
  \small{y \!\!\!\!\!\!}&=&\small{\!\!\!\!\!\! \left[\!
      \begin{array}{cccc}
        1 & 0 & 0 & 0 \\
      \end{array}\!
    \right]\!\!\left[\!
    \begin{array}{c}
      x_1 \\
      x_2 \\
      x_3 \\
      x_4 \\
    \end{array}\!
  \right].}
\end{eqnarray}

The initial data was provided by PID algorithm with $K_P =9.0, K_I=3.0$ and $K_D=7.5$. The parameters of the proposed method were set as $N=30, j=1500, \lambda=0.031, \epsilon_{1}=10^{-15}$ and $L=50$. Both of the discretization period and control period were set as 20 ms The reference output in PID stage was set as 0.2 m. When the control is switched to the DPC stage, the reference output changed to 0.1 m.

\subsubsection{Computation efficiency}

\begin{figure}[!ht]
  \centering
  \includegraphics[width=4in]{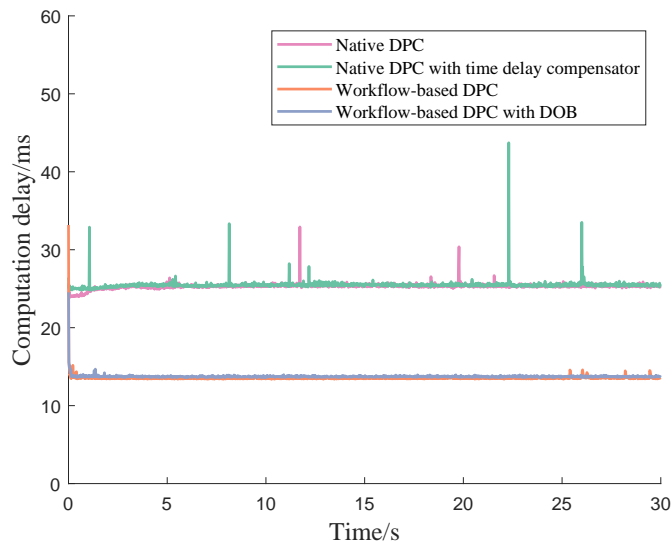}\\
  \caption{Computation Delays in Cloud Controller of the Ball-beam System}
  \label{Computation Delays in Cloud Controller of the Ball-beam System}
\end{figure}

The computation delays in the cloud controller are recorded in Table \ref{Recorded computation times of control example 1}. The results of the computation delay could be divided into two classes. The first class consists of the DPC and the native DPC with time delay compensator. Since the calculation mode of both the two methods are centralized, the computation times of the two methods were similar, which were 25.1917 ms and 25.3186 ms, respectively. The second class is made up of the workflow-based DPC and the workflow-based DPC with DOB. The computation times of the two methods were 13.8064 ms and 13.5918 ms, respectively. Since the computation efficiencies of the same computation mode were similar, we calculated the reduced proportion by the native DPC method without time delay compensator and the workflow-based DPC without DOB-based compensator, of which the result was $45.19\%$. The computation delays were also shown in Fig. \ref{Computation Delays in Cloud Controller of the Ball-beam System}. From this figure, the peak burrs occurred both in the first two methods. But in the last two methods, the recorded computation delays were more stable.

\begin{table}[!ht]
	\centering
	\caption{Recorded computation times of control example 1}
	\begin{tabular}{|c|c|}
		\toprule  
		\textbf{Method}&\textbf{Computation time (ms)} \\
		\hline  
        \hline
        Native DPC& 25.1917 \\
        \hline
        Native DPC with time delay compensator& 25.3186 \\
        \hline
        Workflow-based DPC& 13.8064 \\
        \hline
        Workflow-based DPC with DOB& 13.5918  \\
		\bottomrule  
	\end{tabular}
\label{Recorded computation times of control example 1}
\vspace{-0.5em}
\end{table}

\begin{figure}[!ht]
  \centering
  \includegraphics[width=4in]{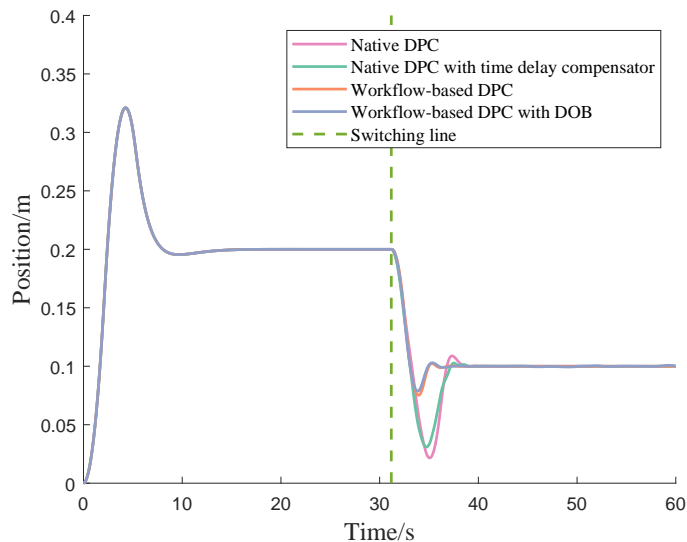}\\
  \caption{Control Results of the Ball-beam System}
  \label{Control Results of the Ball-beam System}
\end{figure}

\subsubsection{Control performances}

The control performances of the ball-beam system is shown in Fig. \ref{Control Results of the Ball-beam System}. The result shows that the two workflow-based methods obtained better performances than the native methods. In detail, the two native DPC had relatively high overshoot because of the affect of high time delays. Compared with the native method, the method with time delay compensator could bring some improvement of the control performance. But since the time delays had not been reduced, the improvement of control performance was limited. The two workflow-based methods actively reduced the computation delay, and so obtained more obvious improvement. In the workflow-based DPC method, some low-value information were lost in the distributed SVD process. Thus, the DOB was deployed to estimate the uncertain brought by the truncation operation. As a result, the workflow-based DPC with DOB achieved the best control performance.

\subsection{Control example 2: vehicle tracking system}

The second control example is the trajectory tracking problem of vehicle system, which is to control the vehicle position as the varying reference signals. In this example, the problem of trajectory tracking control is studied based on the kinematic error model \cite{kong2015kinematic}.

\subsubsection{System model}

\begin{figure}[!htb]
  \centering
  \includegraphics[width=4in]{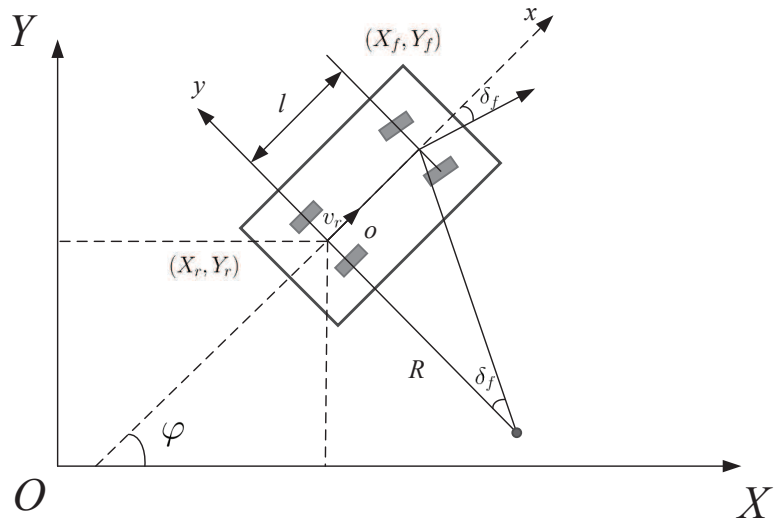}
  \caption{Model of the Vehicle System}
  \label{Model of the Vehicle System}
\end{figure}

As shown in FIg. \ref{Model of the Vehicle System}, we define the X-axis of the inertial coordinate system points to the east, the Y-axis points to the north, and the Z-axis is positive upward. In the meanwhile, we define the \emph{x}-axis of the vehicle coordinate system points to the front of vehicle, and the \emph{y}-axis points to the left of vehicle. Then, the vehicle speed at the rear axle center is defined as $v_r$. The yaw angle of vehicle $\varphi$ is defined as the angle between the X-axis of the inertial coordinate system and the \emph{x}-axis of the vehicle coordinate system, where the counterclockwise direction is positive. $(X_r, Y_r)$ and $(X_f, Y_f)$ represent the coordinates of the rear axle center and the front axle center in the inertial coordinate system, respectively. $l$ is defined as the wheel base and $R$ is the instantaneous steering radius of the rear axle center. $\delta_{f}$ is defined as the drift angle of the front wheel.

The vehicle speed at the rear axis $v_r$ is calculated by
\begin{equation}\label{veh1}
  v_r = \dot{X}_{r}\cos\varphi + \dot{Y}_{r}\sin\varphi.
\end{equation}

The kinematic constraints of the front and rear axles are defined as
\begin{equation}\label{veh2}
  \left\{
    \begin{array}{ll}
        \dot{X}_f\sin(\varphi+\theta_{f}) - \dot{Y}_{f}\cos(\varphi+\theta_{f}) = 0 \\
        \dot{X}_r\sin\varphi - \dot{Y}_{r}\cos\varphi = 0
    \end{array}
  \right.
\end{equation}

Combining the equations (\ref{veh1}) and (\ref{veh2}), we have
\begin{equation}\label{veh3}
  \left\{
    \begin{array}{ll}
        \dot{X}_{r} = v_r\cos\varphi \\
        \dot{Y}_{r} = v_r\sin\varphi
    \end{array}
  \right.
\end{equation}

Based on the geometric relationships of the front and rear wheels, we could obtain
\begin{equation}\label{veh4}
  \left\{
    \begin{array}{ll}
        X_{f} = X_{r} + l\cos\varphi \\
        Y_{f} = Y_{r} + l\sin\varphi
    \end{array}
  \right.
\end{equation}

Substituting (\ref{veh3}) and (\ref{veh4}) to (\ref{veh2}), we could obtain the yaw velocity $\omega$ as
\begin{equation}\label{veh5}
  \omega = \frac{v_r}{l}\tan \delta_{f}.
\end{equation}

In the meanwhile, based on $\omega$ and $v_r$, the steering radius $R$ and the drift angle of the front wheel $\delta_{f}$ could be obtained as
\begin{equation}\label{veh6}
  \left\{
    \begin{array}{ll}
        R = \frac{v_r}{\omega}, \\
        \delta_{f} = \arctan(\frac{l}{R}).
    \end{array}
  \right.
\end{equation}

\begin{figure}[!htb]
  \centering
  \includegraphics[width=4in]{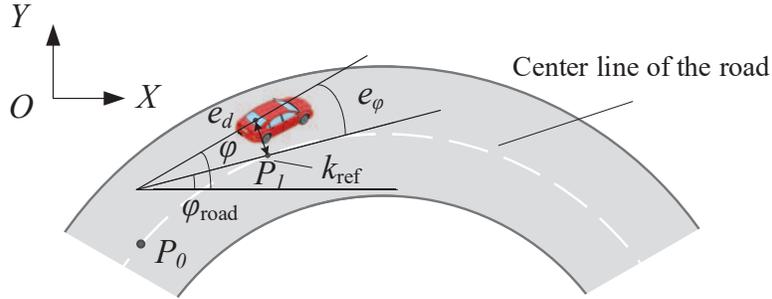}
  \caption{Model of the Vehicle Error System}
  \label{Model of the Vehicle Error System}
\end{figure}

Then, based on the above kinematic relationships of vehicle system, the kinematic error model is derived, of which the schematic diagram is provided in Fig. \ref{Model of the Vehicle Error System}. The distance error $e_d$ is defined as the difference between the rear axle center and its projection point on the center line of the road, where the left side of the road is the positive direction. The the angle error $e_{\varphi}$ is defined as $e_{\varphi} = \varphi - \varphi_{road}$, where $\varphi_{road}$ is the angle between the tangential direction of the center line of the road and the X-axis of the inertial coordinate system and the counterclockwise direction is positive.

Assuming $P_1$ is the projection point of the rear axle center on the center line of the road, $s$ is the arc length between $P_1$ and a reference point $P_0$ on the road, and $k_{ref} = \frac{1}{R}$ is the curvature at the $P_1$ location. Then, the moving speed of the $P_1$ along the center line of the road $\dot{s}$ could be provided as
\begin{equation}
  \dot{s} = \frac{1}{1-k_{ref}e_d}[v_x\cos(e_{\varphi})+v_y\sin(e_{\varphi})]
\end{equation}
where $v_x$ and $v_y$ are the components of the vehicle speed at the rear axle center $v_r$ along the \emph{x}-axis and \emph{y}-axis of the vehicle system, respectively.

The equations of trajectory tracking error could be expressed as
\begin{equation}\label{veh7}
  \left\{
    \begin{array}{ll}
        \dot{e}_{\varphi} = \dot{\varphi} - k_{ref}\dot{s}, \\
        \dot{e}_{d} = v_x\sin(e_{\varphi})+v_y\cos(e_{\varphi}).
    \end{array}
  \right.
\end{equation}

Then, based on the hypothesis of small angle ($\sin(e_{\varphi})\approx e_{\varphi}, \cos(e_{\varphi}) \approx 1$) and letting $k_{ref}e_d \approx 0$, the equations of trajectory tracking error could be simplified as
\begin{equation}\label{veh8}
  \left\{
    \begin{array}{ll}
        \dot{e}_{\varphi} = \dot{\varphi} - \frac{k_{ref}(v_x+v_ye_{\varphi})}{1-k_{ref}e_d} \approx \dot{\varphi} - k_{ref}(v_x+v_ye_{\varphi}), \\
        \dot{e}_d = v_xe_{\varphi} + v_y.
    \end{array}
  \right.
\end{equation}

Ignoring the lateral speed of the vehicle $v_y$, we could obtain
\begin{equation}\label{veh9}
  \left[\!
    \begin{array}{c}
      \dot{e}_{\varphi} \\
      \dot{e}_{d} \\
    \end{array}\!
  \right] \!=\! \left[
              \begin{array}{cc}
                0 & 0 \\
                v_x & 0 \\
              \end{array}
            \right]\!\left[\!
                     \begin{array}{c}
                       e_{\varphi} \\
                       e_{d} \\
                     \end{array}\!
                   \right]\!+\!\left[\!
                             \begin{array}{c}
                               \frac{v_x}{l} \\
                               0 \\
                             \end{array}\!
                           \right]\!\tan(\delta_f)\!+\!\left[\!
                                                   \begin{array}{c}
                                                     -v_x \\
                                                     0 \\
                                                   \end{array}\!
                                                 \right]\!k_{ref}
\end{equation}

Based on (\ref{veh5}) and (\ref{veh6}), we could obtain $k_{ref}=\tan(\frac{\delta_{ref}}{l})$, where $\delta_{ref}$ represents the forward control variable provided by the reference trajectory. Letting $\textbf{u}_1=\tan(\delta_f)\approx \delta_f, \textbf{u}_2=\tan(\delta_{ref})\approx \delta_{ref}, \boldsymbol{\xi} = [e_{\varphi}, e_{d}]^{T}$, the system (\ref{veh9}) could be written as
\begin{equation}\label{veh10}
  \boldsymbol{\dot{\xi}} = A\boldsymbol{\xi} + B_1\textbf{u}_1 + B_2\textbf{u}_2
\end{equation}
where $A\!=\!\left[\!\!
           \begin{array}{cc}
             0 & 0 \\
             v_x & 0 \\
           \end{array}\!\!
         \right], B_1 \!=\! \left[\!\!
                          \begin{array}{cc}
                            \frac{v_x}{l} & 0 \\
                          \end{array}\!\!
                        \right]^{T}\!, B_2 \!= \!\left[\!\!
                                             \begin{array}{cc}
                                               -\frac{v_x}{l} & 0 \\
                                             \end{array}\!\!
                                           \right]^{T}\!.
$

In this control example, two groups of evaluations were conducted with different vehicle speeds $v_r$, which were 20 km/h and 30 km/h. The vehicle is controlled to track the circle trajectory, of which the steering radius $R$ was 42 m. Thus, the curvature of the trajectory $\kappa_{ref}$ was set as 0.024 m$^{-1}$. The wheel base $l$ was set as 0.5 m. The initial data was provided by LQR algorithm with $Q = 20\cdot I_2, R = I$. Both of the discretization period and control period were set as 20 ms. The parameters of the proposed method were set as $N=20, j=1000, \lambda=0.0041$ and $L=-0.2$.

\subsubsection{Computation efficiency}

\begin{figure*}[t]
\subfigure[\scriptsize{Computation delay (20 km/h, $\epsilon_1 \!=\! 10^{-\!4}$)}]{
\begin{minipage}[b]{0.5 \textwidth}
\includegraphics[width=\textwidth]{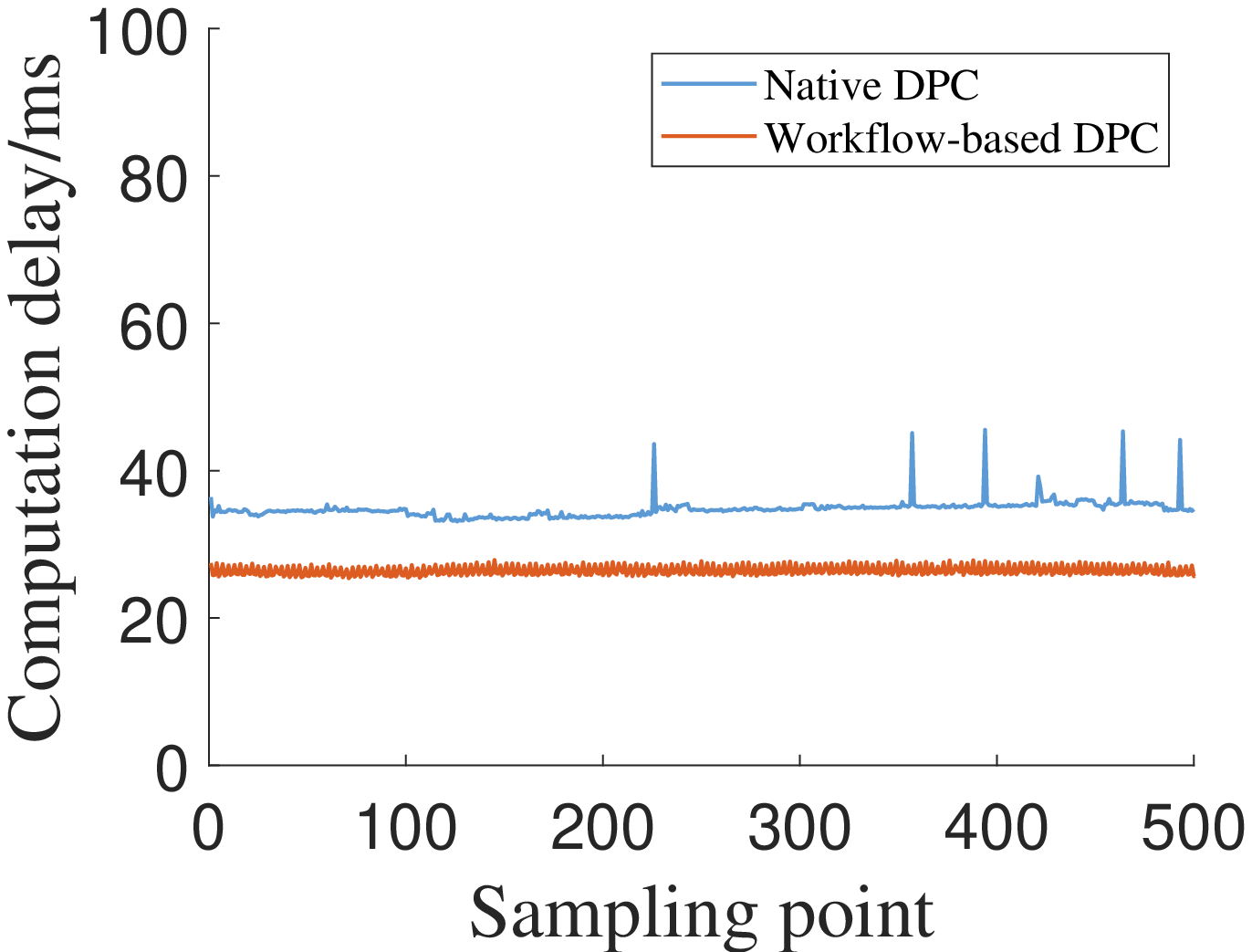}
\label{Comparison of computation delay (20 km/h)}
\end{minipage}
}
\hspace{-0.2in}
\subfigure[\scriptsize{Total delay (20 km/h, $\epsilon_1 \!=\! 10^{-\!4}$)}]{
\begin{minipage}[b]{0.5 \textwidth}
\includegraphics[width=\textwidth]{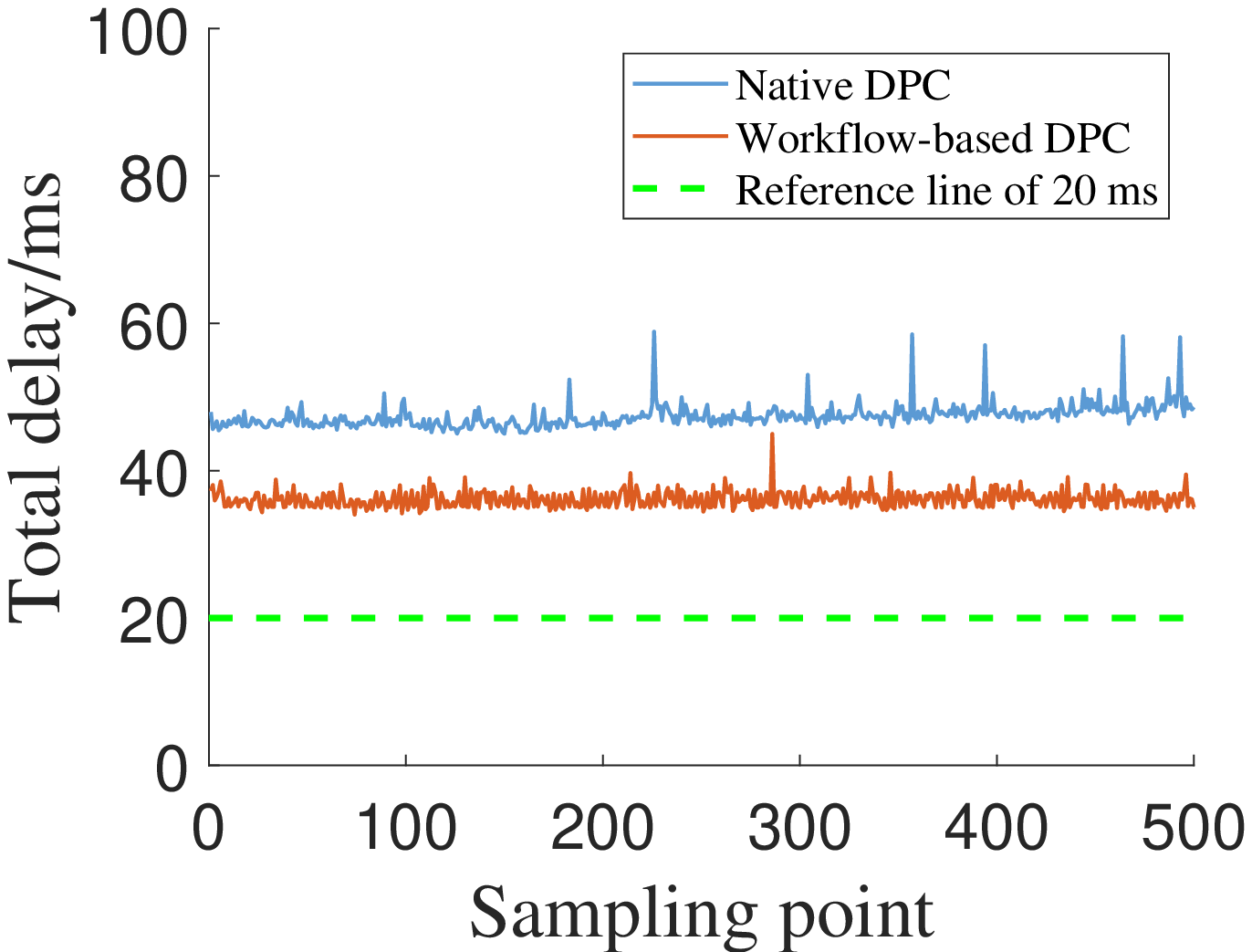}
\label{Comparison of total delay (20 km/h)}
\end{minipage}
}

\hspace{-40.0pt}
\par \vspace{-10.pt}
\hspace{-36.0pt}

\hspace{-0.2in}
\subfigure[\scriptsize{Computation delay (30 km/h, $\epsilon_1 \!= \!10^{-\!2}$)}]{
\begin{minipage}[b]{0.5 \textwidth}
\includegraphics[width=\textwidth]{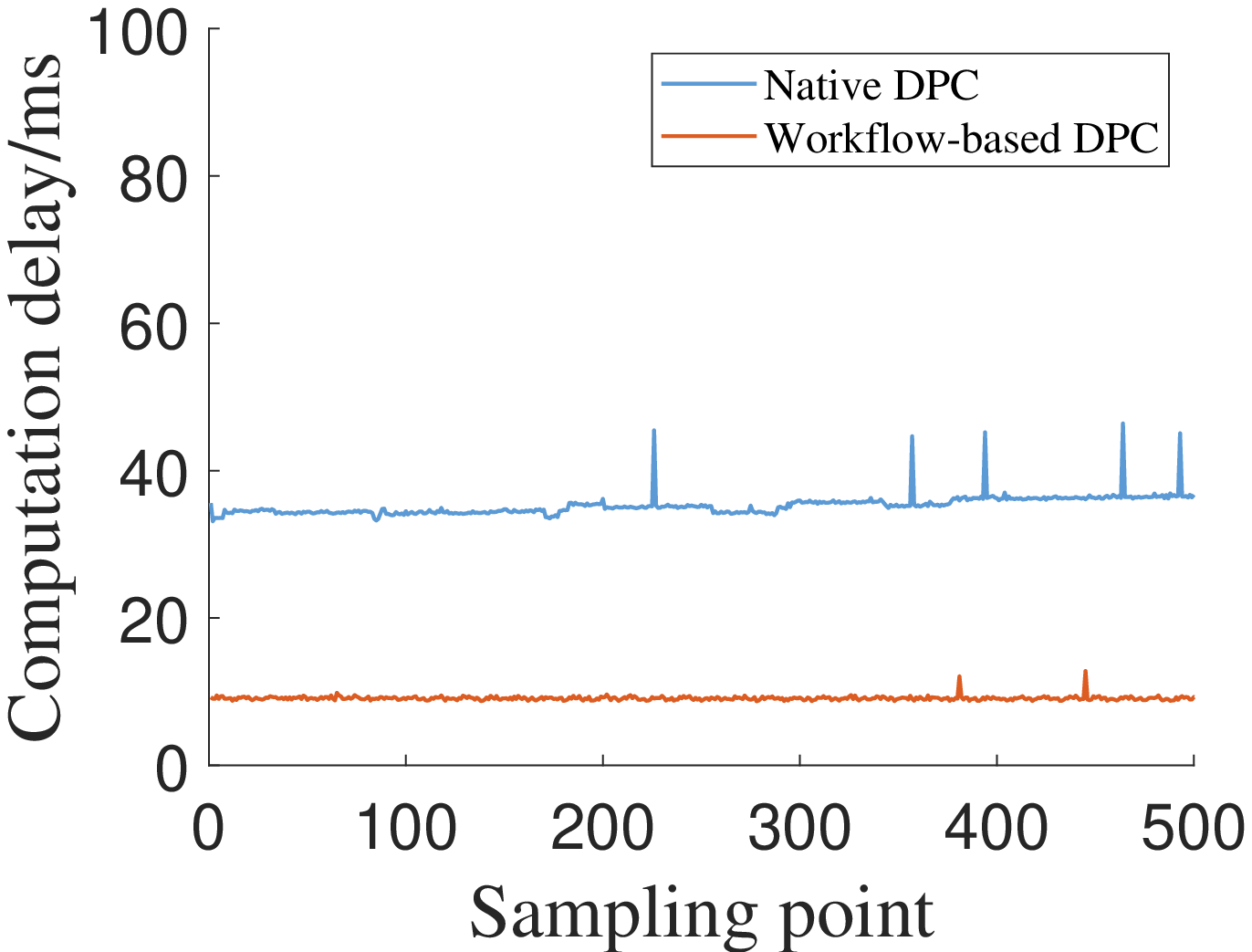}
\label{Comparison of computation delay (30 km/h)}
\end{minipage}
}
\hspace{-0.2in}
\subfigure[\scriptsize{Total delay (30 km/h, $\epsilon_1 \!=\! 10^{-\!2}$)}]{
\begin{minipage}[b]{0.5 \textwidth}
\includegraphics[width=\textwidth]{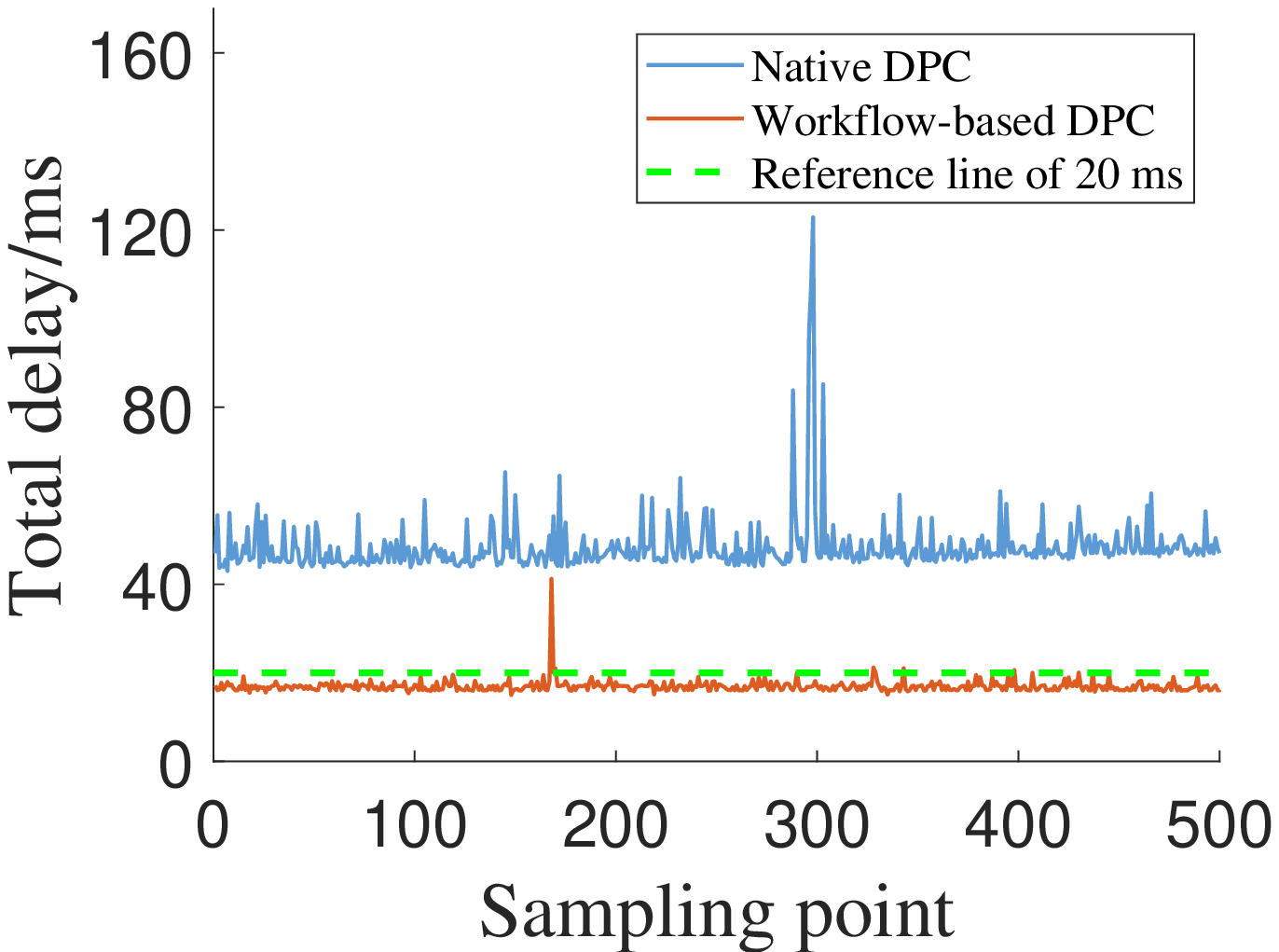}
\label{Comparison of total delay (30 km/h)}
\end{minipage}
}
\caption{Computation Delays and Total Delays of the Workflow-based DPC Method on Vehicle System}
\end{figure*}


Like the control example of the ball-beam system, we recorded the computation times of the four methods on the two groups of evaluations. In the evaluation of $v_r = $ 20 km/s, the truncation precision $\epsilon_{1}$ was set as $10^{-4}$ and the corresponding results are provided in Table \ref{Recorded computation times of control example 2 case 1}. The computation times of the two native DPC methods were similar, which were 35.7293 ms and 36.2084 ms, respectively. The computation times of the two workflow-based DPC methods were 26.4251 ms and 26.0765 ms, respectively. Compared with the native DPC method, the workflow-based DPC method reduced the computation time by $25.43\%$.

Then, the vehicle speed $v_r$ was adjusted to 30 km/s. The truncation precision $\epsilon_{1}$ was reduced to $10^{-2}$ for further improving the computation efficiency. The recorded results of the computation times were provided in Table \ref{Recorded computation times of control example 2 case 2}. The computation times of the two native DPC methods were similar with these of 20 km/s, which were 35.2602 ms and 25.2561 ms, respectively. The computation times of the two workflow-based DPC methods were 9.0432 ms and 9.1726 ms, respectively. Compared with the native DPC method, the workflow-based DPC method reduced the computation time by $74.35\%$.

\begin{table}[!t]
	\centering
	\caption{Recorded computation times of control example 2 \protect\\(20 km/h, $\epsilon_1 \!=\! 10^{-\!4}$)}
	\begin{tabular}{|c|c|}
		\toprule  
		\textbf{Method}&\textbf{Computation time (ms)} \\
		\hline  
        \hline
        Native DPC& 35.7293 \\
        \hline
        Native DPC with time delay compensator& 36.2084 \\
        \hline
        Workflow-based DPC& \textbf{26.4251} \\
        \hline
        Workflow-based DPC with DOB& \textbf{26.0765}  \\
		\bottomrule  
	\end{tabular}
\label{Recorded computation times of control example 2 case 1}
\vspace{-0.5em}
\end{table}
\begin{table}[!t]
	\centering
	\caption{Recorded computation times of control example 2 \protect\\(30 km/h, $\epsilon_1 \!=\! 10^{-\!2}$)}
	\begin{tabular}{|c|c|}
		\toprule  
		\textbf{Method}&\textbf{Computation time (ms)} \\
		\hline  
        \hline
        Native DPC& 35.2602 \\
        \hline
        Native DPC with time delay compensator& 35.2561 \\
        \hline
        Workflow-based DPC& \textbf{9.0432} \\
        \hline
        Workflow-based DPC with DOB& \textbf{9.1726}  \\
		\bottomrule  
	\end{tabular}
\label{Recorded computation times of control example 2 case 2}
\vspace{-0.5em}
\end{table}

In addition, the computation times were also shown in Fig. \ref{Comparison of computation delay (20 km/h)} and Fig. \ref{Comparison of computation delay (30 km/h)}. The blue line represents the computation time of the native DPC method, and the red line means that of the workflow-based DPC method. The computation times of the workflow-based DPC method were more stable than these of the native DPC method. Then, since the evaluations were conducted in the cloud-edge collaborative control scheme, the round-trip communication delays also existed. In Fig. \ref{Comparison of total delay (20 km/h)}, the total delays consisting of the computation and the round-trip communication delays were provided, with the vehicle speed $v_r =$ 20 km/s. The total delay of the native DPC method was about 47 ms, and that of the workflow-based DPC method is about 36 ms. In Fig. \ref{Comparison of total delay (30 km/h)}, the total delays were shown, of which the vehicle speed $v_r$ was set as 30 km/s. The total delay of the native DPC method was similar with that of 20 km/s. The total delay of the workflow-based DPC method was reduced significantly, which was about 19 ms. In the most sampling points, the total delays were lower than 20 ms, the control period. That means the cloud control signals could be received in the required time. The improvement of the control performance would be also brought, of which the detailed results and analyses were provided in the below.

\subsubsection{Control performances}\label{control example 2 control performance}
\begin{figure*}[t]
\subfigure[\scriptsize{Native DPC (20 km/h)}]{
\begin{minipage}[b]{0.5 \textwidth}
\includegraphics[width=\textwidth]{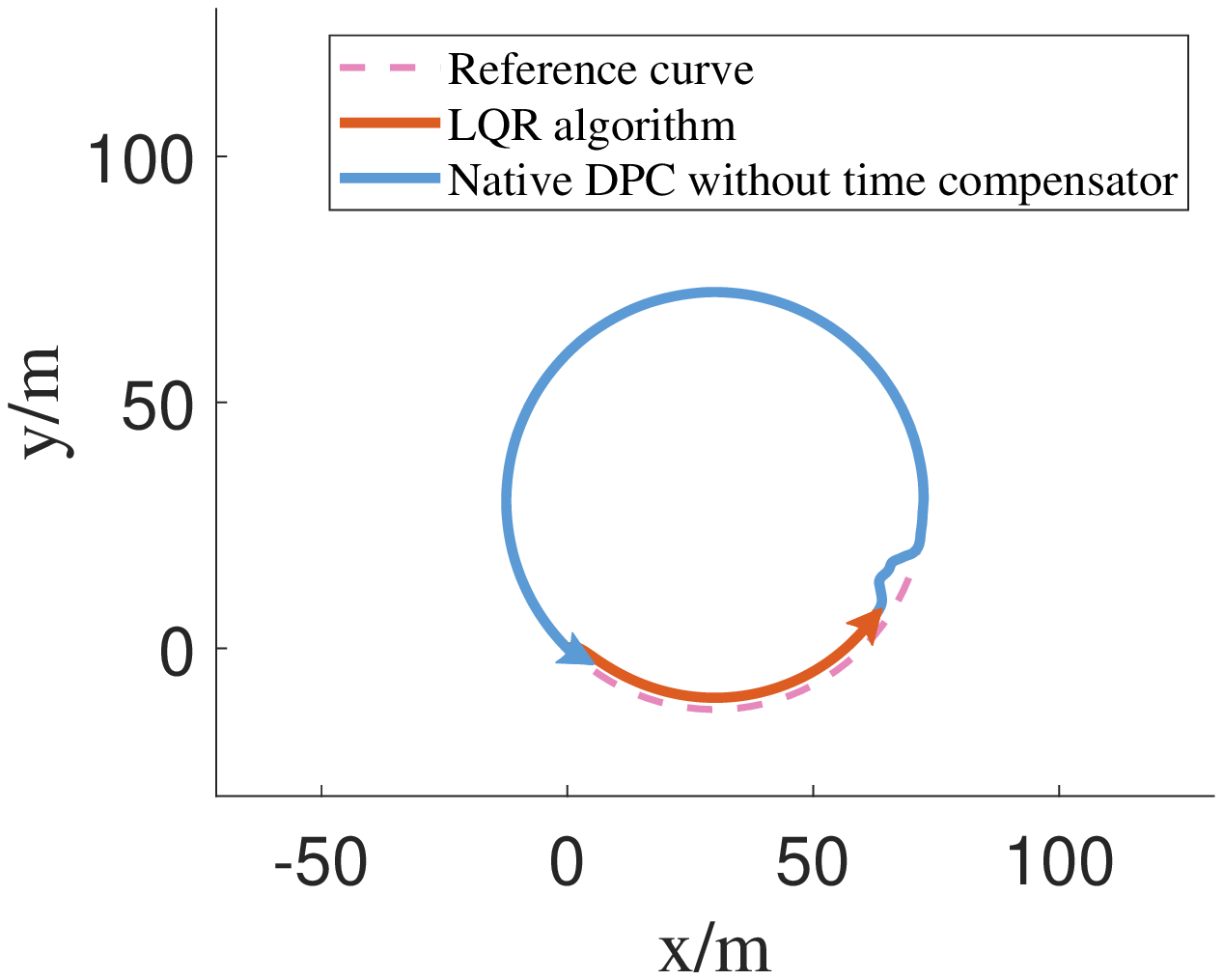}
\label{Native DPC (20 km/h)}
\end{minipage}
}
\hspace{-0.2in}
\subfigure[\scriptsize{Native DPC with time compensator (20 km/h)}]{
\begin{minipage}[b]{0.5 \textwidth}
\includegraphics[width=\textwidth]{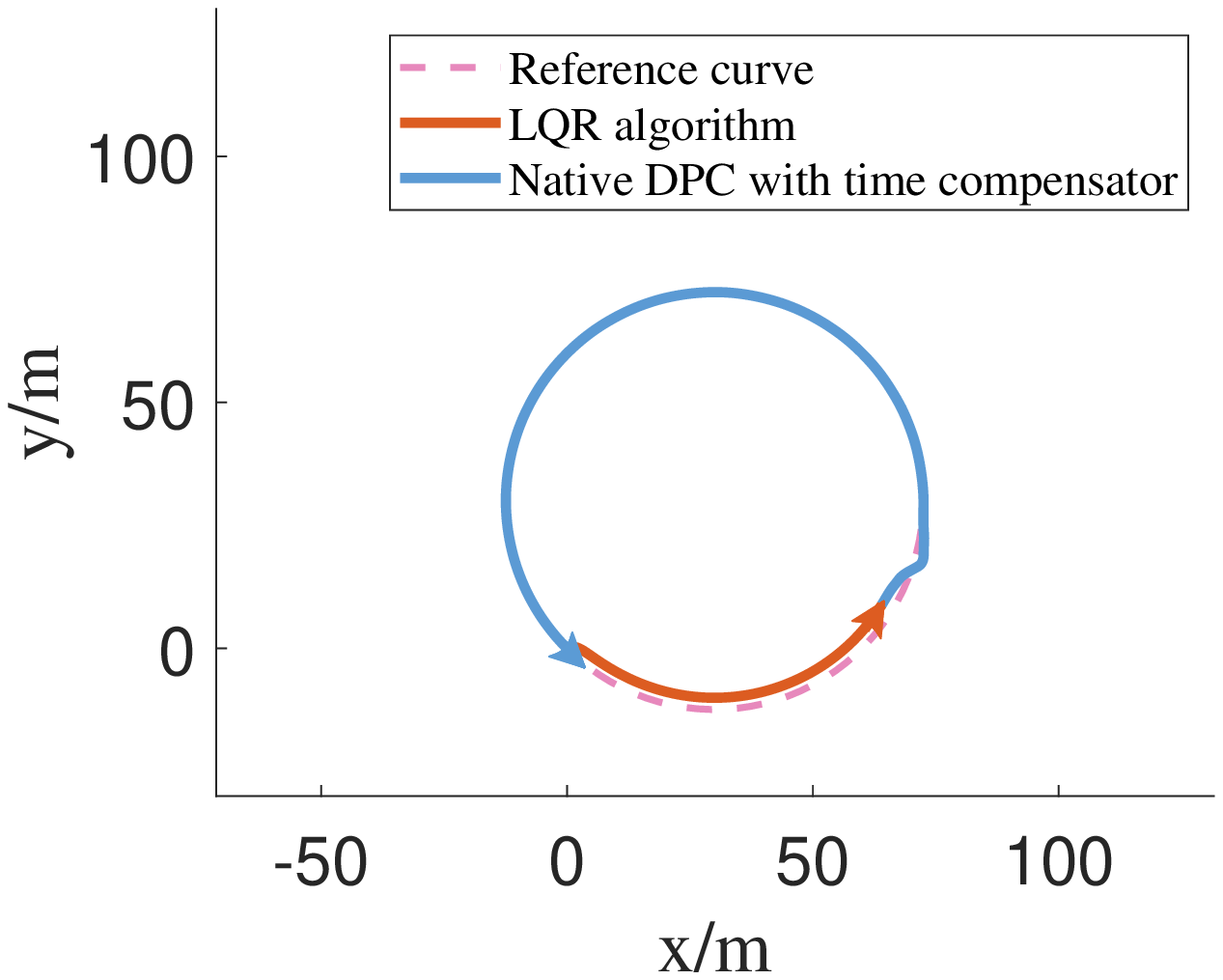}
\label{Native DPC with time compensator (20 km/h)}
\end{minipage}
}

\hspace{-40.0pt}
\par \vspace{-10.pt}
\hspace{-36.0pt}

\hspace{-0.2in}
\subfigure[\scriptsize{Workflow-based DPC (20 km/h)}]{
\begin{minipage}[b]{0.5 \textwidth}
\includegraphics[width=\textwidth]{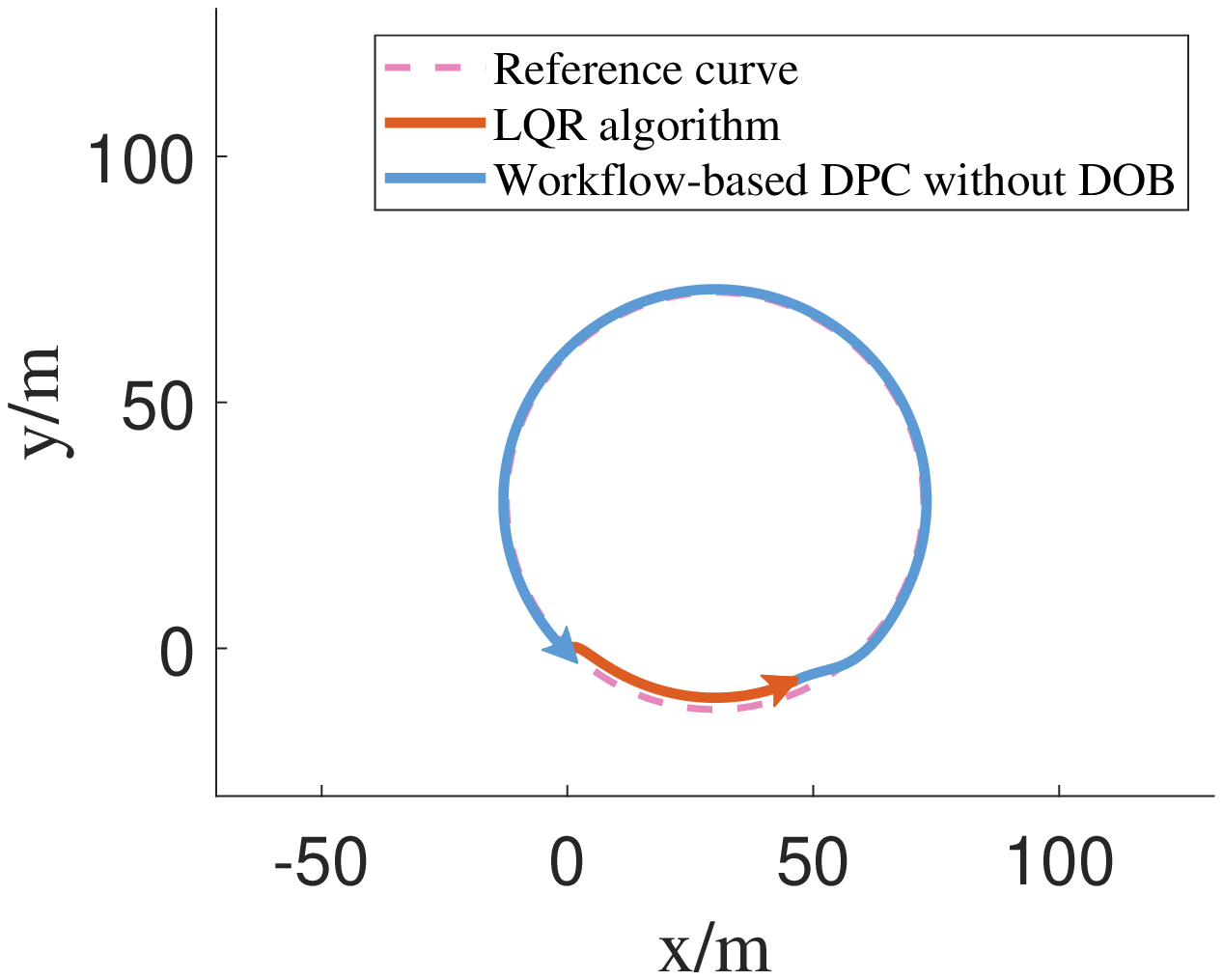}
\label{Workflow-based DPC (20 km/h)}
\end{minipage}
}
\hspace{-0.2in}
\subfigure[\scriptsize{Workflow-based DPC with DOB (20 km/h)}]{
\begin{minipage}[b]{0.5 \textwidth}
\includegraphics[width=\textwidth]{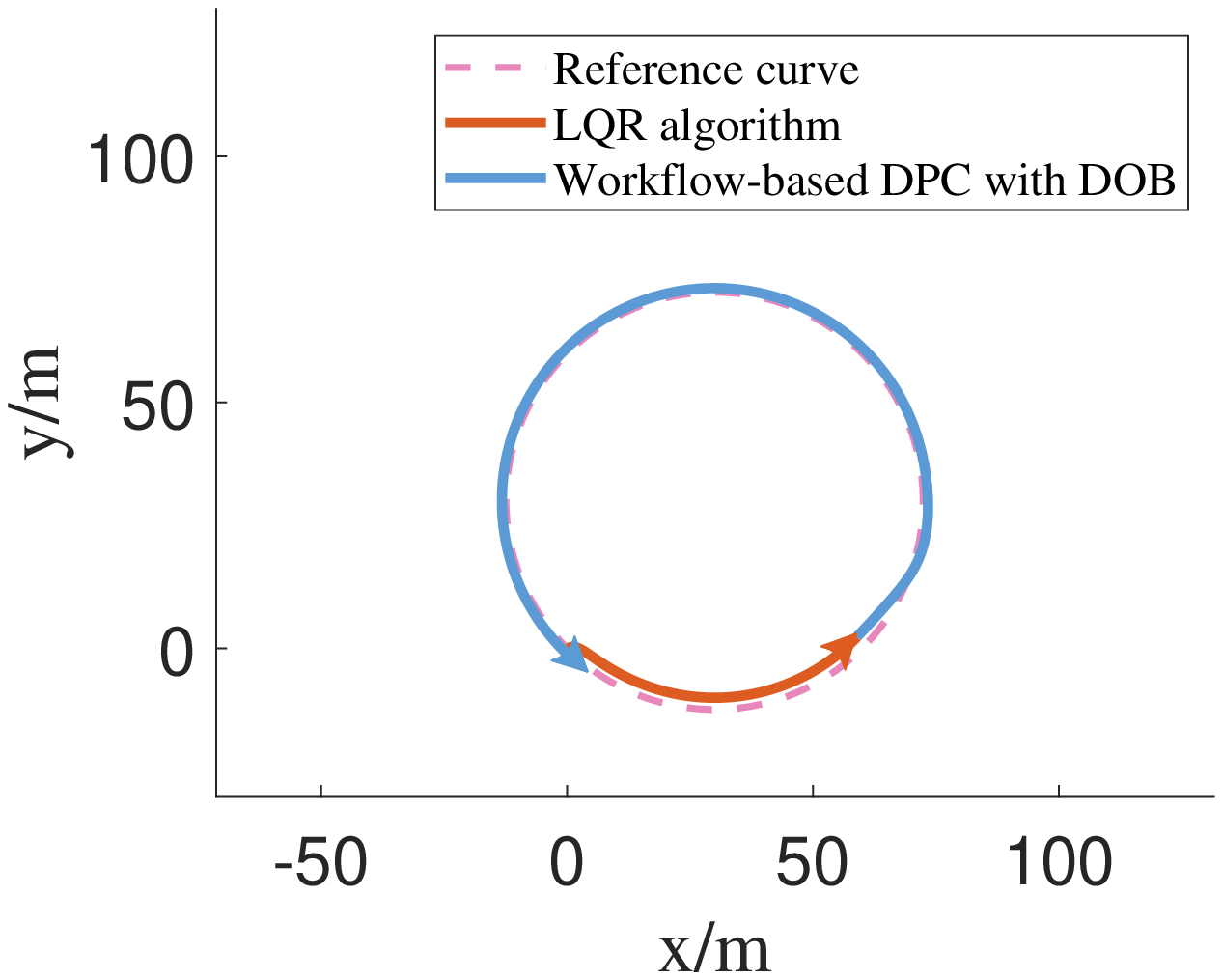}
\label{Workflow-based DPC with DOB (20 km/h)}
\end{minipage}
}
\caption{Evaluations of the Workflow-based DPC Method on Vehicle System (20 km/h)}
\end{figure*}

\begin{figure*}[t]
\subfigure[\scriptsize{Native DPC (30 km/h)}]{
\begin{minipage}[b]{0.5 \textwidth}
\includegraphics[width=\textwidth]{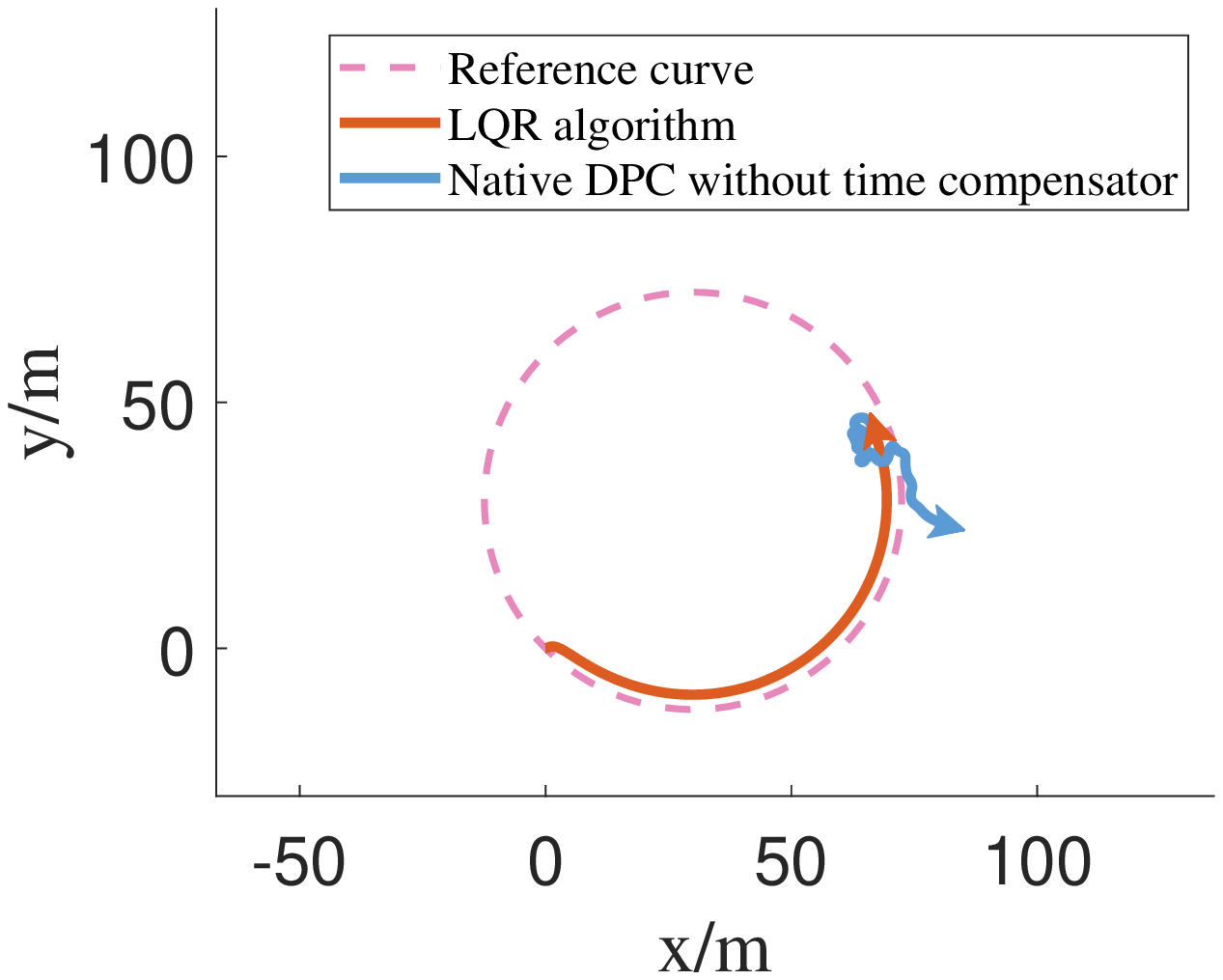}
\label{Native DPC (30 km/h)}
\end{minipage}
}
\hspace{-0.2in}
\subfigure[\scriptsize{Native DPC with time compensator (30 km/h)}]{
\begin{minipage}[b]{0.5 \textwidth}
\includegraphics[width=\textwidth]{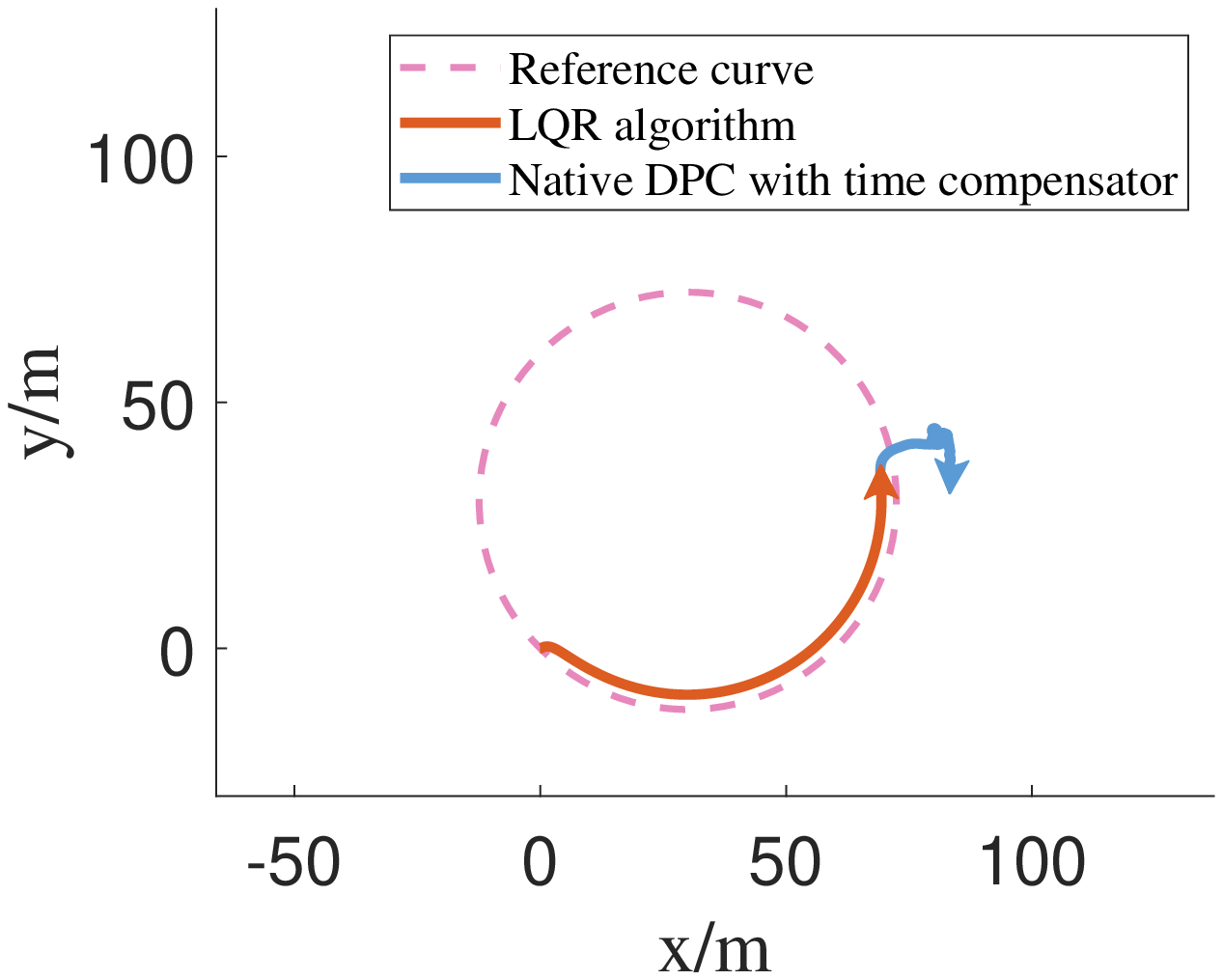}
\label{Native DPC with time compensator (30 km/h)}
\end{minipage}
}

\hspace{-40.0pt}
\par \vspace{-10.pt}
\hspace{-36.0pt}

\hspace{-0.2in}
\subfigure[\scriptsize{Workflow-based DPC  (30 km/h)}]{
\begin{minipage}[b]{0.5 \textwidth}
\includegraphics[width=\textwidth]{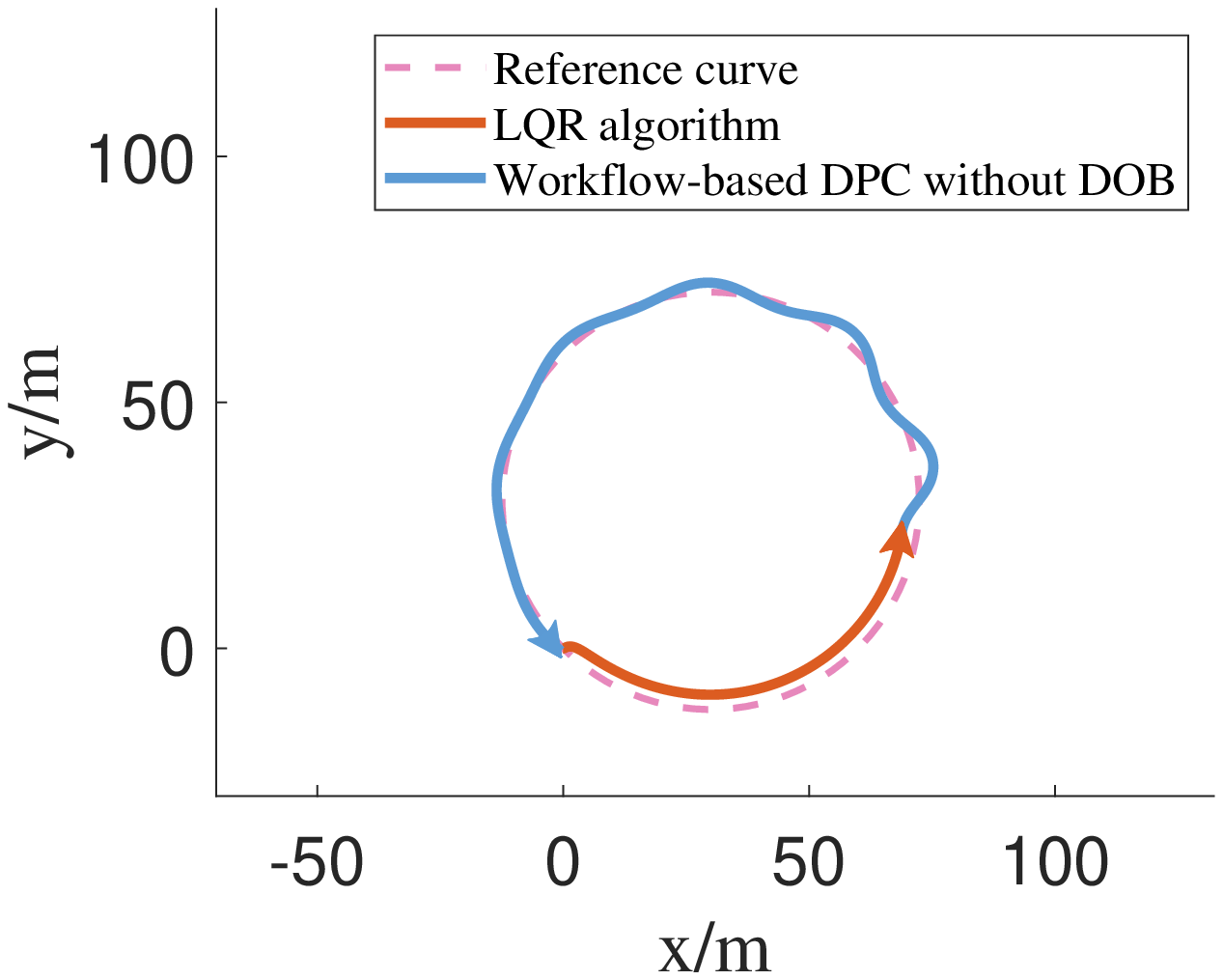}
\label{Workflow-based DPC  (30 km/h)}
\end{minipage}
}
\hspace{-0.2in}
\subfigure[\scriptsize{Workflow-based DPC with DOB (30 km/h)}]{
\begin{minipage}[b]{0.5 \textwidth}
\includegraphics[width=\textwidth]{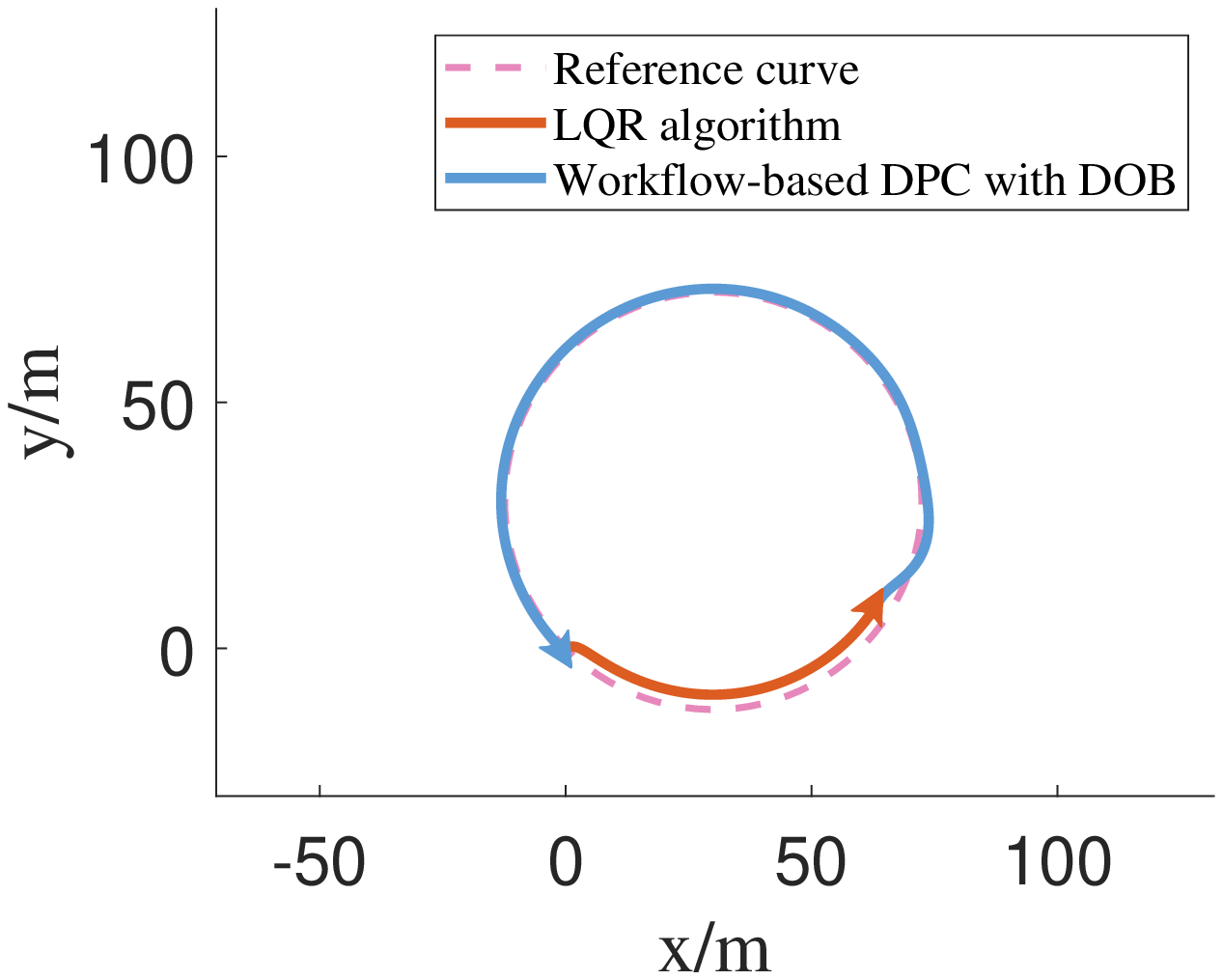}
\label{Workflow-based DPC with DOB (30 km/h)}
\end{minipage}
}
\caption{Evaluations of the Workflow-based DPC Method on Vehicle System (30 km/h)}
\end{figure*}

The evaluations of vehicle trajectory tracking were divided into the groups of 20 km/s and 30 km/s. The control performance of $v_r=$ 20 km/s was provided from Fig. \ref{Native DPC (20 km/h)} to Fig. \ref{Workflow-based DPC with DOB (20 km/h)}. The results show that in 20 km/s, all the methods could obtain desirable control performance. It states that in the relatively low-speed real-time scenario, a certain amount of time delays could be tolerated, and the non-workflow-based DPC controller was enough to carry out the control missions.

However, when the vehicle speed $v_r$ increased to 30 km/s, the native DPC controller could not guarantee a stable and desirable control performance. As shown in Fig. \ref{Native DPC (30 km/h)}, after the data preparation stage being finished, the vehicle became out of control quickly and veered off the circle trajectory. In Fig. \ref{Native DPC with time compensator (30 km/h)}, the native DPC with time delay compensator had not bring improvement on the control performance. As for the workflow-based DPC methods, Fig. \ref{Workflow-based DPC (30 km/h)} shows the cloud controller gradually obtained the descried control performance. The reason is that the cloud control signals could arrived at the controlled plant in the required time. But in the beginning stage, some shakes occurred in the moving curve, since the truncation precision $\epsilon_1$ was set relatively low and the control performance was affected.

Therefore, the DOB was designed here to reduce the negative influence brought by the accuracy loss. As shown in Fig. \ref{Workflow-based DPC with DOB (30 km/h)}, the workflow-based DPC with DOB received the best performance in the whole process. When the cloud controller was switched from LQR algorithm to the workflow-based DPC method with DOB, the shakes were indeed small and even did not appear. That means the switching process was smooth and stable. The cloud-edge composite DPC signals were calculated fast and accurately. As the result, the vehicle was also controlled to track the circle trajectory fast and well.

\subsection{Control example 3: high-dimension numerical system}

The third control example is a 78-dimension numerical system, of which the structure of the state-space model is built based on IEEE 10-generator 39-bus power system \cite{shames2017security, zhang2021structural}.  This control example is applied to detected the computation efficiency for the high-dimension system.

\subsubsection{System model}
We consider the classical linearised synchronous machine model for each node of the power network. The behaviour of bus $i$ could be described by the so-called swing equation:
\begin{equation}
  m_{i}\ddot{\theta}_{i} + d_{i}\dot{\theta}_{i} - P_{mi} = - \sum_{j\in N_{i}}P_{ij},
\end{equation}
where $\theta_{i}$ is the phase angle of bus $i$, $m_i$ and $d_i$ are the inertia and damping coefficients, respectively, $P_{mi}$ is the mechanical input power and $P_{ij}$ is the active power flow from buses $i$ to $j$. Assuming that there are no power losses, neglecting ground admittances, and letting $V_{i}=|V_{i}|e^{j\theta_{i}}$ be the complex voltage of bus $i$, the active power flow between buses $i$ and $j$ is provided by
\begin{equation}
  P_{ij} = k_{ij}\sin(\theta_{i}-\theta_{j}),
\end{equation}
where $k_{ij}=|V_i||V_j|b_{ij}$ and $b_{ij}$ is the susceptance of the power line connecting buses $i$ and $j$. Since the phase angles are close, we could provide the linearised equation of the dynamics of bus $i$:
\begin{equation}
  m_{i}\ddot{\theta}_{i} + d_{i}\dot{\theta}_{i} = -\sum_{j\in N_{i}}k_{ij}(\theta_{i}-\theta_{j}) + P_{mi}.
\end{equation}

Letting $x=[\theta_{1}, \theta_{2}, ..., \theta_{N}, \dot{\theta}_{1}, \dot{\theta}_{2}, ..., \dot{\theta}_{N}]^{T}$ and $u = [P_{m1}, P_{m2}, ..., P_{mN}]^{T}$, we obtain
\begin{equation}
  \dot{x} = Ax+Bu+H\alpha,
\end{equation}
where
\begin{eqnarray*}
  A &=& \left[
          \begin{array}{cc}
            0_N & I_N \\
            -ML & -DM \\
          \end{array}
        \right],
   \\
  B &=& \left[
          \begin{array}{cc}
            0_N & M \\
          \end{array}
        \right]^{T},
   \\
  M &=& diag(\frac{1}{m_1}, \frac{1}{m_2}, ..., \frac{1}{m_N}), \\
  D &=& diag(d_1, d_2, ..., d_N),
\end{eqnarray*}
and $L$ is the Laplacian matrix of graph $\mathcal{G}(\mathcal{V}_P, \mathcal{E})$ with $N = |\mathcal{V}_P|$ nodes. Each node corresponds to a bus in the power network and the undirected edge $(i,j)\in \mathcal{E}_{P}$, if bus $i$ is connected to bus $j$ with edge weight $k_{ij}$ for all $(i,j)\in \mathcal{E}$.

The parameters including the the relationships between the nodes were referred to \cite{hiskens2013ieee, zhang2021structural}. Then, a 78-dimension numerical system with 10-input and 10-output is obtained. As analysed in Remark \ref{remark3}, there exist large amount of the singular values in the same order of magnitudes. If we adjust the truncation accuracy by increasing or decreasing $\epsilon_1$, the change of the upper bounds of the uncertain would be large. Therefore, in this control example, the number of the retained singular values is used to replace $\epsilon_{1}$ to adjust the truncation accuracy. Then, the relationship between the computation efficiency and the number of the retained singular values, and the relationship between the control performance and the number singular values would be discussed. The evaluation was conducted in the non-real-time scenario, which means that only after the calculation of the control variable being finished, the controlled plant would be updated.

\subsubsection{Relationship between the computation efficiency and the number of retained singular values}

\begin{table*}[!t]
	\centering
	\caption{Recorded results of the control example 3: high-dimension numerical system}
    \renewcommand{\multirowsetup}{\centering}
	\begin{tabular}{|c|c|c|c|c|c|c|}
		\hline  
        \cline{1-7}
        \multirow{2}{4.2cm}{\textbf{Number of \\retained singular values}} & \multirow{2}{1.6cm}{\textbf{Baseline}} & \multirow{2}{1.6cm}{\textbf{100}} & \multirow{2}{1.6cm}{\textbf{50}} & \multirow{2}{1.6cm}{\textbf{20}} & \multirow{2}{1.6cm}{\textbf{10}} & \multirow{2}{1.6cm}{\textbf{5}} \cr
        \multirow{2}{*}{} & \multirow{2}{*}{} & \multirow{2}{*}{} & \multirow{2}{*}{} & \multirow{2}{*}{} & \multirow{2}{*}{} & \multirow{2}{*}{} \cr

        \hline
        \hline
        \multirow{2}{*}{\textbf{Computation time (ms)}} & \multirow{2}{*}{367.4199} & \multirow{2}{*}{261.1135} & \multirow{2}{*}{168.9818} & \multirow{2}{*}{151.2747} & \multirow{2}{*}{128.8647} & \multirow{2}{*}{54.7559} \cr
        \multirow{2}{*}{} & \multirow{2}{*}{} & \multirow{2}{*}{} & \multirow{2}{*}{} & \multirow{2}{*}{} & \multirow{2}{*}{} & \multirow{2}{*}{} \cr
        \hline
        \multirow{2}{4.2cm}{\textbf{Reduction of \\computation time}} & \multirow{2}{*}{$\setminus$} & \multirow{2}{*}{28.93\%} & \multirow{2}{*}{54.01\%} & \multirow{2}{*}{58.83\%} & \multirow{2}{*}{64.93\%} & \multirow{2}{*}{85.10\%} \cr
        \multirow{2}{*}{} & \multirow{2}{*}{} & \multirow{2}{*}{} & \multirow{2}{*}{} & \multirow{2}{*}{} & \multirow{2}{*}{} & \multirow{2}{*}{} \cr
        \hline
        \cline{1-7}
	\end{tabular}
\label{Recorded results of the control example 3: high-dimension numerical system}
\end{table*}
\begin{figure*}[t]
\subfigure[\scriptsize{Native DPC}]{
\begin{minipage}[b]{0.5 \textwidth}
\includegraphics[width=\textwidth]{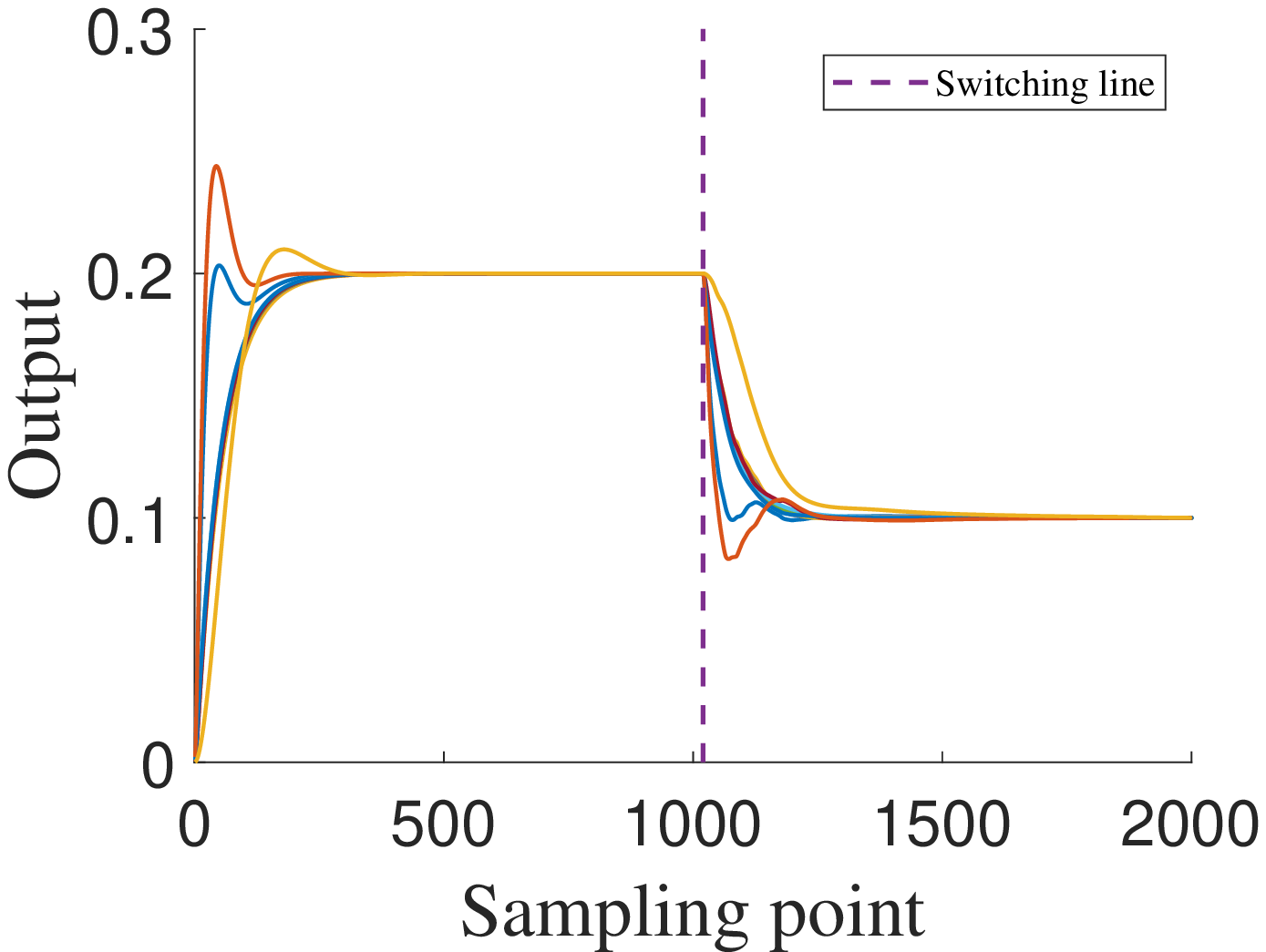}
\label{Native DPC}
\end{minipage}
}
\hspace{-0.2in}
\subfigure[\scriptsize{Retaining 100 singular values}]{
\begin{minipage}[b]{0.5 \textwidth}
\includegraphics[width=\textwidth]{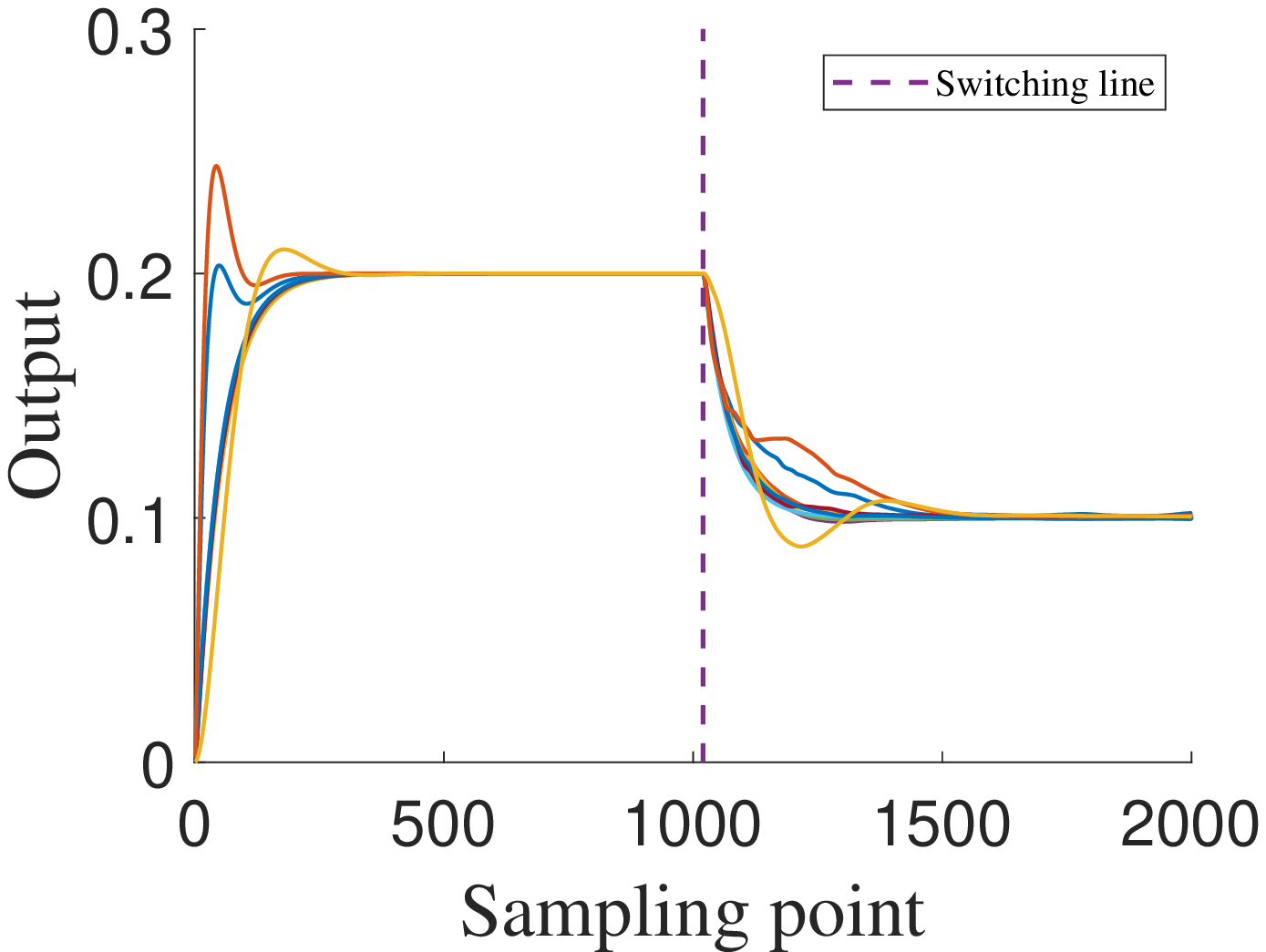}
\label{Workflow-based DPC (100 retained singular values)}
\end{minipage}
}

\hspace{-40.0pt}
\par \vspace{-10.pt}
\hspace{-36.0pt}

\hspace{-0.2in}
\subfigure[\scriptsize{Retaining 20 singular values}]{
\begin{minipage}[b]{0.5 \textwidth}
\includegraphics[width=\textwidth]{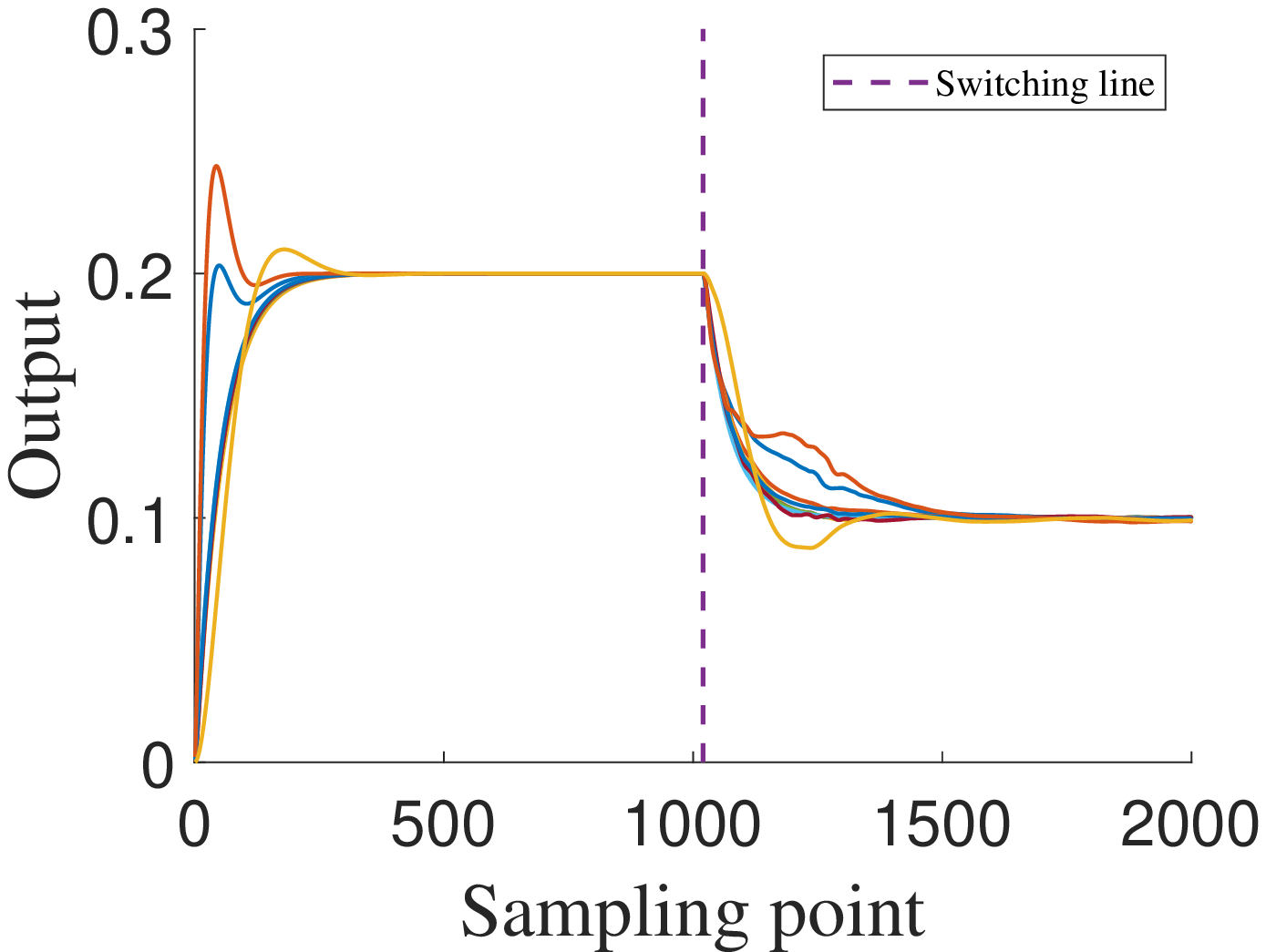}
\label{Workflow-based DPC (20 retained singular values)}
\end{minipage}
}
\hspace{-0.2in}
\subfigure[\scriptsize{Retaining 5 singular values}]{
\begin{minipage}[b]{0.5 \textwidth}
\includegraphics[width=\textwidth]{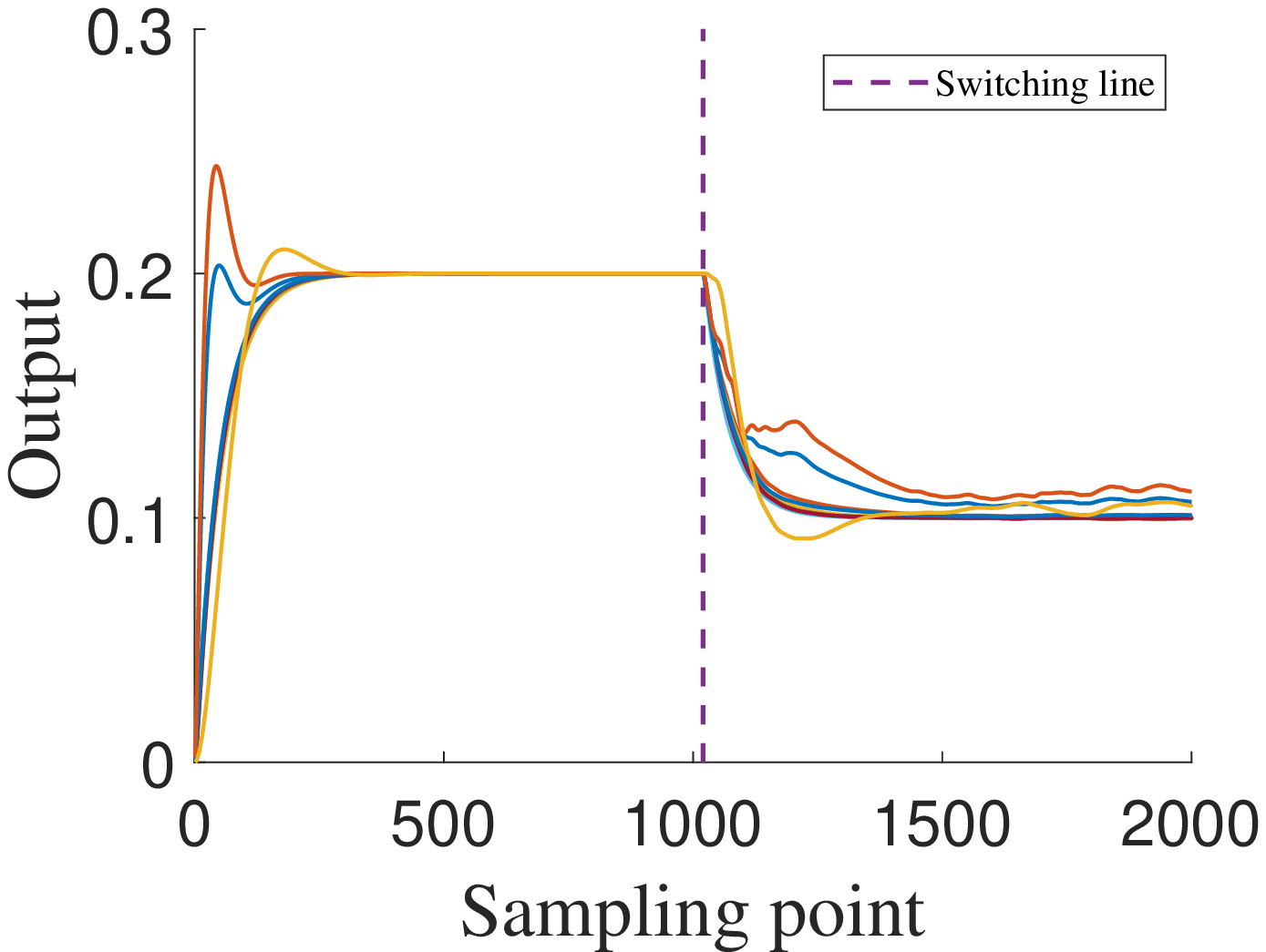}
\label{Workflow-based DPC (5 retained singular values)}
\end{minipage}
}
\caption{Relationship Between Control Performance and the Number of Retained Singular Values}
\label{Relationship between control performance and the number of retained singular values}
\end{figure*}
\begin{figure}[!ht]
  \centering
  \includegraphics[width=4in]{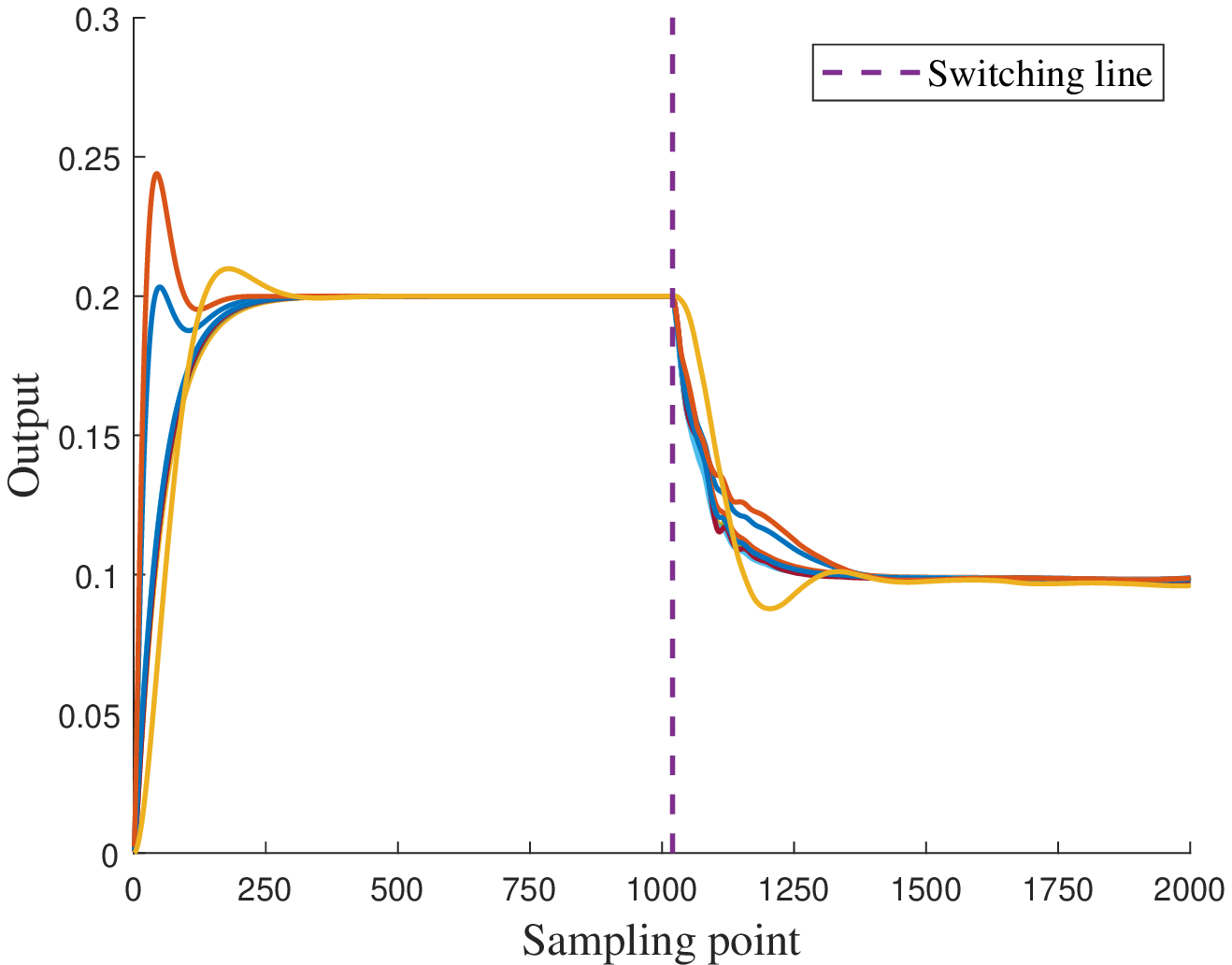}\\
  \caption{Control Result of the Case Retaining 5 Singular Values with DOB}
  \label{Control Result of the case retaining 5 singular values with DOB}
\end{figure}
Six groups of evaluations were carried out to detect the computation efficiency and control performance, when different number of singular values were retained. The recorded results of this control example is provided in Table \ref{Recorded results of the control example 3: high-dimension numerical system}. The baseline represents the native DPC method being conducted in a single standard cloud server with 4 CPU and 8 GB memory. Besides, five groups of experiments were also conducted, of which 100, 50, 20, 10 and 5 singular values were retained, respectively. From the recorded results, the computation time of the baseline was 367.4199 ms. With the decreasing of the retained singular values, the computation efficiency increased. When only 5 singular values were retained, the computation time became 54.7559 ms. Compared with the baseline, the computation time reduced $85.10\%$. It states that the workflow-based DPC could improve the computation efficiency significantly for the high-dimension system.

\subsubsection{Relationship between the control performance and the number of retained singular values}

On the other hand, the decrease of the retained singular values brings the reduction of the control performance. The control performances of the native DPC and the workflow-base DPC with 100, 20, 5 retained singular values are presented in Fig. \ref{Relationship between control performance and the number of retained singular values}. When 100 and 20 singular values were retained, the numerical system still could be controlled to the reference signal relatively accurately and steadily. But the control performances of the two groups were worse than the native DPC because of the losses of computation accuracy. As for the group of retaining 5 singular values, though there existed certain differences from the reference signal, the output had not diverged. The remaining differences could be reduced by the DOB-based compensator, of which the UUB stability has been proved. As shown in Fig. \ref{Control Result of the case retaining 5 singular values with DOB}, the control performance of the case retaining 5 singular values was improved significantly by the edge DOB-based compensator. It states that the computation efficiency, control performance, truncation accuracy and DOB-based compensator should be considered in the meanwhile, to guarantee the fast, stable and high-quality control services in practical applications.

\section{Conclusion}

This paper proposes a cloud-edge collaborative containerised workflow-based fast DPC system with DOB, which could reduce the computation time to within 10 ms for the real-time examples, and by at most $85.10\%$ for the high-dimension example. This paper provides a novel framework of cloud control system, with the combination of cloud-edge collaborative control, workflow-based control and container technology. Although this is a preliminary work, based on its framework, more problems with values would be studied, such as the overall stability of the workflow-based DPC method, workflow-based model predictive control, and co-design of control algorithm and cloud resource elastic scheduling, etc.

\bibliographystyle{ieeetr}
\bibliography{mybib2}

\begin{thebibliography}{10}

\bibitem{xia2012networked}
Y.~Xia, ``From networked control systems to cloud control systems,'' in {\em
  Proceedings of Chinese control conference}, pp.~5878--5883, 2012.

\bibitem{marinescu2017cloud}
D.~C. Marinescu, {\em Cloud computing: theory and practice}.
\newblock Orlando, FL, USA: Morgan Kaufmann, 2017.

\bibitem{senyo2018cloud}
P.~K. Senyo, E.~Addae, and R.~Boateng, ``Cloud computing research: A review of
  research themes, frameworks, methods and future research directions,'' {\em
  International Journal of Information Management}, vol.~38, no.~1,
  pp.~128--139, 2018.

\bibitem{alexandru2020cloud}
A.~B. Alexandru, K.~Gatsis, Y.~Shoukry, S.~A. Seshia, P.~Tabuada, and G.~J.
  Pappas, ``Cloud-based quadratic optimization with partially homomorphic
  encryption,'' {\em IEEE Transactions on Automatic Control}, vol.~66, no.~5,
  pp.~2357--2364, 2021.

\bibitem{tanaka2017directed}
T.~Tanaka, M.~Skoglund, H.~Sandberg, and K.~H. Johansson, ``Directed
  information and privacy loss in cloud-based control,'' in {\em Proceedings of
  American control conference}, pp.~1666--1672, 2017.

\bibitem{liu2017predictive}
G.-P. Liu, ``Predictive control of networked multiagent systems via cloud
  computing,'' {\em IEEE Transactions on Cybernetics}, vol.~47, no.~8,
  pp.~1852--1859, 2017.

\bibitem{xia2020cloud}
Y.~Xia, R.~Gao, M.~Lin, Y.~Ren, and C.~Yan, ``Green energy complementary based
  on intelligent power plant cloud control system,'' {\em Acta Automatica
  Sinica}, vol.~46, no.~9, pp.~1844--1868, 2020.

\bibitem{wang2021cloud}
Z.~Wang, S.~Yang, X.~Xiang, A.~Vasilijevic, N.~Miskovic, and D.~Nad,
  ``Cloud-based mission control of usv fleet: architecture, implementation and
  experiments,'' {\em Control Engineering Practice}, vol.~106, p.~104657, 2021.

\bibitem{xia2015cloud}
Y.~Xia, ``Cloud control systems,'' {\em IEEE/CAA Journal of Automatica Sinica},
  vol.~2, no.~2, pp.~134--142, 2015.

\bibitem{mahmoud2017interaction}
M.~S. Mahmoud and Y.~Xia, ``The interaction between control and computing
  theories: new approaches,'' {\em International Journal of Automation and
  Computing}, vol.~14, no.~3, pp.~254--274, 2017.

\bibitem{xia2022brief}
Y.~Xia, Y.~Zhang, L.~Dai, Y.~Zhan, and Z.~Guo, ``A brief survey on recent
  advances in cloud control systems,'' {\em IEEE Transactions on Circuits and
  Systems II: Express Briefs}, vol.~69, no.~7, pp.~3108--3114, 2022.

\bibitem{gao2017new}
R.~Gao, Y.~Xia, and L.~Ma, ``A new approach of cloud control systems: {CCS}s
  based on data-driven predictive control,'' in {\em Proceedings of Chinese
  control conference}, pp.~3419--3422, 2017.

\bibitem{gao2021design}
R.~Gao, Y.~Xia, L.~Dai, and Z.~Sun, ``Design and implementation of data-driven
  predictive cloud control system,'' {\em arXiv preprint arXiv:2112.14347},
  2021.

\bibitem{ye2021shws}
L.~Ye, Y.~Xia, L.~Yang, and C.~Yan, ``{SHWS}: Stochastic hybrid workflows
  dynamic scheduling in cloud container services,'' {\em IEEE Transactions on
  Automation Science and Engineering}, 2021.

\bibitem{chen2016parallel}
J.~Chen, K.~Li, Z.~Tang, K.~Bilal, S.~Yu, C.~Weng, and K.~Li, ``A parallel
  random forest algorithm for big data in a spark cloud computing
  environment,'' {\em IEEE Transactions on Parallel and Distributed Systems},
  vol.~28, no.~4, pp.~919--933, 2017.

\bibitem{xiao2020malfcs}
G.~Xiao, J.~Li, Y.~Chen, and K.~Li, ``Mal{FCS}: An effective malware
  classification framework with automated feature extraction based on deep
  convolutional neural networks,'' {\em Journal of Parallel and Distributed
  Computing}, vol.~141, pp.~49--58, 2020.

\bibitem{gao2021fast}
R.~Gao, Y.~Xia, G.~Wang, L.~Yang, and Y.~Zhan, ``Fast subspace identification
  method based on containerised cloud workflow processing system,'' {\em arXiv
  preprint arXiv:2112.14349}, 2021.

\bibitem{chen2015disturbance}
W.-H. Chen, J.~Yang, L.~Guo, and S.~Li, ``Disturbance-observer-based control
  and related methods¡ªan overview,'' {\em IEEE Transactions on industrial
  electronics}, vol.~63, no.~2, pp.~1083--1095, 2015.

\bibitem{zhu2018scheduling}
J.~Zhu, X.~Li, R.~Ruiz, and X.~Xu, ``Scheduling stochastic multi-stage jobs to
  elastic hybrid cloud resources,'' {\em IEEE Transactions on Parallel and
  Distributed Systems}, vol.~29, no.~6, pp.~1401--1415, 2018.

\bibitem{mao2019learning}
H.~Mao, M.~Schwarzkopf, S.~B. Venkatakrishnan, Z.~Meng, and M.~Alizadeh,
  ``Learning scheduling algorithms for data processing clusters,'' in {\em
  Proceedings of ACM SIGCOMM}, pp.~270--288, 2019.

\bibitem{kaur2017container}
K.~Kaur, T.~Dhand, N.~Kumar, and S.~Zeadally, ``Container-as-a-service at the
  edge: Trade-off between energy efficiency and service availability at fog
  nano data centers,'' {\em IEEE Wireless Communications}, vol.~24, no.~3,
  pp.~48--56, 2017.

\bibitem{goldschmidt2018container}
T.~Goldschmidt, S.~Hauck-Stattelmann, S.~Malakuti, and S.~Gr{\"u}ner,
  ``Container-based architecture for flexible industrial control
  applications,'' {\em Journal of Systems Architecture}, vol.~84, pp.~28--36,
  2018.

\bibitem{mellado2020container}
J.~Mellado and F.~N{\'u}{\~n}ez, ``A container-based iot-oriented programmable
  logical controller,'' in {\em Proceedings of Industrial Cyberphysical
  Systems}, vol.~1, pp.~55--61, 2020.

\bibitem{ranjan2020energy}
R.~Ranjan, I.~Thakur, G.~S. Aujla, N.~Kumar, and A.~Y. Zomaya,
  ``Energy-efficient workflow scheduling using container based virtualization
  in software defined data centers,'' {\em IEEE Transactions on Industrial
  Informatics}, vol.~16, no.~12, pp.~7646--7657, 2020.

\bibitem{xia2013data}
Y.~Xia, W.~Xie, B.~Liu, and X.~Wang, ``Data-driven predictive control for
  networked control systems,'' {\em Information Sciences}, vol.~235,
  pp.~45--54, 2013.

\bibitem{huang2008dynamic}
B.~Huang and R.~Kadali, {\em Dynamic modeling, predictive control and
  performance monitoring: a data-driven subspace approach}.
\newblock Edmonton, AB, Canada: Springer, 2008.

\bibitem{bjorck2015numerical}
{\AA}.~Bj{\"o}rck, {\em Numerical methods in matrix computations}.
\newblock New York, NY, USA: Springer, 2015.

\bibitem{ginoya2015delta}
D.~Ginoya, P.~Shendge, and S.~Phadke, ``Delta-operator-based extended
  disturbance observer and its applications,'' {\em IEEE Transactions on
  Industrial Electronics}, vol.~62, no.~9, pp.~5817--5828, 2015.

\bibitem{berberich2020data}
J.~Berberich, J.~K{\"o}hler, M.~A. M{\"u}ller, and F.~Allg{\"o}wer,
  ``Data-driven model predictive control with stability and robustness
  guarantees,'' {\em IEEE Transactions on Automatic Control}, vol.~66, no.~4,
  pp.~1702--1717, 2020.

\bibitem{nelson2016mastering}
J.~Nelson, {\em Mastering redis}.
\newblock Birmingham, West Midlands, UK: Packt Publishing Ltd, 2016.

\bibitem{suo2018analysis}
K.~Suo, Y.~Zhao, W.~Chen, and J.~Rao, ``An analysis and empirical study of
  container networks,'' in {\em Proceedings of IEEE INFOCOM Conference on
  Computer Communications}, pp.~189--197, 2018.

\bibitem{rahmat2017application}
M.~F. Rahmat, H.~Wahid, and N.~A. Wahab, ``Application of intelligent
  controller in a ball and beam control system,'' {\em International journal on
  smart sensing and intelligent systems}, vol.~3, no.~1, 2017.

\bibitem{kong2015kinematic}
J.~Kong, M.~Pfeiffer, G.~Schildbach, and F.~Borrelli, ``Kinematic and dynamic
  vehicle models for autonomous driving control design,'' in {\em Proceedings
  of IEEE intelligent vehicles symposium}, pp.~1094--1099, 2015.

\bibitem{shames2017security}
I.~Shames, F.~Farokhi, and T.~H. Summers, ``Security analysis of cyber-physical
  systems using $\mathcal{H}_2$ norm,'' {\em IET Control Theory \&
  Applications}, vol.~11, no.~11, pp.~1749--1755, 2017.

\bibitem{zhang2021structural}
Y.~Zhang, Y.~Xia, and D.-H. Zhai, ``Structural controllability of networked
  relative coupling systems,'' {\em Automatica}, vol.~128, p.~109547, 2021.

\bibitem{hiskens2013ieee}
I.~Hiskens, ``{IEEE} {PES} task force on benchmark systems for stability
  controls,'' {\em Technical Report}, 2013.

\end{thebibliography}

\section{APPENDIX}
\subsection{Self-defined Functions in the DPC Workflow}\label{Self-defined Functions in the DPC Workflow}

The four self-defined functions are provided in Algorithm \ref{Distributed truncated SVD algorithm}, which are used to construct the DPC workflow.
\begin{breakablealgorithm}
    \renewcommand{\algorithmicrequire}{\textbf{Input:}}
	\renewcommand{\algorithmicensure}{\textbf{Output:}}
    \caption{\textbf{Distributed Truncated SVD Algorithm}}
    \label{Distributed truncated SVD algorithm}
    \textbf{Input:} A matrix to be decomposed $A_{m\times n}$, block width $col$.\\
    \textbf{Output:} Decomposed results $U$, $S$ and $V$.

    \begin{algorithmic}[1]
         \STATE \textbf{\textrm{function} \textsc{ParallelSVDbyCols}($A_{m\times n}$, $col$):}
         \STATE \quad Calculate the number of the truncated blocks
         \begin{eqnarray*}
            \quad N_c = round(n / col + 0.45)
         \end{eqnarray*}
         \quad where the \emph{round} command means taking the nearest integer.
         \STATE \quad Build a series of lists to store the truncated blocks and decomposed results
         $l_A \!= list(), l_U \!= list(), l_\Sigma \!= list(), l_V \!= list()$ where the \emph{list()} command means creating an empty list.
         \STATE \quad Fill the list $l_A$ by the $N_c$ column blocks of $A$.
         \STATE \quad \textbf{\textrm{for}} each column block $A_{block}$ in the list $l_A$ \textbf{\textrm{do}}
         \STATE \quad \quad Do SVD operation on $A_{block}$ and obtain the decomposed results $U_{block}$, $\Sigma_{block}$ and $V_{block}$
         \begin{eqnarray*}
            \quad \quad A_{block} = U_{block}\Sigma_{block}V_{block}^{T}.
         \end{eqnarray*}
         \STATE \quad \quad Add $U_{block}$, $\Sigma_{block}$,  $V_{block}$ into $l_U$, $l_\Sigma$, $l_V$, respectively.
         \STATE \quad \textbf{\textrm{end for}}
         \STATE \quad Conduct \textsc{DoMergeOfBlocks} function on $l_U$, $l_\Sigma$, $l_V$ and return the final decomposed results $\hat{U}$, $\hat{\Sigma}$, $\hat{V}$.
         \STATE \textbf{\textrm{end function}}
         \STATE
         \STATE \textbf{\textrm{function}} \textsc{DoMergeOfBlocks}($l_U$, $l_\Sigma$, $l_V$):
         \STATE \quad Calculate the degree of parallelism
         \begin{eqnarray*}
            \quad Nl = len(l_U)
         \end{eqnarray*}
         \quad where the \emph{len} command means obtaining the length of a list.
         \STATE \quad Calculate the workflow depth of SVD stage
         \begin{eqnarray*}
            \quad level = ceil(\log_{2}Nl)
         \end{eqnarray*}
         \quad where the \emph{ceil} means taking the upper nearest integer.
         \STATE \quad \textbf{\textrm{for}} $i \leftarrow 1$ to $level$ \textbf{\textrm{do}}
         \STATE \quad \quad Create a series of copies of $l_U$, $l_\Sigma$ and $l_V$
         \begin{eqnarray*}
            \quad l_{Ut} \leftarrow l_{U};\; l_{\Sigma t} \leftarrow l_{\Sigma};\; l_{Vt} \leftarrow l_{V}.
         \end{eqnarray*}
         \STATE \quad \quad Build new empty lists $l_U = list(), l_\Sigma = list(), l_V = list()$.
         \STATE \quad \quad \textbf{\textrm{for}} $j \leftarrow 1$ to $Nl$ with the step length being 2 \textbf{\textrm{do}}
         \STATE \quad \quad \quad Take two adjacent groups of elements of $l_{Ut}$, $l_{\Sigma t}$, $l_{Vt}$:
         \begin{eqnarray*}
            \quad \quad \quad l_{Ut}(j), l_{\Sigma t}(j), l_{Vt}(j), l_{Ut}(j+1), l_{\Sigma t}(j+1), l_{Vt}(j+1)).
         \end{eqnarray*}
         \STATE \quad \quad \quad Conduct \textsc{BlockMerge} function on the above variables and return the merge results $U_j$, $\Sigma_j$, $V_j$.
         \STATE \quad \quad \quad Add $U_j$, $\Sigma_j$, $V_j$ into $l_U$, $l_\Sigma$, $l_V$, respectively.
         \STATE \quad \quad \textbf{\textrm{end for}}
         \STATE \quad \quad \textbf{\textrm{if}} $Nl$ is odd \textbf{\textrm{then}}
         \STATE \quad \quad \quad Add the last elements of $l_{Ut}$, $l_{\Sigma t}$, $l_{Vt}$ into $l_{U}$, $l_{\Sigma}$, $l_{V}$, respectively.
         \STATE \quad \quad \textbf{\textrm{end if}}
         \STATE \quad \textbf{\textrm{end for}}
         \STATE \quad Return the final merged results which are defined as $\hat{U}$, $\hat{\Sigma}$, $\hat{V}$.
         \STATE \textbf{\textrm{end function}}
         \STATE
         \STATE \textbf{\textrm{function}} \textsc{BlockMerge}($U_1$, $\Sigma_1$, $V_1$, $U_2$, $\Sigma_2$, $V_2$):
         \STATE \quad Conduct \textsc{DoTruncate} function on $U_1$, $\Sigma_1$, $V_1$ and return $U_{1_{k}}$, $\Sigma_{1_{k}}$, $V_{1_{k}}$.
         \STATE \quad Conduct \textsc{DoTruncate} function on $U_2$, $\Sigma_2$, $V_2$ and return $U_{2_{l}}$, $\Sigma_{2_{l}}$, $V_{2_{l}}$.
         \STATE \quad Do SVD operation on $[U_{1_{k}}\Sigma_{1_{k}}\ U_{2_{l}}\Sigma_{2_{l}}]$ and obtain the decomposed results $U_{r}$, $\Sigma_{r}$, $\hat{V}_{r}$
         \begin{eqnarray*}
            [U_{1_{k}}\Sigma_{1_{k}}\ U_{2_{l}}\Sigma_{2_{l}}] = U_{r}\Sigma_{r}\hat{V}_{r}^{T}.
         \end{eqnarray*}
         \STATE \quad Calculate $V_{r}$ = $\hat{V}*blkdiag(V_{1_{r}}, V_{2_{l}})$.
         \STATE \quad Return $U_{r}$, $\Sigma_{r}$, $V_{r}$.
         \STATE \textbf{\textrm{end function}}
         \STATE
         \STATE \textbf{\textrm{function}} \textsc{DoTruncate}($U$, $\Sigma$, $V$):
         \STATE \quad Calculate the rank of $\Sigma$, which is $k$ = $rank$($\Sigma$).
         \STATE \quad Truncate the decomposed matrices as the rank:
         \begin{eqnarray*}
            \quad U_{k} = U(:,1:k);\Sigma_{k} = \Sigma(1:k,1:k); V_{k} = V(1:k,1:k).
         \end{eqnarray*}
         \STATE \quad Return $U_{k}$, $\Sigma_{k}$, $V_{k}$.
         \STATE \textbf{\textrm{end function}}
    \end{algorithmic}
\end{breakablealgorithm}

\end{document}